\documentclass[12pt]{article}

\usepackage{amsmath,amsfonts,amssymb}
\usepackage{amscd}
\usepackage{graphicx}    
\usepackage{hyperref}    
\usepackage{color}

\makeatletter
\renewcommand{\theequation}{\arabic{section}.\arabic{equation}}
\@addtoreset{equation}{section}
\usepackage{braket}
\usepackage{stmaryrd}
\usepackage{mathrsfs}

\newlength{\minitwocolumn}
\def\wha#1{\widehat{#1}}
\def\wtil#1{\widetilde{#1}}

\begin{document}
\begin{titlepage}
\begin{flushright}
\null \hfill Preprint TU-1078\\[3em]
\end{flushright}

\begin{center}
{\Large \bf
DFT in supermanifold formulation and group manifold as background geometry
}
\vskip 1.2cm

Ursula Carow-Watamura${}^{a,}$\footnote{E-mail:\
ursula@tuhep.phys.tohoku.ac.jp}, Noriaki Ikeda${}^{b,}$\footnote{E-mail:\
nikeda@se.ritsumei.ac.jp}, Tomokazu Kaneko${}^{a,}$\footnote{E-mail:\
t\textunderscore kaneko@tuhep.phys.tohoku.ac.jp
} 
\\~and Satoshi Watamura${}^{a,}$\footnote{E-mail:\ watamura@tuhep.phys.tohoku.ac.jp}

\vskip 0.4cm
{

\it
${}^a$
Particle Theory and Cosmology Group, \\
Department of Physics, Graduate School of Science, \\
Tohoku University \\
Aoba-ku, Sendai 980-8578, Japan \\ 

\vskip 0.4cm
${}^b$
Department of Mathematical Sciences,
Ritsumeikan University \\
Kusatsu, Shiga 525-8577, Japan \\

}
\vskip 0.4cm


\begin{abstract}
We develop the formulation of DFT on pre-QP-manifold. The consistency conditions like section condition and closure constraint are unified by a weak master equation. The Bianchi identities are also characterized by the pre-Bianchi identity. Then, the background metric and connections are formulated by using covariantized pre-QP-manifold. An application to the analysis of the DFT on group manifold is given.

\end{abstract}


\end{center}
\end{titlepage}

\newpage

\setcounter{tocdepth}{3}
\tableofcontents

\section{Introduction}
Since Siegel proposed to formulate spacetime duality in superspace \cite{Siegel:1993th,Siegel:1993xq} double field theory (DFT) has been developed and investigated by many authors \cite{Hull:2006va,Hull:2009mi,Hull:2009zb,Hohm:2010jy,Hohm:2010pp,Aldazabal:2013sca} with the aim to construct an effective theory of string with manifest T-duality symmetry. 

  The gauge transformation and diffeomorphism are unified into the generalized Lie derivative of doubled space \cite{Hull:2009zb} similarly as done in generalized geometry \cite{2002math9099H,2004math1221G,Hitchin:2005in,2010arXiv1008.0973H}.
  However, unlike generalized geometry, the gauge algebra of DFT generated by the generalized Lie derivative does not close. To achieve closure we need to constrain the algebra. One way to do so is to impose the section condition (strong constraint), a particular solution of which reduces the DFT action to the standard action of supergravity. 
On the other hand, the section condition is not the only possibility to close the DFT algebra \cite{Aldazabal:2013sca,Berman:2014jba}. 
This was demonstrated via a Scherk-Schwarz \cite{Scherk:1979zr} type compactification of DFT to the double torus, now known as generalized Scherk-Schwarz (GSS) compactification \cite{Aldazabal:2011nj,Grana:2012rr,Aldazabal:2013sca,Berman:2013cli}.  In the GSS compactification, the fields depend also on the internal space coordinate.  
 The doubled coordinates of the internal space enter in a very particular way through the GSS ansatz. It has been shown that the resulting fluxes can be identified with the structure constants of the gaugings of gauged supergravity theories, giving a geometric interpretation to them all, including the non-geometric fluxes. This is very interesting, since in the compactified direction 
the section condition is not imposed, i.e., the GSS compactification gives an alternative solution to the closure constraint.   
    
One aim of this paper is to provide an unified and geometric characterization of the closure constraint based on the supermanifold formulation.  The supermanifold considered here is a non-negatively graded QP-manifold, i.e. a supermanifold 
with a graded Poisson structure (P-structure) and 
a nilpotent vector field (Q-structure), and its generalization called a pre-QP-manifold \cite{Ikeda:2012pv, Bessho:2015tkk, 2016arXiv160801585B, Deser:2016qkw,Heller:2016abk}. 
The usefulness of the graded manifold approach to DFT has been first pointed out by Deser and Stasheff \cite{Deser:2014mxa} and has been applied to the generalized Lie derivative. 
After their work, the canonically transformed Hamiltonian functions of DFT were introduced and the T-duality transformation and Bianchi identities in DFT, including geometric and non-geometric fluxes, were formulated \cite{Bessho:2015tkk,Heller:2016abk,Heller:2017mwz}.  
Here we elaborate the closure constraint for the DFT algebra from, what we call here, the weak master equation and also formulate the Bianchi identity on a pre-QP-manifold.
  We develop a $O(D,D)$ covariant formulation of the canonical transformation and give a geometrical characterization of the Bianchi identities by introducing the pre-Bianchi identity on the doubled space. This yields a geometrical characterization of the Bianchi identities given earlier in \cite{Heller:2017mwz}.  
  
As we shall see, using the graded manifold approach, it becomes possible to characterize the closure constraint without referring explicitly to local coordinates. Thus, it is natural to consider its application to the formulation of DFT on a non-trivial background. For this purpose, we first give the covariantization of the pre-QP-manifold. 
Using this formulation, we examine the closure constraint obtained by the weak master equation.  and the corresponding Bianchi identities obtained from the pre-Bianchi identities. Then we give an application for DFT on the group manifold.  
      
This paper is organized as follows.  In section 2, we introduce the definition of pre-QP-manifolds and reformulate the QP-structures of DFT in a $O(D,D)$ covariant form. Then, we discuss the $O(D,D)$ covariant form of the canonical transformation and work out the closure constraint.  
In section 3, we introduce the generalized vielbein of DFT via a canonical transformation, thus providing a unified picture for the generalized flux and giving a systematic formulation of the Bianchi identities.  Then, we analyze the GSS compactification of DFT by $O(D,D)$ covariant canonical transformation. 
In section 4, we investigate DFT on the group manifold using the pre-QP-structures. 
The last section is devoted to conclusions and discussion.

\section{DFT on pre-QP-manifold}
In this section, we first briefly recall the original formulation of DFT
and the generalized Scherk-Schwarz compactification \cite{Aldazabal:2011nj,Geissbuhler:2011mx,Grana:2012rr,Aldazabal:2013sca}.
Then, we review the supermanifold formulation of DFT.
 The advantage of the supermanifold approach is that its QP-structure gives a concise characterization of the underlying algebra/algebroid structure.

After this preparation, we formulate the DFT
on a pre-QP-manifold and elaborate the closure condition of the generalized Lie derivative and the meaning of the section condition in this wider structure. We shall see that the failure of the closure of the algebra can be understood as a failure of the classical master equation on the QP-manifold. 
However, unlike the original supermanifold approach \cite{Heller:2016abk,Heller:2017mwz} we do not make use of the section condition, since it is not the only possibility to achieve closure of the algebra \cite{Deser:2014mxa,Aldazabal:2013sca,Geissbuhler:2013uka}. 
From the pre-QP structure formulation we obtain new criteria for the closure of the algebra of the generalized Lie derivative. 
We will also see that the $O(D,D)$ transformation is still realized as a canonical transformation. 
\subsection{Structure of DFT}
\subsubsection{The double field theory (DFT)}
DFT was developed with the aim to construct a manifest T-duality invariant formulation of the effective theory of superstring, and thus has manifest $O(D,D)$ symmetry.
In order to realize this symmetry geometrically, the winding sector is included as a Fourier transform of extra dimensions and the theory is defined on a $2D$-dimensional doubled space. 
We denote coordinates of this doubled space as $X^{\hat{M}}=(\tilde{X}_{M}, X^{M})$ where the index $M$ runs from 1 to $D$ and hatted indices are used for the doubled indices, so that $\hat{M}$ runs from 1 to $2D$. The partial derivative $\partial_{\hat{M}}=(\tilde{\partial}^{M}, \partial_{M})$ is also doubled correspondingly. 

The dynamical fields in DFT are the generalized metric ${\cal H}_{\hat{M}\hat{N}}$ and the generalized dilaton $d$. The generalized metric ${\cal H}_{\hat{M}\hat{N}}$ is a symmetric $O(D,D)$ tensor, satisfying
\begin{align}
\eta_{\hat{M}\hat{P}}{\cal H}^{\hat{P}\hat{Q}}\eta_{\hat{Q}\hat{N}}={\cal H}_{\hat{M}\hat{N}}, \quad {\cal H}^{\hat{M}\hat{N}}{\cal H}_{\hat{N}\hat{P}}=\delta_{\hat{P}}^{\hat{M}}
\end{align}
where $\eta_{\hat{M}\hat{N}}$ is an $O(D,D)$ invariant metric,
\begin{align}
\eta_{\hat{M}\hat{N}}=\begin{pmatrix}0&\delta^{M}{}_{N}\\\delta_{M}{}^{N}&0\end{pmatrix}, \label{OriginalODDinvMetric}
\end{align}
and the 2D-dimensional indices are lowered and raised by  $\eta_{\hat{M}\hat{N}}$
and its inverse $\eta^{\hat{M}\hat{N}}$, respectively.  
The generalized metric 
can be parametrized by the $D$-dimensional metric $g_{MN}$ and the 2-form field $b_{MN}$ as
\begin{align}
{\cal H}_{\hat{M}\hat{N}}=\begin{pmatrix}g^{MN}&-g^{MP}b_{PN}\\b_{MP}g^{PN}&g_{MN}-b_{MP}g^{PQ}b_{QN}\end{pmatrix}.\label{O(D)O(D)decomp}
\end{align}

We introduce the generalized vielbein ${\cal E}_{\hat{\bar{A}}}^{\ \hat{M}}$ with index $\hat{\bar{A}}$ for the local flat frame of doubled space, (i.e., $\hat{\bar{A}}$ runs from 1 to $2D$), the $O(D,D)$ invariant metric $\eta^{\hat{\bar A}\hat{\bar B}}$ to raise and lower indices in local flat frame, and an $O(1,D-1)\times O(1,D-1)$ invariant metric  
\begin{align}
S_{\hat{\bar{A}}\hat{\bar{B}}} = \begin{pmatrix}s^{{\bar{A}}{\bar{B}}}&0\\0&s_{{\bar{A}}{\bar{B}}}\end{pmatrix}~,
\end{align}
with $D$-dimensional Minkowski metrics $s^{\bar{A}\bar{B}}$ and $s_{\bar{A}\bar{B}}$.
 For the $O(D,D)$ covariance of DFT, the generalized vielbein ${\cal E}_{\hat{\bar{A}}}{}^{\hat{M}}$ is supposed to be an element of $O(D,D)$.  
With the metric $S_{\hat{\bar{A}}\hat{\bar{B}}}$, the generalized metric ${\cal H}_{\hat{M}\hat{N}}$ is written as
\begin{align}
{\cal H}_{\hat{M}\hat{N}}={\cal E}_{\ \hat{M}}^{\hat{\bar{A}}}S_{\hat{\bar{A}}\hat{\bar{B}}}{\cal E}_{\ \hat{N}}^{\hat{\bar{B}}}~.
\end{align}

With the generalized metric and generalized dilaton, the DFT action of the NS-NS sector \cite{Hohm:2010pp} can be written as: 
\begin{align}
S=\int d^{2D}Xe^{-2d}\Big(\frac{1}{8}{\cal H}^{\hat{M}\hat{N}}\partial_{\hat{M}}{\cal H}^{\hat{K}\hat{L}}\partial_{\hat{N}}{\cal H}_{\hat{K}\hat{L}}-\frac{1}{2}{\cal H}^{\hat{M}\hat{N}}\partial_{\hat{N}}{\cal H}^{\hat{K}\hat{L}}\partial_{\hat{L}}{\cal H}_{\hat{M}\hat{K}}
\cr
-2\partial_{\hat{M}}d\partial_{\hat{N}}{\cal H}^{\hat{M}\hat{N}}+4{\cal H}^{\hat{M}\hat{N}}\partial_{\hat{M}}d\partial_{\hat{N}}d\Big).\label{DFTaction}
\end{align}

In DFT the diffeomorphisms and gauge transformations in the $D$-dimensional theory are unified into the generalized Lie derivative \cite{Hull:2009zb}, 
similarly as in generalized geometry \cite{2002math9099H,2004math1221G}, and the algebra of the generalized Lie derivative characterizes the symmetry of DFT.
The above action is invariant under the gauge transformation generated by the generalized Lie derivative ${\cal L}$. 
The generalized Lie derivative of an $O(D,D)$ vector $V^{\hat M}$ with weight $w$ is defined as
\begin{align}
{\cal L}_{\Lambda}V^{\hat{M}}=\Lambda^{\hat{N}}\partial_{\hat{N}}V^{\hat{M}}+(\eta^{\hat{M}\hat{P}}\eta_{\hat{N}\hat{Q}}\partial_{\hat{P}}\Lambda^{\hat{Q}}-\partial_{\hat{N}}\Lambda^{\hat{M}})V^{\hat{N}} + w\partial_{\hat{N}}\Lambda^{\hat{N}}V^{\hat{M}}. \label{GeneralizedLieDerivative}
\end{align}
Here, $\Lambda$ is a generalized vector corresponding to the gauge parameter depending on the 2D dimensional coordinates, 
\begin{align}
\Lambda^{\hat{M}}=\begin{pmatrix}\tilde{\lambda}_{M}\cr
\lambda^{M}\end{pmatrix}. 
\end{align}
The weight of both generalized metric ${\cal H}_{\hat{M}\hat{N}}$ and generalized vielbein ${\cal E}_{\hat{\bar{A}}}{}^{\hat{M}}$ is 0, 
and the generalized Lie derivative acts on them as
\begin{align}
{\cal L}_{\Lambda}{\cal H}_{\hat{M}\hat{N}}
&=\Lambda^{\hat{P}}\partial_{\hat{P}}{\cal H}_{\hat{M}\hat{N}}-(\partial^{\hat{P}}\Lambda_{\hat{M}}-\partial_{\hat{M}}\Lambda^{\hat{P}}){\cal H}_{\hat{P}\hat{N}}
-(\partial^{\hat{P}}\Lambda_{\hat{M}}-\partial_{\hat{N}}\Lambda^{\hat{P}}){\cal H}_{\hat{M}\hat{P}},
\\ {\cal L}_{\Lambda}{\cal E}_{\hat{\bar{A}}}{}^{\hat{M}}&=\Lambda^{\hat{N}}\partial_{\hat{N}}{\cal E}_{\hat{\bar{A}}}{}^{\hat{M}}+(\partial^{\hat{M}}\Lambda_{\hat{N}}-\partial_{\hat{N}}\Lambda^{\hat{M}}){\cal E}_{\hat{\bar{A}}}{}^{\hat{N}}.
\end{align}

By construction, the generalized Lie derivative contains the usual gauge transformation of the $B$-field and the $D$-dimensional diffeomorphism: 
When we parametrize the generalized metric as \eqref{O(D)O(D)decomp} and take a special solution of the section condition, called the supergravity frame, by requiring that all fields are  independent of the dual coordinate $\tilde X_M$, the DFT action reduces to the NS-NS sector of the supergravity action. In the supergravity frame the generalized Lie derivative reduces to the $D$-dimensional diffeomorphism and the gauge transformation of the $B$-field:
\begin{align}
{\cal L}_{\Lambda}g_{MN}|_{\tilde{\partial}=0}=&L_{\lambda}g_{MN}, \\
{\cal L}_{\Lambda}b_{MN}|_{\tilde{\partial}=0}=&L_{\lambda}b_{MN}+2\partial_{[M}\tilde{\lambda}_{N]}, 
\end{align}
where $L_\lambda$ is the standard Lie derivative with respect to a $D$-dimensional vector $\lambda$.
The antisymmetrization of the generalized Lie derivative of DFT is known as C-bracket
\begin{align}
[V_{1}, V_{2}]_{C}\equiv&{\cal L}_{V_{1}}V_{2}-{\cal L}_{V_{2}}V_{1}.
\end{align}
In the supergravity frame, the C-bracket reduces to the Courant bracket. 

The obstruction for defining a generalization of the diffeomorphism by the generalized Lie derivative is that ${\cal L}_{\Lambda}$ does not satisfy the Leibniz rule. That is
\begin{align}
\Delta^M(\Lambda_1,\Lambda_2,V)
\equiv{\cal L}_{\Lambda_1}({\cal L}_{\Lambda_2}V^M)
-{\cal L}_{{\cal L}_{\Lambda_1}\Lambda_2}V^M
-{\cal L}_{\Lambda_2}{\cal L}_{\Lambda_1}V^M
\end{align}
does not vanish, which is apparently satisfied by the standard Lie derivative $L_\lambda$. 
Vanishing of the above $\Delta^M(\Lambda_1,\Lambda_2,V)$ means 
that the commutator of the
generalized Lie derivative satisfies
\begin{align}
[{\cal L}_{\Lambda_{1}}, {\cal L}_{\Lambda_{2}}]={\cal L}_{{\cal L}_{\Lambda_{1}}\Lambda_{2}}
\end{align}
which is also called the closure condition. 
Closure is always guaranteed, when the section condition is imposed
\begin{align}
\eta^{\hat{M}\hat{N}}(\partial_{\hat{M}}\Phi)(\partial_{\hat{N}}\Psi)=0
\end{align}
where $\Phi$ and $\Psi$ denote any fields and gauge parameters of DFT. {However, the section condition is not necessary to achieve} closure for the algebra.

\subsubsection{Generalized Scherk-Schwarz (GSS) 
compactification\label{GSSsec}}

In this paper, we also analyze the generalized Scherk-Schwarz (GSS) compactification of DFT where the $2D$-dimensional DFT is reduced to $(D-n)$-dimensional gauged supergravity. The generalized Lie derivative of the $2D$-dimensional DFT is twisted by the GSS ansatz and the constraint of its internal space is relaxed compared to the section condition \cite{Aldazabal:2011nj,Grana:2012rr,Aldazabal:2013sca,Berman:2013cli}. 

For the GSS compactification, it is convenient to start with the DFT action in flux formulation, which is equivalent to the DFT action in generalized metric formulation \eqref{DFTaction} up to terms which vanish under the section condition \cite{Aldazabal:2011nj}: 

\begin{align}
S=\int dX e^{-2d}{\cal R}~,\label{FluxDFTaction}
\end{align}
with
\begin{align}
{\cal R}={\cal F}_{\hat{\bar{A}}\hat{\bar{B}}\hat{\bar{C}}}{\cal F}_{\hat{\bar{D}}\hat{\bar{E}}\hat{\bar{F}}}\Big(\frac{1}{4}S^{\hat{\bar{A}}\hat{\bar{D}}}\eta^{\hat{\bar{B}}\hat{\bar{E}}}\eta^{\hat{\bar{C}}\hat{\bar{F}}}-\frac{1}{12}S^{\hat{\bar{A}}\hat{\bar{D}}}S^{\hat{\bar{B}}\hat{\bar{E}}}S^{\hat{\bar{C}}\hat{\bar{F}}}-\frac{1}{6}\eta^{\hat{\bar{A}}\hat{\bar{D}}}\eta^{\hat{\bar{B}}\hat{\bar{E}}}\eta^{\hat{\bar{C}}\hat{\bar{F}}}\Big)
\cr
+{\cal F}_{\hat{\bar{A}}}{\cal F}_{\hat{\bar{B}}}\Big(\eta^{\hat{\bar{A}}\hat{\bar{B}}}-S^{\hat{\bar{A}}\hat{\bar{B}}}\Big)~,
\end{align}
where $\eta^{\hat{\bar{A}}\hat{\bar{B}}}$ is an $O(D,D)$ invariant metric in the local frame. 
The generalized fluxes ${\cal F}_{\hat{\bar{A}}\hat{\bar{B}}\hat{\bar{C}}}$ and ${\cal F}_{\hat{\bar{A}}}$ are defined as
\begin{align}
{\cal F}_{\hat{\bar{A}}\hat{\bar{B}}\hat{\bar{C}}}:=&{\cal E}_{\hat{\bar{C}}\hat{M}}{\cal L}_{{\cal E}_{\hat{\bar{A}}}}{\cal E}_{\hat{\bar{B}}}^{\ \hat{M}}=3\Omega_{[\hat{\bar{A}}\hat{\bar{B}}\hat{\bar{C}}]}, \label{GeneralizedFlux}\\
{\cal F}_{\hat{\bar{A}}}:=&-e^{2d}{\cal L}_{{\cal E}_{\hat{\bar{A}}}}e^{-2d}=\Omega^{\hat{\bar{B}}}_{\ \ \hat{\bar{B}}\hat{\bar{A}}}+2{\cal E}_{\hat{\bar{A}}}^{\ \hat{M}}\partial_{\hat{M}}d,\label{GeneralizedFluxForScalar}
\end{align}
where $\Omega_{\hat{\bar{A}}\hat{\bar{B}}\hat{\bar{C}}}:={\cal E}_{\hat{\bar{A}}}^{\ \hat{M}}\partial_{\hat{M}}{\cal E}_{\hat{\bar{B}}}^{\ \hat{N}}{\cal E}_{\hat{\bar{C}}\hat{N}}$ is the 
generalized Weitzenb\"ock connection. Note that, if we parametrize the $O(D,D)$ vielbein by the $D$-dimensional metric $g_{MN}$ and 2-form $b_{MN}$, we find the $H$-flux and the geometric $f$-flux in ${\cal F}_{\hat{\bar{A}}\hat{\bar{B}}\hat{\bar{C}}}$. Alternatively, if we take the parametrization by the fields $\tilde{g}^{MN}$ and the two-vector $\beta^{MN}$ we obtain the so-called non-geometric fluxes. This result provides a unified picture of fluxes discussed in \cite{Heller:2016abk}. 

As in the original Scherk-Schwarz compactification ansatz \cite{Scherk:1979zr}, the generalized Scherk-Schwarz (GSS) compactification of DFT splits the $2D$-dimensional target space with coordinate $X$ into $2d$-dimensional external space with coordinate $\mathbb{X}$ and $2(D-d)$-dimensional internal space with coordinate $\mathbb{Y}$, as $X=(\mathbb{X}, \mathbb{Y})$. 
Then, the ansatz for the generalized vielbein ${\cal E}^{\hat{\bar{A}}}_{\ \hat{M}}(X)$, generalized dilaton $d(X)$ and gauge parameter $\Lambda^{\hat{M}}(X)$ are \cite{Geissbuhler:2011mx,Aldazabal:2011nj},
\begin{align}
{\cal E}^{\hat{\bar{A}}}_{\ \hat{M}}(X)&=\widehat{E}^{\hat{\bar{A}}}{}_{\hat{I}}(\mathbb{X})U^{\hat{I}}{}_{\hat{M}}(\mathbb{Y}), \quad\label{GSSansatz} 
\\d(X)&=\widehat{d}(\mathbb{X})+\bar{d}(\mathbb{Y}), \quad  \\ \Lambda^{\hat{M}}(X)&=\widehat{\Lambda}^{\hat{I}}(\mathbb{X})U_{\hat{I}}{}^{\hat{M}}(\mathbb{Y}). 
\end{align}
where the fields with hat, $\widehat{E}^{\hat{\bar{A}}}{}_{\hat{I}}(\mathbb{X})$, $\widehat{d}(\mathbb{X})$ and $\widehat{\Lambda}^{\hat{I}}(\mathbb{X})$ are the vielbein, dilaton and gauge parameter of the reduced theory, respectively.  
We use the characters $\hat{I},\hat{J},\hat{K},\hat{L}$ and $\hat{H}$ for the indices of the reduced theory, and the corresponding $O(D,D)$ metric $\eta^{\hat{I}\hat{J}}$. 
The matrix $U_{\hat{I}}{}^{\hat{M}}(\mathbb{Y})$ and its inverse $U^{\hat{I}}{}_{\hat{M}}(\mathbb{Y})$ which give the GSS twist are elements of $O(D,D)$ and $\bar{d}(\mathbb{Y})$ is a scalar. The twist matrices are assumed to satisfy 
\begin{align}
U_{\hat{I}}{}^{\hat{M}}\partial_{\hat{M}}g(\mathbb{X})=\partial_{\hat{I}}g(\mathbb{X})~~,\label{GSScondition}
\end{align}
for any field of the reduced theory, which means that the twist occurs only in the internal space, and preserves the Lorentz invariance in the external spacetime of the reduced theory.

By substituting the above GSS ansatz \eqref{GSSansatz} into the generalized fluxes \eqref{GeneralizedFlux} and \eqref{GeneralizedFluxForScalar}, 
we obtain the twisted generalized fluxes:
\begin{align}
{\cal F}_{\hat{\bar{A}}\hat{\bar{B}}\hat{\bar{C}}}=&\widehat{F}_{\hat{\bar{A}}\hat{\bar{B}}\hat{\bar{C}}}+f_{\hat{I}\hat{J}\hat{K}}\widehat{E}_{\hat{\bar{A}}}{}^{\hat{I}}\widehat{E}_{\hat{\bar{B}}}{}^{\hat{J}}\widehat{E}_{\hat{\bar{C}}}{}^{\hat{K}},\label{GSStwistedFlux}\\
{\cal F}_{\hat{\bar{A}}}=&\widehat{F}_{\hat{\bar{A}}}+f_{\hat{I}}\widehat{E}_{\hat{\bar{A}}}{}^{\hat{I}}
\end{align}
where $\widehat{F}_{\hat{\bar{A}}\hat{\bar{B}}\hat{\bar{C}}}$ and $\widehat{F}_{\hat{\bar{A}}}$ are generalized fluxes obtained from $\widehat{E}_{\hat{\bar{A}}}{}^{\hat{I}}$ and $\widehat{d}$ containing the dynamical field in the external spacetime. They have the same structure as the original generalized fluxes \eqref{GeneralizedFlux} and \eqref{GeneralizedFluxForScalar}, namely
\begin{align}
\widehat{F}_{\hat{\bar{A}}\hat{\bar{B}}\hat{\bar{C}}}=&3\widehat{\Omega}_{[\hat{\bar{A}}\hat{\bar{B}}\hat{\bar{C}}]}, \label{ExtFlux}\\
\widehat{F}_{\hat{\bar{A}}}=&\widehat{\Omega}^{\hat{\bar{B}}}{}_{\hat{\bar{B}}\hat{\bar{A}}}+2\widehat{E}_{\hat{\bar{A}}}{}^{\hat{I}}\partial_{\hat{I}}\widehat{d},
\end{align}
where $\widehat{\Omega}_{\hat{\bar{A}}\hat{\bar{B}}\hat{\bar{C}}}=\widehat{E}_{\hat{\bar{A}}}{}^{\hat{I}}\partial_{\hat{I}}\widehat{E}_{\hat{\bar{B}}}{}^{\hat{J}}\widehat{E}_{\hat{\bar{C}}\hat{J}}$. 
The effect of the twist in the internal space is confined to the generalized fluxes $f_{\hat{M}\hat{N}\hat{P}}$ and $f_{\hat{M}}$ 
defined by the matrix $U_{\hat{I}}{}^{\hat{M}}(\mathbb{Y})$ as
\begin{align}
f_{\hat{I}\hat{J}\hat{K}}:=&3\wtil{\Omega}_{[\hat{I}\hat{J}\hat{K}]},\label{IntFlux}\\
f_{\hat{I}}:=&\wtil{\Omega}^{\hat{J}}{}_{\hat{J}\hat{I}}+2U_{\hat{I}}{}^{\hat{M}}\partial_{\hat{M}}\bar{d},
\end{align}
where $\wtil{\Omega}_{\hat{I}\hat{J}\hat{K}}=U_{\hat{I}}{}^{\hat{M}}\partial_{\hat{M}}U_{\hat{J}}{}^{\hat{N}}U_{\hat{K}\hat{N}}$. In the GSS compactification, these fluxes $f_{\hat{I}\hat{J}\hat{K}}, f_{\hat{I}}$ are assumed to be constant in the same way as in the original Scherk-Schwarz compactification. Imposing this ansatz the resulting theory becomes independent of the internal coordinate and the DFT action is reduced to the effective action of the so-called gauged double field theory (GDFT) \cite{Grana:2012rr}. 

The gauge algebra of GDFT is inherited from the original DFT. The corresponding generalized Lie derivative of a generalized vector $\hat V(\mathbb{X})$ in the reduced theory can be derived from the one of the original DFT by substituting the GSS ansatz as
\begin{align}
\widehat{{\cal L}}_{\wha{\Lambda}(\mathbb{X})}\wha{V}^{\hat{I}}(\mathbb{X})={\cal L}_{\wha{\Lambda}(\mathbb{X})}\wha{V}^{\hat{I}}(\mathbb{X})+f^{\hat{I}}{}_{\hat{J}\hat{K}}\wha{\Lambda}^{\hat{J}}(\mathbb{X})\wha{V}^{\hat{K}}(\mathbb{X}). \label{GDFTLieDer}
\end{align}
Thus, the generalized Lie derivative $\widehat{\cal L}$ of GDFT is defined on the reduced fields and gauge parameter which depend only on the external spacetime coordinate $\mathbb{X}$. 

The algebra of $\widehat{\cal L}$ closes by the closure constraint for GDFT fields and the Jacobi identity of the structure constant $f_{\hat{I}\hat{J}}{}^{\hat{K}}$, 
\begin{align}
\partial_{\hat{I}}\wha{V}(\mathbb{X})\partial^{\hat{I}}\wha{W}(\mathbb{X})=0, \quad f_{[\hat{I}\hat{J}}{}^{\hat{H}}f_{\hat{K}]\hat{L}\hat{H}}=0. 
\end{align}
Note that the condition for closure for the internal space  
is relaxed compared to the solution by using the section condition.

\subsection{Supermanifold formulation of DFT}

In this section, first we summarize briefly the supermanifold formulation of DFT given in the refs. \cite{Deser:2014mxa, Bessho:2015tkk, Heller:2016abk}. 
Then, we define a pre-QP-manifold which is a generalization of QP-manifold
that can describe the algebraic structure of DFT. On the pre-QP-manifold the closure condition is relaxed, which fits to the generalized Lie derivative and gives a new understanding of the section condition. 

The algebra on a QP-manifold is a kind of simplified graded algebra as used in the BV- and BRST- formalism. See e.g. \cite{Ikeda:2012pv} and references therein. There are two structures on this supermanifold. One is called P-structure, which specifies a graded Poisson bracket and thus defines the derivation and the vector field on the supermanifold. The other is called Q-structure, which is a nilpotent Hamiltonian vector field $\cal Q$, an analogue of the BRST charge, the nilpotency of which is imposed by the classical master equation. It is known that a QP-manifold of degree 2 gives a concise definition of a Courant algebroid \cite{Roytenberg:1999} and thus fits to the formulation of the generalized geometry \cite{2002math9099H,2004math1221G}.

The QP-manifold, however, is too strict to apply to DFT, as one can imagine from the relation between the double geometry and generalized geometry or, more concretely, from the relation between the C-bracket and Courant bracket. 

We shall see that on a pre-QP-manifold we have more freedom to accommodate the 
DFT algebra, and we can obtain a new closure condition for the generalized Lie derivative, 
which gives another alternative to the section condition used in the original DFT.

\subsubsection{Pre-QP-manifold and derived bracket\label{PreQPMfdAndDerivedBracketSec}}
A P-manifold of degree $n$ is a pair $({\cal M}, \omega)$, where
${\cal M}$ is a graded manifold with $\mathbb{Z}$ graded coordinates
and $\omega$ is a graded symplectic form of degree $n$ defining a P-structure.
In our context, we always consider the case with non-negative $n$.
The graded Poisson bracket $\{-,-\}$ is calculated from the graded symplectic form $\omega$ as
\begin{align}
\{f,g\}\equiv (-1)^{|f|+n+1}\iota_{X_{f}}\iota_{X_{g}}\omega~,
\end{align}
for $f,g\in C^{\infty}({\cal M})$, where $|\cdot |$ denotes the degree. Here $X_{f}$ is a Hamiltonian vector field defined by $\iota_{X_f} \omega = - df$. Using graded Darboux coordinates $(q^a, p_a)$, the graded Poisson bracket has a local coordinate representation given by
\begin{align}
\{f, g\}=\frac{f\overleftarrow{\partial}}{\partial q^{a}}\frac{\overrightarrow{\partial}g}{\partial p_{a}}-(-1)^{|q||p|}\frac{f\overleftarrow{\partial}}{\partial p_{a}}\frac{\overrightarrow{\partial}g}{\partial q^{a}}~.
\end{align}
The graded Poisson bracket is of degree $-n$ and satisfies the following 
graded version of skew symmetricity, Leibniz rule and Jacobi identity,
\begin{align}
\{f,g\}=&(-1)^{(|f|-n)(|g|-n)+1}\{g,f\},\label{PBSymmetry}\\
\{f,gh\}=&\{f,g\}h+(-1)^{(|f|-n)|g|}g\{f,h\},\label{PBLeibniz}\\
\{f,\{g,h\}\}=&\{\{f,g\},h\}+(-1)^{(|f|-n)(|g|-n)}\{g,\{f,h\}\}. \label{PBJacobi}
\end{align}

In the P-manifold we can define the following canonical transformation. Let $\alpha\in C^{\infty}({\cal M})$ be a degree $n$ function. Then, a canonical transformation generated by $\alpha$ is defined by the following exponential adjoint action,
\begin{align}
e^{\delta_{\alpha}}f=f+\{f,\alpha\}+\frac{1}{2!}\{\{f,\alpha\},\alpha\}+\cdots\label{CanonicalTransf}
\end{align} 
where $f\in C^{\infty}({\cal M})$ is any smooth function. Since the degree of the Poisson bracket $\{\cdot,\cdot\}$ is $-n$, the operation $e^{\delta_{\alpha}}$ is degree preserving. Moreover, one can prove that this operation is a canonical transformation satisfying
\begin{align}
e^{\delta_{\alpha}}\{f,g\}=\{e^{\delta_{\alpha}}f, e^{\delta_{\alpha}}g\}. 
\label{canonicaltransformationtwisting}
\end{align}
for any smooth function $f,g$ on ${\cal M}$.

On the P-manifold, we can also define a Q-structure by
specifying a degree $n+1$ function $\Theta\in C^{\infty}({\cal M})$. 
 This function $\Theta$ defines a degree $1$ graded vector field ${\cal Q}$ as
\begin{align}
{\cal Q}f=\{\Theta, f\}. \label{Homologicalvectorfield}
\end{align}
If this vector is nilpotent, i.e., ${\cal Q}^{2}=0$, it is called a homological vector field, and the corresponding $\Theta$ is called a Hamiltonian function or a homological function associated to ${\cal Q}$.
The condition ${\cal Q}^{2}=0$ is equivalent to the classical master equation,
\begin{align}
\{\Theta, \Theta\}=0. \label{ClassicalMasterEquation}
\end{align}
A triple $({\cal M}, \omega, \Theta)$, i.e., a P-manifold with a homological function $\Theta$ is called a QP-manifold.
The algebra on the $n=2$ QP-manifold is equivalent to a Courant algebroid and as we mentioned, it is too restrictive for DFT.

The pre-QP-manifold which is applicable to DFT is obtained by relaxing the classical master equation.
The pair $(\omega, {\cal Q})$ is called a pre-QP-structure and the triple $({\cal M}, \omega, {\cal Q})$ is called a pre-QP-manifold 
without requiring the nilpotency ${\cal Q}^2=0$ of the vector field 
${\cal Q}$ in (\ref{Homologicalvectorfield}). Consequently, the classical master equation does not vanish,
\begin{equation}
\{ \Theta,\Theta \}\not=0~.
\end{equation} 
Nevertheless, also in this case we call $\Theta$ the Hamiltonian function of the pre-QP-manifold.  As we shall see, in the pre-QP-manifold approach, the classical master equation is replaced by another condition, and the section condition is just one of the possible solutions to that condition.

Since the definition of a canonical transformation is independent of the Q-structure, the equation \eqref{canonicaltransformationtwisting} still holds and thus, 
\begin{align}
e^{\delta_{\alpha}}\{\Theta,\Theta\}=\{e^{\delta_{\alpha}}\Theta, e^{\delta_{\alpha}}\Theta\}.
\end{align}
This means that the classical master equation is preserved 
by the canonical transformation,
i.e., if $\Theta$ satisfies the classical master equation, 
then so does 
$e^{\delta_{\alpha}}\Theta$. 

There is another important object in the supermanifold formulation, namely, the derived bracket.  
For any degree $n+1$ function $\Theta$ on a P-manifold, the derived bracket is defined as a bilinear operation for $f,g\in C^{\infty}({\cal M})$:
\begin{align}
[f,g]:=-\{\{f,\Theta\},g\}~.
\end{align}
The derived bracket $[-,-]$ is not necessarily graded antisymmetric.
Using the graded anti-symmetry \eqref{PBSymmetry} and the Jacobi identity \eqref{PBJacobi} of the Poisson bracket, we obtain 
the following identity of the derived bracket
for any $f, g, h \in C^{\infty}({\cal M})$,
\begin{align}
[f,[g,h]]=&\{\{f,\Theta\},\{\{g,\Theta\},h\}\}\nonumber\\
=&\{\{\{\{f,\Theta\},g\},\Theta\},h\}+(-1)^{(|f|+n+1)(|g|+n+1)}\{\{g,\Theta\},\{\{f,\Theta\},h\}\}\nonumber\\
&~~~~~~~~~~~~~~~~~~~~~~~~~~~~~
+(-1)^{(|f|+n+1)(|g|+n)}\frac{1}{2}\{\{g,\{f,\{\Theta,\Theta\}\}\},h\}\cr
=&[[f,g],h]+(-1)^{(|f|+n+1)(|g|+n+1)}[g,[f,h]]
\cr~~~~&~~~~~~~~~~~~~~~~~~~~~~
+(-1)^{|g|+n}\frac{1}{2}\{\{\{\{\Theta,\Theta\},f\},g\},h\}.
\label{Jacobiofderivedbracket01}
\end{align}

If the classical master equation $\{\Theta,\Theta\}=0$ is satisfied,
the derived bracket $[\cdot, \cdot]$ satisfies the following Leibniz identity
\footnote{If the bracket $[\cdot , \cdot]$ is graded antisymmetric, \eqref{Jacobiofderivedbracket02} is the graded Jacobi identity. In general, it is called the graded Leibniz identity. In this paper we simply call it the Leibniz identity} of
degree $-n+1$,
\begin{align}
[f,[g,h]] =&[[f,g],h]+(-1)^{(|f|-n+1)(|g|-n+1)}[g,[f,h]].
\label{Jacobiofderivedbracket02}
\end{align}
An important point to apply the pre-QP-manifold to DFT is the fact that the condition
\footnote{This condition is also derived in \cite{2016arXiv160801585B} as the condition on the pre-QP-manifold, and later in the context of an $L_{\infty}$ algebra in \cite{Deser:2016qkw}. 
}
\begin{align}
\{\{\{\{\Theta,\Theta\},f\},g\},h\}=0 \label{WeakMasterEQ}
\end{align}
is sufficient to satisfy the Leibniz identity \eqref{Jacobiofderivedbracket02}.
Thus, we can relax the classical master equation to the equation 
(\ref{WeakMasterEQ}) for closure of the derived bracket. 
We call the condition (\ref{WeakMasterEQ}) the weak master equation.
In general, $\{\{\{\{\Theta,\Theta\},f\},g\},h\} \neq 0$, however, $\{\{\{\{\Theta,\Theta\},f\},g\},h\}=0$ can be satisfied by restricting the space of functions instead of requiring the classical master equation. 
This observation is important to understand a unified picture for the closure condition in DFT and the role of the section condition.

The weak master equation \eqref{WeakMasterEQ} is changed by a twist.
We consider a twist of a Hamiltonian function by a function $\alpha$, $\Theta \rightarrow \Theta^{\prime} = e^{\delta_{\alpha}} \Theta$.
Then, the closure condition changes to
\begin{eqnarray}
&& 
\{\{\{\{\Theta^{\prime}, \Theta^{\prime}\}, f\}, g\}, h\} =0,
\end{eqnarray}
Using Eq.\eqref{canonicaltransformationtwisting}, this condition is written as 
\begin{eqnarray}
&& 
e^{\delta_{\alpha}} \{\{\{\{\Theta, \Theta\}, e^{- \delta_{\alpha}} f\}, e^{- \delta_{\alpha}} g\}, e^{- \delta_{\alpha}} h\} =0,
\end{eqnarray}
If functions $f, g, h$ satisfying the weak master equation \eqref{WeakMasterEQ} are elements of $C^{\infty}({\cal N})$, where ${\cal N}$ is a graded submanifold of the pre-QP-manifold ${\cal M}$, then the closure condition is satisfied by the functions twisted by $e^{- \delta_{\alpha}}$.
This is demonstrated in the following.
Especially, this gives the proper condition of the structure constant in the GSS twist in section \ref{GSSasCanonicalTransf}.

\subsubsection{Generalized Lie derivative on pre-QP-manifold\label{QPDFT}}
In the following we give the formulation of the generalized Lie derivative of DFT on the pre-QP-manifold. It gives a characterization of the closure of the gauge algebra
using the weak master equation.

In order to construct the supermanifold formulation of DFT, we take a $2D$ dimensional doubled space, $\widetilde{M}\times M$, and
an $n=2$ graded symplectic manifold ${\cal M}=T^{*}[2]T[1](\widetilde{M}\times M)$. 
We introduce local coordinates on ${\cal M}$ as $(X^{\hat{M}},Q^{\hat{M}}, P_{\hat{M}}, \Xi_{\hat{M}})$ of degree $(0,1,1,2)$, respectively. $X^{\hat{M}}=(\tilde{X}_{M}, X^{M})$ is a general coordinate on the base manifold $\wtil{M}\times M$. $Q^{\hat{M}}=(\tilde{q}_{M}, q^{M})$ is a degree one fiber coordinate of the tangent bundle $T[1](\wtil{M}\times M)$. 
$\Xi_{\hat{M}}=(\tilde{\xi}^{M}, \xi_{M} )$ and $P_{\hat{M}}=(\tilde{p}^{M}, p_{M})$ are the fiber coordinates of the cotangent bundle, which are conjugate to $X^{\hat{M}}$ and $Q^{\hat{M}}$, respectively. 

The symplectic structure on ${\cal M}$ is 
\begin{align}
\omega=\delta X^{\hat{M}}\wedge \delta \Xi_{\hat{M}} + \delta Q^{\hat{M}}\wedge \delta P_{\hat{M}},
\end{align}
which leads to the following graded Poisson brackets,
\begin{align}
&\{X^{\hat{M}},\Xi_{\hat{N}}\}= -\{\Xi_{\hat{N}} ,X^{\hat{M}}\}=\delta^{\hat{M}}_{\hat{N}},\\
&\{Q^{\hat{M}},P_{\hat{N}}\}= \{P_{\hat{N}} ,Q^{\hat{M}}\}=\delta^{\hat{M}}_{\hat{N}}. 
\end{align}
{In order to formulate DFT on the pre-QP-manifold, it is necessary to introduce the following coordinates $(Q',P')$ which we call the DFT basis:}
\begin{equation}
Q'^{\hat{M}}:=\frac{1}{\sqrt[]{2}}(Q^{\hat{M}}-\eta^{\hat{M}\hat{N}}P_{\hat{N}})~~~,~~
P'_{\hat{M}}:=\frac{1}{\sqrt[]{2}}(P_{\hat{M}}+\eta_{\hat{M}\hat{N}}Q^{\hat{N}}),\label{DFTbasis}
\end{equation}
where $\eta_{\hat{M}\hat{N}}$ and $\eta^{\hat{M}\hat{N}}$ are the $O(D,D)$ invariant metrics defined in \eqref{OriginalODDinvMetric}.
In the DFT basis, the Poisson brackets are
\begin{align}
&\{Q'^{\hat{M}},Q'^{\hat{N}}\}=\eta^{\hat{M}\hat{N}},\quad \{P'_{\hat{M}},P'_{\hat{N}}\}=\eta_{\hat{M}\hat{N}},\quad \{Q'^{\hat{M}},P'_{\hat{N}}\}=0.
\end{align}

The relation between the supermanifold formulation and 
the standard DFT description is established 
by an injection map $j'$ from the generalized tangent and cotangent bundle
over the doubled spacetime $\wha{M}=\wtil{M}\times M$
to the graded manifold $\cal{M}$,
$j': \wha{M} \oplus T{\wha{M}} \oplus T{\wha{M}} \oplus T^*{\wha{M}} 
\to {\cal M}$ with degree shifting
 as follows,
\begin{align}
j': \Big(X^{\hat{M}}, \partial_{\hat{M}}, \partial_{\hat{M}}, d X^{\hat{M}}\Big)\longmapsto (X^{\hat{M}}, \Xi_{\hat{M}}, P'_{\hat{M}},  Q'^{\hat{M}}). 
\end{align}
Then, the push forward $j'_{*}$ and pullback $j'^{*}$ 
of a generalized vector field $V = V^{\hat{M}}\partial_{\hat{M}}$ is defined as 
\begin{align}
&j'_{*}:V = V^{\hat{M}}\partial_{\hat{M}} \longmapsto  
V^{\hat{M}}P'_{\hat{M}},
\\
&j'^{*}:V^{\hat{M}}P'_{\hat{M}}  \longmapsto 
V^{\hat{M}}\partial_{\hat{M}},
\end{align}
where $V^{\hat{M}}P'_{\hat{M}} \in C^{\infty}({\cal M})$.
We can also define the similar map for a $1$-form on $\wha M$.
In the following, we identify the generalized vectors $V^{\hat{M}}\partial_{\hat{M}}$ with $V^{\hat{M}}P'_{\hat{M}}$ and omit to write $j'_{*}$ and $j'^{*}$ for simplicity. 
The $O(D,D)$ invariant inner product for the generalized vector fields
 $V$ and $V'$ can be defined by using the graded Poisson bracket of
the corresponding functions on the supermanifold,
\begin{align}
\braket{V,V'}=\{V,V'\}. 
\end{align}

In order to formulate the generalized Lie derivative of DFT by using the pre-QP-manifold, we take the following degree $3$ function 
constructed by $\Xi_{\hat M}$ and $P'_{\hat M}$,
\begin{align}
\Theta_{0}=\eta^{\hat{M}\hat{N}}\Xi_{\hat{M}}P'_{\hat{N}},
\label{QstructureOfDFT}
\end{align}
It is straightforward to show that the derived bracket using this $\Theta_{0}$ gives the 
generalized Lie derivative on a generalized vector field $V$,
\begin{align}
{\cal L}_{\Lambda}V=
[\Lambda, V]_D =-\{\{\Lambda,\Theta_{0}\},V\}.
\end{align}
as well as the generalized anchor map for a function $f$
\begin{align}
\rho_{\Lambda}(f)=-\{\{\Lambda, \Theta_{0}\},f\}. 
\end{align}
Thus, the $\Theta_0$ in (\ref{QstructureOfDFT}) is a suitable 
Hamiltonian for the supermanifold formulation of DFT.

On the other hand, it is easy to see that the classical master equation
 of  
 $\Theta_{0}$, \eqref{ClassicalMasterEquation} does not vanish:
\begin{align}
\{\Theta_{0}, \Theta_{0}\}=\eta^{\hat{M}\hat{N}}\Xi_{\hat{M}}\Xi_{\hat{N}}.
\label{Poissonbracketofthetazero}
\end{align}
This means that for the algebra of the derived bracket with $\Theta_0$, i.e., the generalized Lie derivative defined on a pre-QP-manifold, the closure condition  
should be generalized as discussed in the previous subsection.  There, we have shown that the condition for the Leibniz identity of the derived bracket on a pre-QP-manifold is given by the weak master equation,  
\eqref{WeakMasterEQ}. In the DFT case, the weak master equation is 
\begin{align}
{\{\{\{\{\Theta_{0},\Theta_{0}\},f\},g\},h\}}=0\label{closureconstraint}. 
\end{align}

Eq.\eqref{closureconstraint} gives the closure constraint of the generalized 
Lie derivative as follows:
The gauge parameter of generalized Lie derivative is a generalized vector, thus on the pre-QP-manifold, we require the weak master equation for the generalized vectors $V_{1},V_{2},V_{3}$:
\begin{align}
{\{\{\{\{\Theta_{0},\Theta_{0}\},V_{1}\},V_{2}\},V_{3}\}}=0. 
\end{align}
This condition is equivalent to the closure constraint of the generalized Lie derivative in DFT:  
\begin{align}
\partial^{\hat{M}}V_{1}^{\hat{N}}V_{2\hat{N}}\partial_{\hat{M}}V_{3}^{\hat{Q}}-2\partial^{\hat{M}}V_{1}^{[\hat{P}}\partial_{\hat{M}}V_{2}^{\hat{Q}]}V_{3\hat{P}}=0.
\end{align}

\section{Bianchi identity and GSS twist in {pre-}QP-formalism\label{CanonicalTransfSec}}

In this section, first we analyze the canonical transformation on pre-QP-manifold. We show that the generalized vielbein can be introduced by a canonical transformation from the general tangent frame to the local flat frame. 
The definition
 of the generalized vielbein as an $O(D,D)$ covariant canonical transformation provides a unified picture for the generalized flux
 and gives a systematic way to formulate the Bianchi identities.
 
We also show that the twist operation in the GSS compactification of DFT can be analyzed by an $O(D,D)$ covariant canonical transformation. We show that the canonical transformation of the
 Hamiltonian function of DFT gives the twisted Hamiltonian function which realizes the generalized Lie derivative of gauged DFT (GDFT). 

\subsection{Canonical transformation on supermanifold}

To define the generalized vielbein and flux of DFT, we need to introduce a local flat frame.
Correspondingly, we generalize the supermanifold to 
${\mathcal M}=T^*[2](T[1](\widetilde{M}\times M)+V[1])$ where V[1] is a vector space corresponding to the local flat frame with degree $1$ shift. 
We introduce the degree $1$ coordinates $(\bar Q^{\hat{\bar A}}, \bar P_{\hat{\bar A}})$ of $T^*[2]V[1]$, in adition to the  coordinates $(X^{\hat M},\Xi_{\hat M}, Q^{\hat{M}}, P_{\hat{M}})$ of $T^*[2]T[1](\widetilde{M}\times M)$.
We also define the corresponding DFT basis coordinates $Q'^{\hat{\bar A}}$ and $P'_{\hat{\bar A}}$, \begin{align}
\bar{Q}'^{\hat{\bar{A}}}:=\frac{1}{\sqrt[]{2}}(\bar{Q}^{\hat{\bar{A}}}-\eta^{\hat{\bar{A}}\hat{\bar{B}}}\bar{P}_{\hat{\bar{B}}})~~,~~\bar{P}'_{\hat{\bar{A}}}:=\frac{1}{\sqrt[]{2}}(\bar{P}_{\hat{\bar{A}}}+\eta_{\hat{\bar{A}}\hat{\bar{B}}}\bar{Q}'^{\hat{\bar{B}}})
\end{align}
 in the same way as the DFT basis, $Q'^{\hat{M}}$ and $P'_{\hat{M}}$ defined in (\ref{DFTbasis}). Here, $\eta_{\hat{\bar{A}}\hat{\bar{B}}}$ and its inverse $\eta^{\hat{\bar{A}}\hat{\bar{B}}}$ are $O(D,D)$ invariant metrics. 
The Poisson brackets for $\bar{Q}'^{\hat{\bar{A}}}$ and $\bar{P}'_{\hat{\bar{A}}}$ are
\begin{align}
\{\bar{Q}'^{\hat{\bar{A}}},\bar{Q}'^{\hat{\bar{B}}}\}=\eta^{\hat{\bar{A}}\hat{\bar{B}}}, \quad \{\bar{P}'_{\hat{\bar{A}}},\bar{P}'_{\hat{\bar{B}}}\}=\eta_{\hat{\bar{A}}\hat{\bar{B}}}, \quad \{\bar{Q}'^{\hat{\bar{A}}}, \bar{P}'_{\hat{\bar{B}}}\}=0. 
\end{align}

Canonical transformations are generated by degree 2 functions on the $n=2$ supermanifold.  All possible degree 2 functions made from $P'$ and $\bar{P}'$ are given by linear combinations of the following functions
\begin{align}
{\cal A}:={\cal A}_{\hat{\bar{A}}}^{\ \hat{M}}\eta^{\hat{\bar{A}}\hat{\bar{B}}}P'_{\hat{M}}\bar{P}'_{\hat{\bar{B}}}, \quad t:=t_{\hat{P}}^{\ \hat{M}}\eta^{\hat{N}\hat{P}}P'_{\hat{M}}P'_{\hat{N}}, \quad \bar{t}:=\bar{t}_{\hat{\bar{A}}}^{\ \hat{\bar{B}}}\eta^{\hat{\bar{C}}\hat{\bar{A}}}\bar{P}'_{\hat{\bar{B}}}\bar{P}'_{\hat{\bar{C}}},\label{CanonicalTransfFunc}
\end{align}
where the parameters ${\cal A}_{\hat{\bar{A}}}^{\ \hat{M}}, t_{\hat{P}}^{\ \hat{M}}$ and $\bar{t}_{\hat{\bar{A}}}^{\ \hat{\bar{B}}}$ are functions of $X^{\hat{M}}=(\tilde{X}_{M}, X^{M})$. The function ${\cal A}$ generates $GL(2D)$ transformations in general, and $t$ and $\bar{t}$ generate $O(D,D)$ transformations in each frame. 
We can, in principle, also consider the degree 2 functions including $Q'$ and $\bar{Q}'$. However, we do not need them in the following discussions. We could also introduce the degree 2 function made from $\Xi_{\hat{M}}$, but they do not generate the canonical transformations of the fiber directions. 

We discuss the canonical transformations generated by each function in \eqref{CanonicalTransfFunc} in the following.

\subsubsection{Canonical transformation by ${\cal A}$ and generalized flux\label{CanonicaTransfByE}}
The canonical transformation generated by ${\cal A}$ is 
\begin{align}
e^{\delta_{{\cal A}}}P'_{\hat{M}}=&\sum_{n=0}^{\infty}\frac{(-1)^{{n}}}{(2n)!}(K^{n})_{\hat{M}}{}^{\hat{N}}P'_{\hat{N}}+\sum_{n=0}^{\infty}\frac{(-1)^{{n}}}{(2n+1)!}(K^{n})_{\hat{M}}^{\ \hat{N}}\eta_{\hat{N}\hat{P}}{\cal A}_{\hat{\bar{A}}}{}^{\hat{P}}\eta^{\hat{\bar{A}}\hat{\bar{B}}}\bar{P}'_{\hat{\bar{B}}},\\
e^{\delta_{{\cal A}}}\bar{P}'_{\hat{\bar{A}}}=&\sum_{n=0}^{\infty}\frac{(-1)^{{n}}}{(2n)!}(L^{n})_{\hat{\bar{A}}}{}^{\hat{\bar{B}}}\bar{P}'_{\hat{\bar{B}}}-\sum_{n=0}^{\infty}\frac{(-1)^{{n}}}{(2n+1)!}(L^{n})_{\hat{\bar{A}}}{}^{\hat{\bar{B}}}{\cal A}_{\hat{\bar{B}}}{}^{\hat{M}}P'_{\hat{M}},\\
e^{\delta_{{\cal A}}}\Xi_{\hat{M}}=&\Xi_{\hat{M}}+\frac{1}{2}\partial_{\hat{M}}{\cal A}_{\hat{\bar{A}}}{}^{\hat{N}}(A^{-1})^{\hat{\bar{B}}}{}_{\hat{N}}\bar{P}'^{\hat{\bar{A}}} \bar{P}'_{\hat{\bar{B}}}-\frac{1}{2}\partial_{\hat{M}}{\cal A}_{\hat{\bar{A}}}{}^{\hat{N}}({\cal A}^{-1})^{\hat{\bar{B}}}{}_{\hat{N}}e^{\delta_{{\cal A}}}(\bar{P}'^{\hat{\bar{A}}})e^{\delta_{{\cal A}}}( \bar{P}'_{\hat{\bar{B}}}),
\end{align}
where $K$ and $L$ are matrices defined as $K_{\hat{M}}^{\ \hat{N}}=\eta_{\hat{M}\hat{P}}{\cal A}_{\hat{\bar{A}}}{}^{\hat{P}}\eta^{\hat{\bar{A}}\hat{\bar{B}}}{\cal A}_{\hat{\bar{B}}}{}^{\hat{N}}$ and $L_{\hat{\bar{A}}}{}^{\hat{\bar{B}}}={\cal A}_{\hat{\bar{A}}}{}^{\hat{N}}\eta_{\hat{N}\hat{P}}{\cal A}_{\hat{\bar{C}}}{}^{\hat{P}}\eta^{\hat{\bar{C}}\hat{\bar{B}}}$, correspondingly. 
Since the coordinates $P'_{\hat{M}}$ and $\bar{P}'_{\hat{\bar{A}}}$ correspond to the basis of the generalized tangent vector and the local flat frame, respectively, we can identify the matrix ${\cal A}_{\hat{\bar{A}}}^{\ \hat{M}}$
with the generalized vielbein.
When we identify ${\cal A}_{\hat{\bar{A}}}{}^{\hat{M}}$ with
 the generalized vielbein, $K_{\hat{M}}^{\ \hat{N}}$ is reduced to $\delta_{\hat{M}}^{\ \hat{N}}$ and $L_{\hat{\bar{A}}}{}^{\hat{\bar{B}}}$ to $\delta_{\hat{\bar{A}}}{}^{\hat{\bar{B}}}$, respectively. 

For the DFT, we do not need the most general $GL(2D)$ transformation of type $e^{\delta_{\cal A}}$, but it is sufficient to consider the canonical transformation of the subgroup $O(2)\times O(D,D)$ of $GL(2D)$. We parametrize $O(2)$ by an angle $\theta$ and $O(D,D)$ by an element ${\cal E}_{\hat{\bar{A}}}{}^{\hat{M}}$ which is identified with the vielbein. 
Then, a canonical transformation generated by ${\cal A}=\theta {\cal E}$ is 
\begin{align}
e^{\theta\delta_{{\cal E}}}P'_{\hat{M}}=& P'_{\hat{M}}\cos \theta+ {\cal E}^{\hat{\bar{A}}}_{\ \hat{M}}\bar{P}'_{\hat{\bar{A}}}\sin\theta, \\
e^{\theta\delta_{{\cal E}}}\bar{P}'_{\hat{\bar{A}}}=&- {\cal E}_{\hat{\bar{A}}}{}^{\hat{M}}P'_{\hat{M}}\sin\theta+ \bar{P}'_{\hat{\bar{A}}}\cos\theta,\\
e^{\theta\delta_{{\cal E}}}\Xi_{\hat{M}}
=&\Xi_{\hat{M}}-\frac{1}{2}\Omega_{\hat{M}\hat{N}\hat{P}}P'^{\hat{N}} P'^{\hat{P}}\sin^{2}\theta+{\cal E}_{\hat{\bar{A}}}{}^{\hat{P}}\Omega_{\hat{M}\hat{N}\hat{P}}P'^{\hat{N}}\bar{P}'^{\hat{\bar{A}}} \sin\theta\cos\theta\nonumber\\
&\hspace{50mm}+\frac{1}{2}\Omega_{\hat{M}\hat{N}\hat{P}}{\cal E}_{\hat{\bar{A}}}{}^{\hat{N}}{\cal E}^{\ \hat{P}}_{\hat{\bar{B}}}\bar{P}'^{\hat{\bar{A}}} \bar{P}'^{\hat{\bar{B}}}\sin^{2}\theta.\label{ECanonicalTransfOfXi}
\end{align}
where $\Omega_{\hat{M}\hat{N}\hat{P}}={\cal E}^{\hat{\bar{A}}}{}_{\hat{M}}{\cal E}^{\hat{\bar{B}}}{}_{\hat{N}}{\cal E}^{\hat{\bar{C}}}{}_{\hat{P}}\Omega_{\hat{\bar{A}}\hat{\bar{B}}\hat{\bar{C}}}$ with $\Omega_{\hat{\bar{A}}\hat{\bar{B}}\hat{\bar{C}}}={\cal E}_{\hat{\bar{A}}}{}^{\hat{M}}\partial_{\hat{M}}{\cal E}_{\hat{\bar{B}}}{}^{\hat{N}}{\cal E}_{\hat{\bar{C}}\hat{N}}$. 
In the following, we consider the transformation with parameter $\theta=\frac{\pi}{2}$. In this case, the transformation rules simplify as 
\begin{gather}
e^{\frac{\pi}{2}\delta_{{\cal E}}}P'_{\hat{M}}={\cal E}^{\hat{\bar{A}}}{}_{\hat{M}}\bar{P}'_{\hat{\bar{A}}},\quad e^{\frac{\pi}{2}\delta_{{\cal E}}}\bar{P}'_{\hat{\bar{A}}}=- {\cal E}_{\hat{\bar{A}}}{}^{\hat{M}}P'_{\hat{M}},\\
e^{\frac{\pi}{2}\delta_{{\cal E}}}\Xi_{\hat{M}}=\Xi_{\hat{M}}-\frac{1}{2}\Omega_{\hat{M}\hat{N}\hat{P}}P'^{\hat{N}} P'^{\hat{P}}+\frac{1}{2}\Omega_{\hat{M}\hat{N}\hat{P}}{\cal E}_{\hat{\bar{A}}}{}^{\hat{N}}{\cal E}^{\ \hat{P}}_{\hat{\bar{B}}}\bar{P}'^{\hat{\bar{A}}} \bar{P}'^{\hat{\bar{B}}}.
\end{gather}
Then, the twisted Hamiltonian function becomes, 
\begin{align}
e^{\frac{\pi}{2}\delta_{{\cal E}}}\Theta_{0}=&{\cal E}^{\ \hat{M}}_{\hat{\bar{A}}}\Xi_{\hat{M}}\bar{P}'^{\hat{\bar{A}}}+\frac{1}{3!}{\cal F}_{\hat{\bar{A}}\hat{\bar{B}}\hat{\bar{C}}}\bar{P}'^{\hat{\bar{A}}} \bar{P}'^{\hat{\bar{B}}}\bar{P}'^{\hat{\bar{C}}}-\frac{1}{2}\Omega_{\hat{\bar{C}}\hat{\bar{A}}\hat{\bar{B}}}{\cal E}^{\hat{\bar{A}}}{}_{\hat{P}}{\cal E}^{\hat{\bar{B}}}{}_{\hat{Q}}P'^{\hat{P}} P'^{\hat{Q}}\bar{P}'^{\hat{\bar{C}}}\label{TwistedHamiltonian}
\end{align}
where ${\cal F}_{\hat{\bar{A}}\hat{\bar{B}}\hat{\bar{C}}}$ and $\Omega_{\hat{\bar{A}}\hat{\bar{B}}\hat{\bar{C}}}$ are the generalized flux and generalized Weitzenb\"ock connection introduced in \eqref{GeneralizedFlux}. Thus, we have obtained the generalized flux and the Weitzenb\"ock connection as a flux generated by a canonical transformation of Hamiltonian $\Theta_0$. 

We have concluded that the second term in \eqref{TwistedHamiltonian} is in fact a generalized flux by comparing with the explicit form of the vielbein given by ${\cal E}_{\hat{\bar{A}}}^{\ \hat{M}}$ in \eqref{GeneralizedFlux}. This can be proven also
 directly using the representation of the generalized flux defined by the derived bracket as follows:
\begin{align}
{\cal F}_{\hat{\bar{A}}\hat{\bar{B}}\hat{\bar{C}}}=
\braket{{\cal E}_{\hat{\bar{C}}}, {\cal L}_{{\cal E}_{\hat{\bar{A}}}}{\cal E}_{\hat{\bar{B}}}}=&-\{{\cal E}_{\hat{\bar{C}}},\{\{{\cal E}_{\hat{\bar{A}}},\Theta_{0}\},{\cal E}_{\hat{\bar{B}}}\}\}\cr
=&-\{{\cal E}_{\hat{\bar{C}}}^{\ \hat{P}}P'_{\hat{P}},\{\{{\cal E}_{\hat{\bar{A}}}^{\ \hat{M}}P'_{\hat{M}},\Theta_{0}\},{\cal E}_{\hat{\bar{B}}}^{\ \hat{N}}P'_{\hat{N}}\}\}\cr
=&-\{e^{-\frac{\pi}{2}\delta_{{\cal E}}}\bar{P}'_{\hat{\bar{C}}},\{\{e^{-\frac{\pi}{2}\delta_{{\cal E}}}\bar{P}'_{\hat{\bar{A}}},\Theta_{0}\},e^{-\frac{\pi}{2}\delta_{{\cal E}}}\bar{P}'_{\hat{\bar{B}}}\}\}\cr
=&-e^{-\frac{\pi}{2}\delta_{{\cal E}}}\{\bar{P}'_{\hat{\bar{C}}},\{\{\bar{P}'_{\hat{\bar{A}}},e^{\frac{\pi}{2}\delta_{{\cal E}}}\Theta_{0}\},\bar{P}'_{\hat{\bar{B}}}\}\}
\label{GeneralizedFluxQP}. 
\end{align}
We regard the generalized vielbein as a set of the generalized vectors ${\cal E}_{\hat{\bar{A}}}={\cal E}_{\hat{\bar{A}}}{}^{\hat{M}}P'_{\hat{M}}$ on the pre-QP manifold.
The first line is the definition of the generalized flux by the derived bracket. After applying the canonical transformation,
we obtain the last line, where the Poisson bracket with
$\bar{P}'_{\hat{\bar{A}}}, \bar{P}'_{\hat{\bar{B}}},\bar{P}'_{\hat{\bar{C}}}$ picks up the coefficient of 
$\bar{P}'^{\hat{\bar{A}}}, \bar{P}'^{\hat{\bar{B}}},\bar{P}'^{\hat{\bar{C}}}$
 in the transformed Hamiltonian $e^{\frac{\pi}{2}\delta_{E}}\Theta_{0}$. 
 Thus, we obtain the representation of the generalized flux in pre-QP-manifold
  as
  \begin{align}
  {\cal F}_{\hat{\bar{A}}\hat{\bar{B}}\hat{\bar{C}}}=
  -\{\{\{e^{\frac{\pi}{2}\delta_{{\cal E}}}\Theta_{0},\bar{P}'_{\hat{\bar{A}}}\},\bar{P}'_{\hat{\bar{B}}}\},\bar{P}'_{\hat{\bar{C}}}\}
  \label{GeneralizedFluxQP1}. 
  \end{align}

 This explains why the generalized flux is generated by a canonical transformation of the Hamiltonian.

\subsubsection{Canonical transformation by $t$ and $\bar{t}$}
Another possibility of an $O(D,D)$ covariant canonical transformation is given by the function $t$ in \eqref{CanonicalTransfFunc}. The matrix $t_{\hat{M}}{}^{\hat{N}}$ is an element of the Lie algebra of $O(D,D)$ and $(e^{t})_{\hat{M}}{}^{\hat{N}}=T_{\hat{M}}{}^{\hat{N}}$ is a $O(D,D)$ matrix.
The canonical transformations for coordinates $P'_{\hat M}$ and $\Xi_{\hat M}$ are computed as follows. 
\begin{align}
e^{\delta_{t}}P'_{\hat{M}}=&(T^{-1})_{\hat{M}}{}^{\hat{N}}P'_{\hat{N}},\label{TransfRule1}\\
e^{\delta_{t}}\Xi_{\hat{M}}=&\Xi_{\hat{M}}+\frac{1}{2}(\partial_{\hat{M}}TT^{-1})_{\hat{N}}{}^{\hat{P}}P'^{\hat{N}}P'_{\hat{P}}.\label{TransfRule2}
\end{align}
Note that $\partial_{\hat{M}}TT^{-1}$ is a Maurer-Cartan form on $O(D,D)$. The above transformation rules lead to the twisted Hamiltonian function
\begin{align}
e^{\delta_{t}}\Theta_{0}=&T_{\hat{M}}{}^{\hat{N}}\Xi_{\hat{N}}P'^{\hat{M}}+\frac{1}{3!}f_{\hat{M}\hat{N}\hat{P}}P'^{\hat{M}}P'^{\hat{N}}P'^{\hat{P}}, \label{UTwistedHamiltonian}
\end{align}
where we have defined the flux $f_{\hat{M}\hat{N}\hat{P}}=3T_{[\hat{M}|}{}^{\hat{Q}}(\partial_{\hat{Q}}TT^{-1})_{|\hat{N}\hat{P}]}=3T_{[\hat{M}}{}^{\hat{Q}}\partial_{|\hat{Q}|}T_{\hat{N}}{}^{\hat{R}}T_{ \hat{P}]\hat{R}}$. 

{Finally, we consider the $O(D,D)$ covariant canonical transformation generated by} the function $\bar{t}$ in \eqref{CanonicalTransfFunc}. This transformation leads to the same formula as in \eqref{TransfRule1} and \eqref{TransfRule2} with $P'_{\hat M}$ replaced by $\bar{P}'_{\hat{\bar A}}$ and $T$ by $\bar{T}$, respectively:
\begin{align}
e^{\delta_{\bar{t}}}\bar{P}'_{\hat{\bar{A}}}=&(\bar{T}^{-1})_{\hat{\bar{A}}}{}^{\hat{\bar{B}}}\bar{P}'_{\hat{\bar{B}}},\\
e^{\delta_{\bar{t}}}\Xi_{\hat{M}}=&\Xi_{\hat{M}}+\frac{1}{2}(\partial_{\hat{M}}\bar{T}\bar{T}^{-1})_{\hat{\bar{B}}}{}^{\hat{\bar{C}}}\bar{P}'^{\hat{\bar{B}}}\bar{P}'_{\hat{\bar{C}}}.
\end{align}
For the Hamiltonian function we obtain,
\begin{align}
e^{\delta_{\bar{t}}}\Theta_{0}=&\Xi_{\hat{M}}P'^{\hat{M}}+\frac{1}{2}(\partial_{\hat{M}}\bar{T}\bar{T}^{-1})_{\hat{\bar{B}}}{}^{\hat{\bar{C}}}\bar{P}'^{\hat{\bar{B}}}\bar{P}'_{\hat{\bar{C}}}P'^{\hat{M}}.
\end{align}

These transformations generated by $t$ and $\bar{t}$ yield the $O(D,D)$ transformations in each frame.

We comment on the closure condition.
The closure condition changes by twists as explained in section \ref{PreQPMfdAndDerivedBracketSec}. 
Since twist functions here do not include $\Xi_{{\hat M}}$, 
the closure condition is not changed on a function $f(X)$ of degree $0$.
A $D$-dimensional physical subspace is not changed by a twist, however 
generalized connections are introduced by twists in the closure condition of a generalized Lie derivative on generalized vector fields.
We discuss a concrete example in Eq. \eqref{CovariantQPClosure} for a general case 
in section \ref{CovariantizedPreQPSec}.

\subsection{Generalized Bianchi identity on pre-QP-manifold \label{PreBianchi}}

In this subsection, we propose a formulation of the Bianchi identity on the pre-QP-manifold and derive the DFT Bianchi identities satisfied by the generalized flux and Weitzenb\"ock connection.   
On a QP-manifold, the Bianchi identity can be obtained from the classical master equation, $\{\Theta_F, \Theta_F\} = 0$ for the Hamiltonian function  $\Theta_F$ twisted by a flux as discussed in \cite{Heller:2016abk}.
Especially,  the $n=2$ QP-manifold can be applied to the generalized geometry and the $H$-twisted Hamiltonian defines the Courant bracket
with $H$-flux as a derived bracket. There, the classical master equation
requires the Bianchi identity $dH=0$.
On the other hand, on a pre-QP-manifold the classical master equation is not satisfied,  i.e.,
$\{\Theta, \Theta\} \neq 0$.
Thus, for this case we need a new formulation of the Bianchi identity.

For this purpose we introduce the most general Hamiltonian function $\Theta_{F}$ which includes
 all possible fluxes. Since the DFT on pre-QP-manifold can be formulated by using only $P'$ and $\bar P'$, we need to consider the degree 3 Hamiltonian function consisting of $(X^{\hat{M}}, \Xi_{\hat{M}}, P'^{\hat{M}}, \bar{P}'^{\hat{\bar{C}}})$.
 It can be written by using six arbitrary tensors on $\wha M=\tilde M\times M$, denoted by
 $\bar\rho,\rho,{\mathscr F}, \Phi, \Delta, \Psi$:
\begin{align}
\Theta_{F}=&\bar{\rho}^{\ \hat{M}}_{\hat{\bar{A}}}(X)\Xi_{\hat{M}}\bar{P}'^{\hat{\bar{A}}}+\rho_{\hat{N}}^{\ \hat{M}}(X)\Xi_{\hat{M}}P'^{\hat{N}}+\frac{1}{3!}{\mathscr F}_{\hat{\bar{A}}\hat{\bar{B}}\hat{\bar{C}}}(X)\bar{P}'^{\hat{\bar{A}}} \bar{P}'^{\hat{\bar{B}}}\bar{P}'^{\hat{\bar{C}}}+\frac{1}{2}\Phi_{\hat{\bar{A}}\hat{M}\hat{N}}(X)P'^{\hat{M}} P'^{\hat{N}}\bar{P}'^{\hat{\bar{A}}}\nonumber\\
&+\frac{1}{2}\Delta_{\hat{\bar{A}}\hat{\bar{B}}\hat{M}}(X)P'^{\hat{M}}\bar{P}'^{\hat{\bar{A}}}\bar{P}'^{\hat{\bar{B}}}+\frac{1}{3!}\Psi_{\hat{M}\hat{N}\hat{P}}(X)P'^{\hat{M}}P'^{\hat{N}}P'^{\hat{P}}\label{ThetawithFlux1}
\end{align}
We have seen above that we can generate some of the fluxes
by canonical transformation of the Hamiltonian $\Theta_0$ \eqref{QstructureOfDFT} 
without flux. We show that the Bianchi identities
for the corresponding fluxes can be obtained by using the two Hamiltonians $\Theta_0, \Theta_F$ combined with a canonical transformation as follows:

The Bianchi identity on pre-QP-manifold can be defined by introducing the following degree 4 function ${\cal B}\in C^{\infty}({\cal M})$,
\begin{align}
{\cal B}(\Theta_{F},\Theta_{0},e^{\delta_{\alpha}})=\{\Theta_{F},\Theta_{F}\}-e^{\delta_{\alpha}}\{\Theta_{0}, \Theta_{0}\},\label{BianchiFunction}
\end{align}
where $\alpha$ is a canonical transformation function of degree $2$.

Then, the condition on the pre-QP-manifold for the Bianchi identity is 
the vanishing of the function $\cal B$, which we call pre-Bianchi identity: 
\begin{align}
{\cal B}(\Theta_{F},\Theta_{0}, e^{\delta_{\alpha}})=0.
\label{preBianchiidentity}
\end{align}
 Since for the QP-manifold case the classical master equation is satisfied, the second term of ${\cal B}(\Theta_{F},\Theta_{0},e^{\delta_{\alpha}})$ vanishes and the pre-Bianchi identity reduces to the classical master equation $\{\Theta_{F},\Theta_{F}\}=0$, which gives the Bianchi identities for the 
 fluxes introduced in $\Theta_F$. 

The generalized Bianchi identity of DFT can be given by the pre-Bianchi identity
\eqref{preBianchiidentity} in the following way.
Here, we take the Hamiltonian function: 
\begin{align}
\Theta_{F}=&{\cal E}^{\ \hat{M}}_{\hat{\bar{A}}}\Xi_{\hat{M}}\bar{P}'^{\hat{\bar{A}}}+\frac{1}{3!}{\mathscr F}_{\hat{\bar{A}}\hat{\bar{B}}\hat{\bar{C}}}\bar{P}'^{\hat{\bar{A}}} \bar{P}'^{\hat{\bar{B}}}\bar{P}'^{\hat{\bar{C}}}+\frac{1}{2}\Phi_{\hat{\bar{A}}\hat{M}\hat{N}}P'^{\hat{M}} P'^{\hat{N}}\bar{P}'^{\hat{\bar{A}}}\label{ThetawithFlux2}
\end{align}
where fields ${\cal E},\Phi$ and ${\mathscr F}$ are considered as independent objects. We take $\Theta_{0}$ the standard Hamiltonian function of DFT introduced in \eqref{QstructureOfDFT} and $\alpha$ the transformation function ${\cal A}=\theta {\cal E}$ in \eqref{CanonicalTransfFunc}. Then, for $\theta = {\pi \over 2}$ we obtain ${\cal B}(\Theta_{F},\Theta_{0},e^{\frac{\pi}{2}\delta_{{\cal E}}})$ as
\begin{align}
{\cal B}(\Theta_{F},\Theta_{0}&,e^{\frac{\pi}{2}\delta_{{\cal E}}})\cr
=&(2\partial_{\hat{N}}{\cal E}^{\ \hat{M}}_{\hat{\bar{A}}}{\cal E}_{\hat{\bar{B}}}{}^{\hat{N}}+{\cal E}^{\hat{\bar{C}}\hat{M}}{\mathscr F}_{\hat{\bar{C}}\hat{\bar{A}}\hat{\bar{B}}}-\Omega^{\hat{M}}{}_{\hat{P}\hat{Q}}{\cal E}_{\hat{\bar{A}}}{}^{\hat{P}}{\cal E}_{\hat{\bar{B}}}{}^{\hat{Q}})\Xi_{\hat{M}}\bar{P}'^{\hat{\bar{A}}}\bar{P}'^{\hat{\bar{B}}}\nonumber\\
&+({\cal E}^{\hat{\bar{A}}\hat{M}}\Phi_{\hat{\bar{A}}\hat{N}\hat{P}}+\Omega^{\hat{M}}{}_{\hat{N}\hat{P}} )\Xi_{\hat{M}}P'^{\hat{N}} P'^{\hat{P}}\nonumber\\
&-\frac{1}{3}\Big({\cal E}^{\ \hat{M}}_{\hat{\bar{A}}}\partial_{\hat{M}}{\mathscr F}_{\hat{\bar{B}}\hat{\bar{C}}\hat{\bar{D}}}-\frac{3}{4}{\mathscr F}_{\hat{\bar{E}}\hat{\bar{A}}\hat{\bar{B}}}{\mathscr F}^{\hat{\bar{E}}}{}_{\hat{\bar{C}}\hat{\bar{D}}}+\frac{3}{4}\Omega_{\hat{\bar{E}}\hat{\bar{A}}\hat{\bar{B}}}\Omega^{\hat{\bar{E}}}{}_{\hat{\bar{C}}\hat{\bar{D}}} \Big)\bar{P}'^{\hat{\bar{A}}}\bar{P}'^{\hat{\bar{B}}} \bar{P}'^{\hat{\bar{C}}}\bar{P}'^{\hat{\bar{D}}}\nonumber\\
&+\Big(-{\cal E}^{\ \hat{P}}_{\hat{\bar{A}}}\partial_{\hat{P}}\Phi_{\hat{\bar{B}}\hat{M}\hat{N}}+\frac{1}{2}{\mathscr F}_{\hat{\bar{C}}\hat{\bar{A}}\hat{\bar{B}}}\Phi^{\hat{\bar{C}}}{}_{\hat{M}\hat{N}}\nonumber\\
&~~~~~~~~~-\Phi_{\hat{\bar{A}}\hat{M}\hat{Q}} \Phi_{\hat{\bar{B}}\hat{N}}{}^{\hat{Q}}+\frac{1}{2}\Omega_{\hat{P}\hat{M}\hat{N}} \Omega^{\hat{P}}{}_{\hat{Q}\hat{R}}{\cal E}_{\hat{\bar{A}}}{}^{\hat{Q}}{\cal E}_{\hat{\bar{B}}}{}^{\hat{R}} \Big)P'^{\hat{M}} P'^{\hat{N}}\bar{P}'^{\hat{\bar{A}}}\bar{P}'^{\hat{\bar{B}}}\nonumber\\
&+\frac{1}{4}(\Phi_{\hat{R}\hat{M}\hat{N}}\Phi^{\hat{R}}{}_{\hat{P}\hat{Q}}-\Omega_{\hat{R}\hat{M}\hat{N}}\Omega^{\hat{R}}{}_{\hat{P}\hat{Q}})P'^{\hat{M}} P'^{\hat{N}}P'^{\hat{P}} P'^{\hat{Q}}. 
\end{align} 

Therefore, the pre-Bianchi identity gives the following equations
\begin{align}
2\partial_{\hat{N}}{\cal E}_{[\hat{\bar{A}}}{}^{\hat{M}}{\cal E}_{\hat{\bar{B}}]}{}^{\hat{N}}+{\cal E}^{\hat{\bar{C}}\hat{M}}{\mathscr F}_{\hat{\bar{C}}\hat{\bar{A}}\hat{\bar{B}}}-\Omega^{\hat{M}}{}_{\hat{P}\hat{Q}}{\cal E}_{\hat{\bar{A}}}{}^{\hat{P}}{\cal E}_{\hat{\bar{B}}}{}^{\hat{Q}}=&0,\label{preBianchi1}\\
{\cal E}^{\hat{\bar{A}}\hat{M}}\Phi_{\hat{\bar{A}}\hat{N}\hat{P}}+\Omega^{\hat{M}}_{\ \ \hat{N}\hat{P}}=&0,\label{preBianchi2}\\
{\cal E}^{\ \hat{M}}_{[\hat{\bar{A}}}\partial_{\hat{M}}{\mathscr F}_{\hat{\bar{B}}\hat{\bar{C}}\hat{\bar{D}}]}-\frac{3}{4}{\mathscr F}_{\hat{\bar{E}}\hat{[\bar{A}}\hat{\bar{B}}}{\mathscr F}^{\hat{\bar{E}}}{}_{\hat{\bar{C}}\hat{\bar{D}}]}+\frac{3}{4}\Omega_{\hat{\bar{E}}[\hat{\bar{A}}\hat{\bar{B}}}\Omega^{\hat{\bar{E}}}{}_{\hat{\bar{C}}\hat{\bar{D}}]}=&0,\label{preBianchi3}\\
-{\cal E}^{\ \hat{P}}_{[\hat{\bar{A}}}\partial_{\hat{P}}\Phi_{\hat{\bar{B}}]\hat{M}\hat{N}}+\frac{1}{2}{\mathscr F}_{\hat{\bar{C}}\hat{\bar{A}}\hat{\bar{B}}}\Phi^{\hat{\bar{C}}}{}_{\hat{M}\hat{N}}-\Phi_{[\hat{\bar{A}}[\hat{M}\hat{Q}} \Phi_{\hat{\bar{B}}]\hat{N}]}{}^{\hat{Q}}+\frac{1}{2}\Omega_{\hat{P}\hat{M}\hat{N}} \Omega^{\hat{P}}{}_{\hat{Q}\hat{R}}{\cal E}_{\hat{\bar{A}}}{}^{\hat{Q}}{\cal E}_{\hat{\bar{B}}}{}^{\hat{R}}=&0,\label{preBianchi4}\\
\Phi_{\hat{R}[\hat{M}\hat{N}}\Phi^{\hat{R}}_{\ \ \hat{P}\hat{Q}]}-\Omega_{\hat{R}[\hat{M}\hat{N}}\Omega^{\hat{R}}_{\ \ \hat{P}\hat{Q}]}=&0.\label{preBianchi5}
\end{align}
By solving \eqref{preBianchi1}
and \eqref{preBianchi2}, we obtain local expressions for 
${\mathscr{F}}_{\hat{\bar{A}}\hat{\bar{B}}\hat{\bar{C}}}$ and $\Phi_{\hat{\bar{A}}\hat{N}\hat{P}}$.
These local expressions are consistent with the definition of the generalized flux ${\cal F}_{\hat{\bar{A}}\hat{\bar{B}}\hat{\bar{C}}}$ and the Weitzenb\"ock connection $\Omega_{\hat{\bar{A}}\hat{\bar{B}}\hat{\bar{C}}}$ in \eqref{TwistedHamiltonian}. 
The third equation \eqref{preBianchi3} is nothing but the generalized Bianchi identity in \cite{Aldazabal:2013sca,Geissbuhler:2013uka}. 
See also \cite{Chatzistavrakidis:2018ztm}.
The fourth equation \eqref{preBianchi4} gives another generalized Bianchi identity for $\Phi_{\hat{\bar{A}}\hat{M}\hat{N}}$. 
The equation \eqref{preBianchi5} does not give a new condition. 

Note that in the above derivation, we have used the $\Theta_F$ given in \eqref{ThetawithFlux2} for simplicity. However, in principle, we can use the most general Hamiltonian with fluxes given in 
\eqref{ThetawithFlux1}. As a result of the pre-Bianchi identity, we obtain the condition
that the redundant fluxes vanish.

\subsection{GSS twist as canonical transformation\label{GSSasCanonicalTransf}}
 
In this section, we show that the GSS ansatz \eqref{GSSansatz} can be understood as a canonical transformation. The generalized Lie derivative \eqref{GDFTLieDer} and generalized flux \eqref{GSStwistedFlux} after the compactification will be derived by using the canonical transformations
on the pre-QP-manifold. 

For the GSS compactification in the supermanifold formalism, we also 
split the base manifold into internal and external spaces. 
We use the same notation for coordinates $X=(\mathbb{X},\mathbb{Y})$ as in section \ref{GSSsec}, that is $\mathbb{X}$ is used for the $2(D-d)$-dimensional external space and $\mathbb{Y}$ is used for the $2d$-dimensional internal space. 

Furthermore, 
we introduce $2D$-dimensional intermediate coordinates of degree 1 denoted by $(\wha{Q}^{\hat{I}},\wha{P}_{\hat{I}})$ where $\hat I,\hat{J},\hat{K},\hat{L},\hat{H}=1,...,2D$ and we define the corresponding DFT basis as 
\begin{equation}
\widehat{Q}'^{\hat{I}}:=\frac{1}{\sqrt[]{2}}(\widehat{Q}^{\hat{I}}-\eta^{\hat{I}\hat{J}}\widehat{P}_{\hat{J}})~~,~~ \wha{P}'_{\hat{I}}:=\frac{1}{\sqrt[]{2}}(\wha{P}_{\hat{I}}+\eta_{\hat{I}\hat{J}}\wha{Q}^{\hat{J}})
\end{equation}
 in the same way as $Q'^{\hat{M}}$ and $P'_{\hat{M}}$ introduced before. 
The constant matrix $\eta_{\hat{I}\hat{J}}$ and its inverse $\eta^{\hat{I}\hat{J}}$ are $O(D,D)$ invariant metrics. The Poisson brackets for $\wha{Q}'^{\hat{I}}$ and $\wha{P}'_{\hat{I}}$ are
\begin{align}
\{\wha{Q}'^{\hat{I}},\wha{Q}'^{\hat{I}}\}=\eta^{\hat{I}\hat{J}},\quad \{\wha{P}'_{\hat{I}},\wha{P}'_{\hat{J}}\}=\eta_{\hat{I}\hat{J}}, \quad \{\wha{Q}'^{\hat{I}},\wha{P}'_{\hat{J}}\}=0. 
\end{align}
The intermediate coordinate $\wha{P}'_{\hat{I}}$ of the supermanifold is associated to the basis of the generalized vector $\wha{V}(\mathbb{X})$ of the reduced theory with injection $j'$: 
\begin{align}
&j'_{*}:\wha{V}(\mathbb{X})=\wha{V}^{\hat{I}}(\mathbb{X})\partial_{\hat{I}}\longmapsto \wha{V}^{\hat{I}}(\mathbb{X})\wha{P}'_{\hat{I}},\\
&j'^{*}:\wha{V}^{\hat{I}}(\mathbb{X})\wha{P}'_{\hat{I}}\longmapsto \wha{V}^{\hat{I}}(\mathbb{X})\partial_{\hat{I}}. 
\end{align}
We identify the generalized vectors on the reduced theory $\wha{V}^{\hat{I}}(\mathbb{X})\partial_{\hat{I}}$ with $\wha{V}^{\hat{I}}(\mathbb{X})\wha{P}'_{\hat{I}}$ and again omit to write the maps $j'_{*}$ and $j'^{*}$ for simplicity. 

Due to the intermediate coordinate $\wha{P}'_{\hat{I}}$, there are three new types of canonical transformation functions in addition to \eqref{CanonicalTransfFunc}. 
\begin{align}
\wha{A}:=\wha{A}_{\hat{\bar{A}}}{}^{\hat{I}}\eta^{\hat{A}\hat{B}}\wha{P}'_{\hat{I}}\bar{P}'_{\hat{\bar{B}}},\quad A:=A_{\hat{I}}{}^{\hat{M}}\eta^{\hat{I}\hat{J}}\wha{P}'_{\hat{J}}P'_{\hat{M}},\quad \wha{t}:=\wha{t}_{\hat{I}}{}^{\hat{J}}\eta^{\hat{I}\hat{K}}\wha{P}'_{\hat{J}}\wha{P}'_{\hat{K}}. \label{canonicalIntermediate}
\end{align}
Here, the matrices $\wha{A}_{\hat{\bar{A}}}{}^{\hat{I}}$ and  $A_{\hat{I}}{}^{\hat{M}}$ 
are elements of $GL(2D)$ in general, and the matrix $\wha{t}_{\hat{I}}{}^{\hat{J}}$ 
belongs to the Lie algebra of $O(D,D)$.  The canonical transformation generated by these functions are 
given by the same formulae as in the previous section. 

For the GSS compactification, the generalized vielbein of the reduced theory is identified with the matrix $\wha{A}_{\hat{\bar{A}}}{}^{\hat{I}}$ in \eqref{canonicalIntermediate} and taken as a
function of $\mathbb{X}$ and an element of $O(D,D)$. Then, the canonical transformation with $\wha{A}={\pi \over 2}\wha{E}$ 
defines a map from the local flat coordinate to the intermediate coordinate.
\begin{align}
e^{-\frac{\pi}{2}\delta_{\wha{E}}}\bar{P}'_{\hat{\bar{A}}}=\wha{E}_{\hat{\bar{A}}}{}^{\hat{I}}(\mathbb{X})\wha{P}'_{\hat{I}}. \label{VielbeinOfReducedTheory}
\end{align}

The GSS twist \eqref{GSSansatz} is produced by the canonical transformation with $A=\frac{\pi}{2}U$
where the parameter $U_{\hat{I}}{}^{\hat{M}}$ is an element of $O(D,D)$ which depends only on $\mathbb{Y}$, 
and the components of $U_{\hat{I}}{}^{\hat{M}}$ are non-trivial only when both indices lie in the internal directions (condition \eqref{GSScondition}). 
Then, the canonical transformation $e^{-\frac{\pi}{2}\delta_{U}}$ provides the GSS twist of the generalized vielbein $\wha{E}_{\hat{\bar{A}}}{}^{\hat{I}}(\mathbb{X})$ and the gauge parameter $\Lambda^{\hat{I}}(\mathbb{X})$ of the reduced theory. 
\begin{align}
&e^{-\frac{\pi}{2}\delta_{U}}(\wha{E}_{\hat{\bar{A}}}{}^{\hat{I}}(\mathbb{X})\wha{P}'_{\hat{I}})=\wha{E}_{\hat{\bar{A}}}{}^{\hat{I}}(\mathbb{X})U_{\hat{I}}{}^{\hat{M}}(\mathbb{Y})P'_{\hat{M}},\label{QPGSSanzats1}\\
&e^{-\frac{\pi}{2}\delta_{U}}(\wha{\Lambda}^{\hat{I}}(\mathbb{X})\wha{P}'_{\hat{I}})=\wha{\Lambda}^{\hat{I}}(\mathbb{X})U_{\hat{I}}{}^{\hat{M}}(\mathbb{Y})P'_{\hat{M}}. \label{QPGSSanzats2}
\end{align}

The canonical transformation generated by $\wha{t}$ in \eqref{CanonicalTransfFunc} is the $O(D,D)$ transformation of frame $\wha{P}_{\hat{I}}$. 
Using this degrees of freedom, we can define a more general form of GSS twists: 
\begin{align}
&e^{-\frac{\pi}{2}\delta_{U}}e^{\delta_{\wha{t}}}(\wha{E}_{\hat{\bar{A}}}{}^{\hat{I}}(\mathbb{X})\wha{P}'_{\hat{I}})=\wha{E}_{\hat{\bar{A}}}{}^{\hat{I}}(\mathbb{X})\Big(\wha{T}_{\hat{I}}{}^{\hat{J}}U_{\hat{J}}{}^{\hat{M}}(\mathbb{Y})\Big)P'_{\hat{M}},\label{QPGSSanzats1'}\\
&e^{-\frac{\pi}{2}\delta_{U}}e^{\delta_{\wha{t}}}(\wha{\Lambda}^{\hat{I}}(\mathbb{X})\wha{P}'_{\hat{I}})=\wha{\Lambda}^{\hat{I}}(\mathbb{X})\Big(\wha{T}_{\hat{I}}{}^{\hat{J}}U_{\hat{J}}{}^{\hat{M}}(\mathbb{Y})\Big)P'_{\hat{M}}. \label{QPGSSanzats2'}
\end{align}
When we assume that $\wha{T}_{\hat{I}}{}^{\hat{J}}$ depends only on $\mathbb{Y}$, we can regard the matrix $\wha{T}_{\hat{I}}{}^{\hat{J}}U_{\hat{J}}{}^{\hat{M}}(\mathbb{Y})$ as GSS twist matrix. 
In \eqref{QPGSSanzats1} and \eqref{QPGSSanzats2}, the GSS twist is generated by the canonical transformation $e^{\frac{\pi}{2}\delta_{U}}$. 
On the other hand, when we take $U_{\hat{I}}{}^{\hat{M}}=\delta_{\hat{I}}{}^{\hat{M}}$ in \eqref{QPGSSanzats1'} and \eqref{QPGSSanzats2'}, the GSS twist is generated by the canonical transformation $e^{\delta_{\wha{T}}}$. 
In the following, we discuss the GSS twist by the canonical transformation $e^{\frac{\pi}{2}\delta_{U}}$.

The twisted Hamiltonian function is given by the same equation as \eqref{TwistedHamiltonian} where $E_{\hat{\bar{A}}}{}^{\hat{M}}$ is replaced by $U_{\hat{I}}{}^{\hat{M}}(\mathbb{Y})$:
\begin{align}
{\wha{\Theta}}_{\rm GSS}=&e^{-\frac{\pi}{2}\delta_{U}}\Theta_{0}\cr
=&U^{\ \hat{M}}_{\hat{I}}\Xi_{\hat{M}}\wha{P}'^{\hat{I}}+\frac{1}{3!}f_{\hat{I}\hat{J}\hat{K}}\wha{P}'^{\hat{I}}\wha{P}'^{\hat{J}}\wha{P}'^{\hat{K}}-\frac{1}{2}\wtil{\Omega}_{\hat{I}\hat{J}\hat{K}}U^{\hat{J}}{}_{\hat{M}}U^{\hat{K}}{}_{\hat{N}}P'^{\hat{M}}P'^{\hat{N}}\wha{P}'^{\hat{I}}~.\label{GSSTwistedHamiltonian}
\end{align}
Here, the Weitzenb\"ock connection $\wtil{\Omega}_{\hat{I}\hat{J}\hat{K}}=U_{\hat{I}}{}^{\hat{M}}\partial_{\hat{M}}U_{\hat{J}}{}^{\hat{N}}U_{\hat{K}\hat{N}}$ is made from $U_{\hat{I}}{}^{\hat{M}}(\mathbb{Y})$ and the resulting flux $f_{\hat{I}\hat{J}\hat{K}}=3\wtil{\Omega}_{[\hat{I}\hat{J}\hat{K}]}$ is assumed to be constant by the GSS ansatz.

The generalized Lie derivative on the reduced theory is derived from the parent theory as
\begin{align}
{\cal L}_{\Lambda}V=-\{\{\Lambda,\Theta_{0}\},V\}.
\end{align}
The right hand side is calculated using the property of the canonical transformation as
\begin{align}
-\{\{\Lambda,\Theta_{0}\},V\}=&-\{\{e^{-\frac{\pi}{2}\delta_{U}}(\widehat{\Lambda}^{\hat{I}}(\mathbb{X})\wha{P}'_{\hat{I}}),\Theta_{0}\},e^{-\frac{\pi}{2}\delta_{U}}(\widehat{V}^{\hat{J}}(\mathbb{X})\wha{P}'_{\hat{J}})\}\cr
=&-e^{-\frac{\pi}{2}\delta_{U}}\{\{\widehat{\Lambda}^{\hat{I}}(\mathbb{X})\wha{P}'_{\hat{I}},{\wha{\Theta}}_{\rm GSS}\},\widehat{V}^{\hat{J}}(\mathbb{X})\wha{P}'_{\hat{J}}\}\cr
=&e^{-\frac{\pi}{2}\delta_{U}}\Big({{\cal L}}_{\widehat{\Lambda}}\widehat{V}^{\hat{I}}\wha{P}'_{\hat{I}}+f_{\hat{J}\hat{K}}{}^{\hat{I}}\widehat{\Lambda}^{\hat{J}}\widehat{V}^{\hat{K}}\wha{P}'_{\hat{I}}\Big)\cr
=&U_{\hat{I}}^{\ \hat{M}}\Big({{\cal L}}_{\widehat{\Lambda}}\widehat{V}^{\hat{I}}+f_{\hat{J}\hat{K}}{}^{\hat{I}}\widehat{\Lambda}^{\hat{J}}\widehat{V}^{\hat{K}}\Big)P'_{\hat{M}}
\end{align}
From this we can read off that
\begin{align}
\wha{\cal L}_{\wha{\Lambda}}\wha{V}
=&-
\{\{\widehat{\Lambda}(\mathbb{X}),\wha{\Theta}_{\rm GSS}\},\widehat{V}(\mathbb{X})\}
\cr=&\Big({{\cal L}}_{\widehat{\Lambda}}\widehat{V}^{\hat{I}}
+f_{\hat{J}\hat{K}}{}^{\hat{I}}\widehat{\Lambda}^{\hat{J}}\widehat{V}^{\hat{K}}\Big)\wha{P}'_{\hat{I}}
~.
\end{align}
Thus, this derived bracket realizes the generalized Lie derivative of GDFT \eqref{GDFTLieDer}.

The closure condition for the derived bracket is provided by the weak master equation \eqref{WeakMasterEQ} for generalized vectors of the reduced theory. 
The master equation for the twisted Hamiltonian function ${\wha{\Theta}}_{\rm GSS}$ is
{\begin{align}
\{{\wha{\Theta}}_{\rm GSS},{\wha{\Theta}}_{\rm GSS}\}
=&\eta^{\hat{M}\hat{N}}\Xi_{\hat{M}}\Xi_{\hat{N}}+\eta^{\hat{M}\hat{N}}\partial_{\hat{N}}U_{\hat{I}}{}^{\hat{P}}U_{\hat{J}\hat{P}}\Xi_{\hat{M}}\wha{P}'^{\hat{I}}\wha{P}'^{\hat{J}}+\frac{3}{4}f_{\hat{H}\hat{I}\hat{J}}f^{\hat{H}}{}_{\hat{K}\hat{L}}\wha{P}'^{\hat{I}}\wha{P}'^{\hat{J}}\wha{P}'^{\hat{K}}\wha{P}'^{\hat{L}}\cr
&-\wtil{\Omega}_{\hat{I}\hat{J}\hat{K}}U^{\hat{I}\hat{M}}U^{\hat{J}}{}_{\hat{N}}U^{\hat{K}}{}_{\hat{P}}\Xi_{\hat{M}}P'^{\hat{N}}P'^{\hat{P}}\cr
&+\Big(\partial_{\hat{P}}\wtil{\Omega}_{\hat{J}\hat{K}\hat{L}}-\frac{1}{2}f_{\hat{I}\hat{J}\hat{H}}\wtil{\Omega}^{\hat{H}}{}_{\hat{K}\hat{L}}\Big)U^{\ \hat{P}}_{\hat{I}}U^{\hat{K}}{}_{\hat{M}}U^{\hat{L}}{}_{\hat{N}}P'^{\hat{M}}P'^{\hat{N}}\wha{P}'^{\hat{I}}\wha{P}'^{\hat{J}}\cr
&-\wtil{\Omega}_{\hat{I}\hat{K}\hat{L}}\wtil{\Omega}_{\hat{J}\hat{H}}{}^{\hat{L}}U^{\hat{K}}{}_{\hat{M}}U^{\hat{H}}{}_{\hat{N}}P'^{\hat{M}}P'^{\hat{N}}\wha{P}'^{\hat{I}}\wha{P}'^{\hat{J}}\cr
&+\frac{1}{4}\eta^{\hat{S}\hat{R}}\partial_{R}U_{\hat{I}\hat{N}}U^{\hat{I}}{}_{\hat{M}}\partial_{\hat{S}}U_{\hat{J}\hat{Q}}U^{\hat{J}}{}_{\hat{P}}P'^{\hat{M}}P'^{\hat{N}}P'^{\hat{P}}P'^{\hat{Q}}. 
\end{align}}
Then, the weak master equation for the generalized vectors $\wha{V}_{1}(\mathbb{X}), \wha{V}_{2}(\mathbb{X}),\wha{V}_{3}(\mathbb{X})$ of the reduced theory 
\begin{align}
\{\{\{\{{\wha{\Theta}}_{\rm GSS},{\wha{\Theta}}_{\rm GSS}\},\wha{V}_{1}\},\wha{V}_{2}\},\wha{V}_{3}\}=0
\end{align}
leads the following conditions for the generalized vectors and structure constants:
\begin{align}
&\partial^{\hat{I}}\wha{V}_{1}^{\hat{J}}\wha{V}_{2\hat{J}}\partial_{\hat{I}}\wha{V}_{3}^{\hat{K}}-2\partial^{\hat{I}}\wha{V}_{1}^{[\hat{J}}\partial_{\hat{I}}\wha{V}_{2}^{\hat{K}]}\wha{V}_{3\hat{J}}=0,\\
&f_{\hat{H}[\hat{I}\hat{J}}f^{\hat{H}}{}_{\hat{K}\hat{L}]}=0. 
\end{align}

Since the generalized vectors depend only on the external space coordinate $\mathbb{X}$, the first condition is just the closure constraint for the external space. The second condition is satisfied by the Jacobi identity for  the constant flux $f_{\hat{I}\hat{J}\hat{K}}$.

By the canonical transformation given by the dynamical generalized vielbein $\widehat{E}$, the Hamiltonian function ${\wha{\Theta}}_{\rm GSS}$ is transformed as 
\begin{align}
e^{\frac{\pi}{2}\delta_{\wha{E}}}{\wha{\Theta}}_{\rm GSS}
=&{\widehat E}_{\hat{\bar{A}}}{}^{\hat{I}}U^{\ \hat{M}}_{\hat{I}}\Xi_{\hat{M}}\bar{P}'^{\hat{\bar{A}}}
+\frac{1}{3!}(\wha{F}_{\hat{\bar{A}}\hat{\bar{B}}\hat{\bar{C}}}
+f_{\hat{I}\hat{J}\hat{K}}\wha{E}_{\hat{\bar{A}}}{}^{\hat{I}}\wha{E}_{\hat{\bar{B}}}{}^{\hat{J}}\wha{E}_{\hat{\bar{C}}}{}^{\hat{K}})\bar{P}'^{\hat{\bar{A}}}\bar{P}'^{\hat{\bar{B}}}\bar{P}'^{\hat{\bar{C}}}\cr
&-\frac{1}{2}\wha{\Omega}_{\hat{\bar{C}}\hat{\bar{A}}\hat{\bar{B}}}\wha{E}^{\hat{\bar{A}}}{}_{\hat{I}}\wha{E}^{\hat{\bar{B}}}{}_{\hat{J}}\wha{P}'^{\hat{I}}\wha{P}'^{\hat{J}}\bar{P}'^{\hat{\bar{C}}}-\frac{1}{2}\wtil{\Omega}_{\hat{I}\hat{J}\hat{K}}U^{\hat{J}}{}_{\hat{M}}U^{\hat{K}}{}_{\hat{N}}{\wha E}_{\hat{\bar{A}}}{}^{\hat{I}}P'^{\hat{M}}P'^{\hat{N}}\bar{P}'^{\hat{\bar{A}}}.
\end{align}
where $\widehat{F}_{\hat{\bar{A}}\hat{\bar{B}}\hat{\bar{C}}}$ and $f_{\hat{I}\hat{J}\hat{K}}$ are defined in \eqref{ExtFlux} and \eqref{IntFlux}, respectively.

The dynamical field in the reduced effective theory is in ${ \wha E}_{\hat{\bar{A}}}{}^{\hat I}$. Therefore,
the generalized flux of the theory after the GSS compactification is calculated in superspace formalism by applying the canonical transformation as: 
\begin{align}
{\cal F}_{\hat{\bar{A}}\hat{\bar{B}}\hat{\bar{C}}}=\braket{{\cal E}_{\hat{\bar{C}}},{\cal L}_{{\cal E}_{\hat{\bar{A}}}}{\cal E}_{\hat{\bar{B}}}}=&-\{{\cal E}_{\hat{\bar{C}}},\{\{{\cal E}_{\hat{\bar{A}}},\Theta_{0}\},{\cal E}_{\hat{\bar{B}}}\}\}\cr
=&-\{e^{-\frac{\pi}{2}\delta_{U}}e^{-\frac{\pi}{2}\delta_{\widehat{E}}}\bar{P}'_{\hat{\bar{C}}},\{\{e^{-\frac{\pi}{2}\delta_{U}}e^{-\frac{\pi}{2}\delta_{\widehat{E}}}\bar{P}'_{\hat{\bar{A}}},\Theta_{0}\},e^{-\frac{\pi}{2}\delta_{U}}e^{-\frac{\pi}{2}\delta_{\widehat{E}}}\bar{P}'_{\hat{\bar{B}}}\}\}\cr
=&-e^{-\frac{\pi}{2}\delta_{U}}e^{-\frac{\pi}{2}\delta_{\widehat{E}}}\{\bar{P}'_{\hat{\bar{C}}},\{\{\bar{P}'_{\hat{\bar{A}}},e^{\frac{\pi}{2}\delta_{\widehat{E}}}{\wha{\Theta}}_{\rm GSS}\},\bar{P}'_{\hat{\bar{B}}}\}\}\cr
=&\widehat{F}_{\hat{\bar{A}}\hat{\bar{B}}\hat{\bar{C}}}+f_{\hat{M}\hat{N}\hat{P}}\widehat{E}_{\hat{\bar{A}}}{}^{\hat{M}}\widehat{E}_{\hat{\bar{B}}}{}^{\hat{N}}\widehat{E}_{\hat{\bar{C}}}{}^{\hat{P}}. 
\end{align}

Thus, we can obtain the representation of the generalized flux as
\begin{align}
{\cal F}_{\hat{\bar{A}}\hat{\bar{B}}\hat{\bar{C}}}=-
{\{\{\{e^{\frac{\pi}{2}\delta_{\widehat{E}}}\wha{\Theta}_{\rm GSS},\bar{P}'_{\hat{\bar{A}}}\},\bar{P}'_{\hat{\bar{B}}}\},\bar{P}'_{\hat{\bar{C}}}\}}
\end{align}
This shows that the GSS twisted flux appears in the twisted Hamiltonian function in the same way as in \eqref{GeneralizedFluxQP1}. 

We summarize the correspondence between the DFT and the GSS compactified DFT objects on the pre-QP-manifolds in table \ref{Table2}. 
The Q-structure $\Theta_{0}$ is replaced by $\wha{\Theta}_{\rm GSS}$ in the GSS compactified DFT. 
Thus, the deformation of the background of DFT on the pre-QP-manifold is realized by the deformation of the Hamiltonian function. 
\begin{table}[ht]
\begin{center}
\begin{tabular}{c||c|c}
 &DFT&GSS\\
\hline\hline
generalized vector & $V^{\hat{M}}(X)P'_{\hat{M}}$ & $\wha{V}^{\hat{I}}(\mathbb{X})\wha{P}'_{\hat{I}}$\\
\hline
Hamiltonian & $\Theta_{0}$ & $\wha{\Theta}_{\rm GSS}$\\
\hline
\begingroup
\renewcommand{\arraystretch}{0.7}
\begin{tabular}{c}
generalized\\
 Lie derivative 
\end{tabular}
\endgroup
& ${\cal L}_{\Lambda}V=-\{\{\Lambda,\Theta_{0}\},V\}$ & $\wha{{\cal L}}_{\wha{\Lambda}}\wha{V}=-\{\{\wha{\Lambda},\wha{\Theta}_{\rm GSS}\},\wha{V}\}$\\
\hline
\begingroup
\renewcommand{\arraystretch}{0.7}
\begin{tabular}{c}
weak\\ master equation 
\end{tabular}
\endgroup
& $\{\{\{\{\Theta_{0},\Theta_{0}\},V_{1}\},V_{2}\},V_{3}\}=0$ & $\{\{\{\{\wha{\Theta}_{\rm GSS},\wha{\Theta}_{\rm GSS}\},\wha{V}_{1}\},\wha{V}_{2}\},\wha{V}_{3}\}=0$\\
\hline
generalized flux & $-\{\{\{e^{\frac{\pi}{2}\delta_{{\cal E}}}\Theta_{0},\bar{P}'_{\hat{\bar{A}}}\},\bar{P}'_{\hat{\bar{B}}}\},\bar{P}'_{\hat{\bar{C}}}\}$ & $-\{\{\{e^{\frac{\pi}{2}\delta_{\wha{E}}}\wha{\Theta}_{\rm GSS},\bar{P}'_{\hat{\bar{A}}}\},\bar{P}'_{\hat{\bar{B}}}\},\bar{P}'_{\hat{\bar{C}}}\}$
\end{tabular}
\caption{Correspondence of the objects in standard DFT and in GSS twisted case on the pre-QP-manifold.}
\label{Table2}
\end{center}
\end{table}

\section{Covariantized pre-QP-manifold and DFT on group manifold\label{CovariantizedPreQPSec}}

We have seen in previous section that in GSS compactification, the information of the twisted background is confined to the Hamiltonian function ${\wha{\Theta}_{\rm GSS}}$ in \eqref{GSSTwistedHamiltonian}.  The generalized Lie derivative and flux are then obtained simply by replacing the Hamiltonian $\Theta_0$ of original DFT with the Hamiltonian $\wha{\Theta}_{\rm GSS}$ twisted by the GSS ansatz.  We consider here to replace the GSS background with a general non-trivial background. Therefore, we use again the intermediate $O(D,D)$ frame as in GSS case, which is denoted by the indices $\hat I,\hat J\cdots$ together with the general tangent frame and the flat local frame with indices $\hat M, \hat N\cdots$ and $\hat{\bar A}, \hat{\bar B},\cdots$, respectively. For notations, see appendix \ref{Notations}. Then, we take the ansatz analogous to (\ref{GSSansatz}) for the total vielbein as
\begin{equation}
{\cal E}_{\hat{\bar A}}{}^{\hat M}=\widehat E_{\hat{\bar A}}{}^{\hat I}E_{\hat I}{}^{\hat M}~,
\end{equation}
where $E_{\hat I}{}^{\hat M}$ is the background vielbein and $\widehat E_{\hat{\bar A}}{}^{\hat I}$ is the fluctuation vielbein.
The geometry of the background is contained in the background vielbein and
the dynamical fields are in the fluctuation vielbein.

Here, we formulate the double geometry on a non-trivial background with pre-QP-manifold. As we have seen in the previous sections, the structure of the generalized Lie derivative and the consistency conditions are characterized by the pre-QP-structure. For this purpose, we introduce the covariantized pre-QP-manifold
 and analyze the double geometry of the background. The background vielbein is an element of $GL(2D)$ and the metric 
$\eta_{\hat{M}\hat{N}}$ is not constant in general.  As in the standard geometry, we introduce the generalized affine connection and the covariant derivative. The generalized Lie derivative and the fluxes are also considered in the background geometry.  
 Then, the fluctuation is introduced on this background. We apply this formulation to DFT on the group manifold proposed in \cite{Bosque:2015jda}.

\subsection{$GL(2D)$ covariant formulation of pre-QP-manifold\label{DFTonGroupMfdOnQPmfd}}
First, let us consider the pre-QP-manifold on an arbitrary background. 
For this purpose, we introduce the background covariant supercoordinate conjugate to the coordinate of the base manifold. The new Hamiltonian function is given in terms of these covariant coordinates,
 realizing the covariant derivative of generalized vectors analogously to the GSS case. 
The weak master equation for the Hamiltonian function provides several conditions for closure of the covariant generalized Lie derivative. 
We also discuss the canonical transformations in covariant form and derive the pre-Bianchi identitiy.

\subsubsection{Covariant derivative on pre-QP-manifold}
Let $\widetilde{M}\times M$ be a $2D$-dimensional manifold \footnote{The formulation in this paper is applicable to general $2D$-dimensional base manifold without decomposition into $\widetilde{M}\times M$.} with local coordinates $X^{\hat{M}}=(\tilde{x}_{M}, x^{M})$ where
 $\hat{M},\hat{N}, \cdots$ are $GL(2D)$ indices. 
 We consider the graded manifold ${\cal M}=T^{*}[2](T[1](\widetilde{M}\times M)\oplus V[1]\oplus \bar{V}[1])$ with coordinates $(X^{\hat{M}}, Q^{\hat{M}}, P_{\hat{M}}, \Xi_{\hat{M}})$ for $T^{*}[2]T[1](\widetilde{M}\times M)$ as in section \ref{QPDFT} which is associated with the base manifold and the generalized tangent frame. We also introduce the graded coordinates for the vector space as, 
$({\wha{Q}^{\hat{I}}}, {\wha{P}_{\hat{I}}})$ for $T^{*}[2]V[1]$ associated with the intermediate frame in the GSS case, 
and $({\bar{Q}}^{\hat{\bar{A}}},{\bar{P}}_{\hat{\bar{A}}})$ for $T^{*}[2]\bar{V}[1]$ associated with the local flat frame. 
The graded Poisson bracket is defined as
\begin{align}
\{X^{\hat{M}},\Xi_{\hat{N}}\}=&-\{\Xi_{\hat{N}}, X^{\hat{M}}\}=\delta^{\hat{M}}_{\hat{N}}, \\
\{Q^{\hat{M}}, P_{\hat{N}}\}=&\{P_{\hat{N}}, Q^{\hat{M}}\}=\delta^{\hat{M}}_{\hat{N}},\\
\{{\wha{Q}^{\hat{I}}}, {\wha{P}_{\hat{J}}}\}=&\{{\wha{P}_{\hat{J}}}, {\wha{Q}^{\hat{I}}}\}={\delta_{\hat{J}}^{\hat{I}}},\\
\{{\bar{Q}}^{\hat{\bar{A}}}, {\bar{P}}_{\hat{\bar{B}}}\}=&\{{\bar{P}}_{\hat{\bar{B}}}, {\bar{Q}}^{\hat{\bar{A}}}\}=\delta_{\hat{\bar{B}}}^{\hat{\bar{A}}}.
\end{align}

In order to discuss the geometric structures of the curved target space $\wha{M}=\wtil{M}\times M$ using the P-manifold ${\cal M}$, we introduce an injection map $j:\wha{M}\times T\wha{M}\times T\wha{M}\times T^{*}\wha{M}\to {\cal M}$ as
\begin{align}
j:(X^{\hat{M}},\partial_{\hat{M}},\partial_{\hat{M}},dX^{\hat{M}})\longmapsto (X^{\hat{M}},\Xi_{\hat{M}},P_{\hat{M}},Q^{\hat{M}}). 
\end{align}
That is, the $2D$-dimensional vector field $V^{\hat{M}}\partial_{\hat{M}}$ and 1-form field $\alpha_{\hat{M}}dX^{\hat{M}}$ on $\wha{M}$ are identified with $V^{\hat{M}}P_{\hat{M}}$ and $\alpha_{\hat{M}}Q^{\hat{M}}$ by the push forward $j_{*}$ and pullback $j^{*}$, respectively: 
\begin{gather}
j_{*}:V^{\hat{M}}\partial_{\hat{M}}\longmapsto V^{\hat{M}}P_{\hat{M}}, \quad j_{*}:\alpha_{\hat{M}}dX^{\hat{M}}\longmapsto \alpha_{\hat{M}}Q^{\hat{M}}, \\
j^{*}:V^{\hat{M}}P_{\hat{M}}\longmapsto V^{\hat{M}}\partial_{\hat{M}}, \quad j_{*}:\alpha_{\hat{M}}Q^{\hat{M}}\longmapsto \alpha_{\hat{M}}dX^{\hat{M}}. 
\end{gather}
We also use the coordinates $\wha{P}_{\hat{I}}$ and $\wha{Q}^{\hat{I}}$ corresponding to the basis of the intermediate frame and its dual, respectively.
Note that in this subsection, we are not using the DFT basis. 
Since the coordinate $\Xi_{\hat{M}}$ is conjugate to $X^{\hat{M}}$, the Poisson bracket $\{-,\Xi_{\hat{M}}\}$ corresponds to the partial derivative $\partial_{\hat{M}}$ on $\wha{M}$.

In order to introduce the covariant derivative $\nabla_{\hat{M}}$ on the target space $\wha{M}$ by the Poisson bracket, we define the covariantized coordinate $\Xi_{\hat{M}}^{\nabla}$ 
 on the P-manifold with affine connection $\Gamma$ and spin connection $W$ as
\begin{align}
\Xi_{\hat{M}}^{\nabla}:= \Xi_{\hat{M}}+\Gamma_{\hat{M}\hat{N}}{}^{\hat{P}}Q^{\hat{N}}P_{\hat{P}}+W_{\hat{M}\hat{I}}{}^{\hat{J}}\wha{Q}^{\hat{I}}\wha{P}_{\hat{J}}.
\end{align}
With this definition, the Poisson bracket of all coordinates with $\Xi_{\hat{M}}^{\nabla}$ are   
\begin{align}
\{X^{\hat{M}},\Xi^{\nabla}_{\hat{N}}\}=&\delta_{\hat{N}}^{\hat{M}}~,\label{XXiPoissonbracket1}\\
\{P_{\hat{M}},\Xi^{\nabla}_{\hat{N}}\}=&\Gamma_{\hat{N}\hat{M}}{}^{\hat{P}}P_{\hat{P}}~,\label{CovariantPoissonBrakets2}\\
\{Q^{\hat{M}},\Xi^{\nabla}_{\hat{N}}\}=&-\Gamma_{\hat{N}\hat{P}}{}^{\hat{M}}Q^{\hat{P}}~,\\
\{\wha{P}_{\hat{I}},\Xi^{\nabla}_{\hat{N}}\}=&W_{\hat{N}\hat{I}}{}^{\hat{J}}\wha{P}_{\hat{J}}~,\\
\{\wha{Q}^{\hat{I}},\Xi^{\nabla}_{\hat{N}}\}=&-W_{\hat{N}\hat{\bar{J}}}{}^{\hat{\bar{I}}}\wha{Q}^{\hat{\bar{J}}}~,
\\
\{\Xi_{\hat{M}}^{\nabla},\Xi_{\hat{N}}^{\nabla}\}
=&-\frac{1}{2}R_{\hat{M}\hat{N}\hat{S}}{}^{\hat{R}}Q^{\hat{S}}P_{\hat{R}}-\frac{1}{2}\bar{R}_{\hat{M}\hat{N}\hat{I}}{}^{\hat{J}}\wha{Q}^{\hat{I}}\wha{P}_{\hat{J}}~.\label{CovariantPoissonBrakets6}
\end{align}
Here, $R_{\hat{M}\hat{N}\hat{S}}{}^{\hat{R}}$ and $\bar{R}_{\hat{M}\hat{N}\hat{I}}{}^{\hat{J}}$ are curvature tensor defined by $\Gamma_{\hat{M}\hat{N}}{}^{\hat{P}}$ and $W_{\hat{M}\hat{I}}{}^{\hat{J}}$, respectively:
\begin{align}
R_{\hat{M}\hat{N}\hat{S}}{}^{\hat{R}}&=2\partial_{[\hat{M}}\Gamma_{\hat{N}]\hat{S}}{}^{\hat{R}}+2\Gamma_{[\hat{M}|\hat{P}|}{}^{\hat{R}}\Gamma_{\hat{N}]\hat{S}}{}^{\hat{P}},
\\
\bar{R}_{\hat{M}\hat{N}\hat{I}}{}^{\hat{J}}&=2\partial_{[\hat{M}}W_{\hat{N}]\hat{I}}{}^{\hat{J}}+2W_{[\hat{M}|\hat{K}|}{}^{\hat{J}}W_{\hat{N}]\hat{I}}{}^{\hat{K}}.
\end{align}
The Poisson bracket $\{-,\Xi_{\hat{M}}^{\nabla}\}$ with the vector fields $V^{\hat{M}}P_{\hat{M}}, \wha{V}^{\hat{I}}\wha{P}_{\hat{I}}$ and 1-forms $\alpha_{\hat{M}}Q^{\hat{M}}, \wha{\alpha}_{\hat{I}}\wha{Q}^{\hat{I}}$ on the P-manifold realizes their covariant derivative on $\wha{M}$:
\begin{gather}
\{V^{\hat{M}}P_{\hat{M}},\Xi^{\nabla}_{\hat{N}}\}=\nabla_{\hat{N}}V^{\hat{M}}P_{\hat{M}},\quad \{\alpha_{\hat{M}}Q^{\hat{M}}, \Xi_{\hat{N}}^{\nabla}\}=\nabla_{\hat{N}} \alpha_{\hat{M}}Q^{\hat{M}},\\
\{\wha{V}^{\hat{I}}\wha{P}_{\hat{I}},\Xi^{\nabla}_{\hat{N}}\}=\nabla_{\hat{N}}\wha{V}^{\hat{I}}\wha{P}_{\hat{I}},\quad \{\wha{\alpha}_{\hat{I}}\wha{Q}^{\hat{I}}, \Xi_{\hat{N}}^{\nabla}\}=\nabla_{\hat{N}} \wha{\alpha}_{\hat{I}}\wha{Q}^{\hat{I}},
\end{gather}
where the covariant derivative of a tensor $V_{\hat I}{}^{\hat M}$ is defined as
\begin{align}
\nabla_{\hat{P}}V_{\hat{I}}^{\ \hat{M}}=\partial_{\hat{P}}V_{\hat{I}}{}^{\hat{M}}
+\Gamma_{\hat{P}\hat{N}}{}^{\hat{M}}V_{\hat{I}}^{\ \hat{N}}-W_{\hat{P}\hat{I}}{}^{\hat{J}}V_{\hat{J}}^{\ \hat{M}}.
\end{align}

We introduce the background vielbein $E_{\hat{I}}{}^{\hat{M}}$ on $\wha{M}$ by the canonical transformation as in the previous section. 
Let us denote this canonical transformation by $\phi$. It is defined by
\begin{gather}
\phi(P_{\hat{M}})=E^{\hat{I}}{}_{\hat{M}}\wha{P}_{\hat{I}},\quad \phi(\wha{P}_{\hat{I}})=E_{\hat{I}}{}^{\hat{M}}P_{\hat{M}},\\
\phi(Q^{\hat{M}})=E_{\hat{I}}{}^{\hat{M}}\wha{Q}^{\hat{I}},\quad \phi(\wha{Q}^{\hat{I}})=E^{\hat{I}}{}_{\hat{M}}Q^{\hat{M}}. 
\end{gather}
where $E^{\hat{I}}{}_{\hat{M}}$ is the inverse vielbein, $E^{\hat{I}}{}_{\hat{M}}E_{\hat{J}}{}^{\hat{M}}=\delta^I_J$.
Furthermore, we  assume that $\phi$ acts trivially on the base manifold coordinate as $\phi(X^{\hat{M}})=X^{\hat{M}}$.
We also require the condition of Poisson map on the dual coordinate $\Xi_{\hat N}$. Especially,  
\begin{align}
&\phi(\{X^{\hat{M}},\Xi_{\hat{N}}\})=\{\phi(X^{\hat{M}}),\phi(\Xi_{\hat{N}})\}=\delta_{\hat{N}}^{\hat{M}},
\cr&\phi(\{\ \bullet\ ,\Xi_{\hat{N}}\})=\{\phi(\ \bullet\ ),\phi(\Xi_{\hat{N}})\}=0,
\label{PoissonMapcondition1}
\end{align}
should hold, where $\bullet$ represents all other coordinates. 
Then, the consistent canonical transformation of the dual coordinate $\Xi_{\hat{M}}$ is 
\begin{align}
\phi(\Xi_{\hat{M}})=\Xi_{\hat{M}}-\Omega_{\hat{M}\hat{N}}{}^{\hat{P}}Q^{\hat{N}}P_{\hat{P}}+\Omega_{\hat{M}\hat{N}}{}^{\hat{P}}E_{\hat{I}}{}^{\hat{N}}E^{\hat{J}}{}_{\hat{P}}\wha{Q}^{\hat{I}}\wha{P}_{\hat{J}}.
\label{CTofDualcoordinate}
\end{align}
Here $\Omega_{\hat{M}\hat{N}}{}^{\hat{P}}$ is the Weitzenb\"ock connection of the background defined
by
\begin{equation}
\Omega_{\hat{M}\hat{N}}{}^{\hat{P}} =- \partial_{\hat M}E^{\hat I}{}_{\hat N}{} { E}_{\hat{I}}{}^{\hat{P}}~.
\label{Weitzenbackground}
\end{equation}

The transformation rule (\ref{CTofDualcoordinate}) is obtained straightforwardly: 
The canonical transformation of the dual coordinate $\Xi_{\hat{M}}$ should be degree preserving. It can be expanded in general as
\begin{align}
\phi(\Xi_{\hat{M}})=A_{\hat{M}}{}^{\hat{N}}\Xi_{\hat{N}}&+\frac{1}{2}B_{\hat{M}}{}^{\hat{N}\hat{P}}P_{\hat{N}}P_{\hat{P}}+C_{\hat{M}\hat{N}}{}^{\hat{P}}Q^{\hat{N}}P_{\hat{P}}+\frac{1}{2}D_{\hat{M}\hat{N}\hat{P}}Q^{\hat{N}}Q^{\hat{P}}\cr
&+\frac{1}{2}\wha{B}_{\hat{M}}{}^{\hat{I}\hat{J}}\wha{P}_{\hat{I}}\wha{P}_{\hat{J}}+\wha{C}_{\hat{M}\hat{I}}{}^{\hat{J}}\wha{Q}^{\hat{I}}\wha{P}_{\hat{J}}+\frac{1}{2}\wha{D}_{\hat{M}\hat{I}\hat{J}}\wha{Q}^{\hat{I}}\wha{Q}^{\hat{J}},
\end{align}
where the coefficients are functions of $X^{\hat{M}}$. 
Then, the condition of the canonical transformation (\ref{PoissonMapcondition1}) determines the coefficients.

The covariantized coordinate $\Xi^\nabla$ should be consistent with the canonical transformation $\phi$.
Using these transformation rules, the canonical transformation $\phi$ acts on $\Xi_{\hat{M}}^{\nabla}$ as
\begin{align}
\phi(\Xi_{\hat{M}}^{\nabla})=\Xi_{\hat{M}}+(W_{\hat{M}\hat{I}}{}^{\hat{J}}E^{\hat{I}}{}_{\hat{N}}E_{\hat{J}}{}^{\hat{P}}-\Omega_{\hat{M}\hat{N}}{}^{\hat{P}})Q^{\hat{N}}P_{\hat{P}}+(\Omega_{\hat{M}\hat{N}}{}^{\hat{P}}+\Gamma_{\hat{M}\hat{N}}{}^{\hat{P}})E_{\hat{I}}{}^{\hat{N}}E^{\hat{J}}{}_{\hat{P}}\wha{Q}^{\hat{I}}\wha{P}_{\hat{J}}.
\end{align}

If we require the vielbein postulate, 
\begin{align}
\nabla_{\hat{M}}E_{\hat{I}}{}^{\hat{N}}=0,\label{VielbeinPostulate}
\end{align}
that is,
\begin{align}
W_{\hat{M}\hat{I}}{}^{\hat{J}}E^{\hat{I}}{}_{\hat{N}}E_{\hat{J}}{}^{\hat{P}}-\Omega_{\hat{M}\hat{N}}{}^{\hat{P}}-\Gamma_{\hat{M}\hat{N}}{}^{\hat{P}}=0. \label{VielbeinpostulateConnection}
\end{align}
then, this leads to the invariance of the covariant coordinate $\Xi_{\hat{M}}^{\nabla}$ under the canonical transformation $\phi$ as 
\begin{align}
\phi(\Xi_{\hat{M}}^{\nabla})=\Xi_{\hat{M}}^{\nabla}.
\end{align}
Consequently, the Poisson brackets  
(\ref{XXiPoissonbracket1})-(\ref{CovariantPoissonBrakets6}) are invariant under the canonical transformation $\phi$

\subsubsection{$O(D,D)$ structure and DFT basis}
In order to apply the discussion above to DFT, we introduce the $O(D,D)$ structure by defining the standard $O(D,D)$ metric $\eta^{\hat I\hat J}$
on the intermediate frame.
The background vielbein $E_{\hat{I}}{}^{\hat{M}}$ and the metric $\eta^{\hat M\hat N}$ on the generalized tangent frame of the background are related as
\begin{align}
\eta^{\hat{M}\hat{N}}=E_{\hat{I}}{}^{\hat{M}}\eta^{\hat{I}\hat{J}}E_{\hat{J}}{}^{\hat{N}},\label{CurvedEtaDef}
\end{align}
where $\eta^{\hat M\hat N}$ is not constant in general. 
We require compatibility with the $O(D,D)$ metric in the intermediate frame:
\begin{align}
\nabla_{\hat{M}}\eta_{\hat{I}\hat{J}}=0. \label{IntermediateEtaCompatibility}
\end{align}
 The equation \eqref{IntermediateEtaCompatibility} gives the following 
condition on the spin connection:
\begin{align}
{W_{\hat{M}\hat{J}}{}^{\hat{L}}\eta_{\hat{L}\hat{K}}+W_{\hat{M}\hat{K}}{}^{\hat{L}}\eta_{\hat{J}\hat{L}}}=
{W_{\hat{M}\hat{J}\hat{K}}+W_{\hat{M}\hat{K}\hat{J}}}=
0.
\end{align}

In the present formulation, we also require the vielbein postulate \eqref{VielbeinPostulate}, then  
the covariant derivative of $\eta_{\hat{M}\hat{N}}$ in \eqref{CurvedEtaDef}  automatically vanishes as a consequence of \eqref{IntermediateEtaCompatibility},
\begin{align}
\nabla_{\hat{M}}\eta_{\hat{N}\hat{P}}	=& \partial_{\hat{M}}\eta_{\hat{N}\hat{P}}-\Gamma_{\hat{M}\hat{N}}^{\ \ \ \ \hat{Q}}\eta_{\hat{Q}\hat{P}}-\Gamma_{\hat{M}\hat{P}}^{\ \ \ \ \hat{Q}}\eta_{\hat{N}\hat{Q}}=0.
\end{align}

Now we are ready to introduce the DFT basis on the covariantized pre-QP-manifold as in the section \ref{QPDFT}. Explicitly, we have
\begin{align}
P'_{\hat{M}}:=\frac{1}{\sqrt{2}}(P_{\hat{M}}+\eta_{\hat{M}\hat{N}}Q^{\hat{N}}),\quad &Q'^{\hat{M}}:=\frac{1}{\sqrt[]{2}}(Q^{\hat{M}}-\eta^{\hat{M}\hat{N}}P_{\hat{N}}),\label{primeP}\\
\wha{P}'_{\hat{I}}:=\frac{1}{\sqrt{2}}(\wha{P}_{\hat{I}}+\eta_{\hat{I}\hat{J}}\wha{Q}^{\hat{J}}),\quad &\wha{Q}'^{\hat{I}}:=\frac{1}{\sqrt[]{2}}(\wha{Q}^{\hat{I}}-\eta^{\hat{I}\hat{J}}\wha{P}_{\hat{J}}),\label{primeQ}\\
\bar{P}'_{\hat{\bar{A}}}:=\frac{1}{\sqrt{2}}(\bar{P}_{\hat{\bar{A}}}+\eta_{\hat{\bar{A}}\hat{\bar{B}}}\bar{Q}^{\hat{\bar{B}}}),\quad &\bar{Q}'^{\hat{\bar{A}}}:=\frac{1}{\sqrt[]{2}}(\bar{Q}^{\hat{\bar{A}}}-\eta^{\hat{\bar{A}}\hat{\bar{B}}}\bar{P}_{\hat{\bar{B}}}).
\end{align}
where $\eta^{\hat{\bar A}\hat{\bar B}}$ is the $O(D,D)$ metric on the local flat frame.

The Poisson brackets among the DFT basis coordinates $P'_{\hat{M}},\wha{P}'_{\hat{I}}, Q'^{\hat{M}}, \wha{Q}'^{\hat{I}}$ are given by 
\begin{align}
\{P'_{\hat{M}},P'_{\hat{N}}\}&=\eta_{\hat{M}\hat{N}},\quad \{Q'^{\hat{M}},Q'^{\hat{N}}\}=\eta^{\hat{M}\hat{N}}, 
\label{PBprimePQ}
\cr
\{{\wha{P}'_{\hat{I}}},{\wha{P}'_{\hat{J}}}\}&={\eta_{\hat{I}\hat{J}}},\quad \{{\wha{Q}'^{\hat{I}}},{\wha{Q}'^{\hat{J}}}\}={\eta^{\hat{I}\hat{J}}}, \cr
\{{\bar{P}}'_{\hat{\bar{A}}},{\bar{P}}'_{\hat{\bar{B}}}\}&=\eta_{\hat{\bar{A}}\hat{\bar{B}}},\quad \{{\bar{Q}}'^{\hat{\bar{A}}},{\bar{Q}}'^{\hat{\bar{B}}}\}=\eta^{\hat{\bar{A}}\hat{\bar{B}}}, 
\end{align}
and with the covariant coordinate $\Xi_{\hat{M}}^{\nabla}$ 
\begin{align}
\{P'_{\hat{M}}, \Xi_{\hat{N}}^{\nabla}\}=&\Gamma_{\hat{N}\hat{M}}^{\ \ \ \ \hat{P}}P'_{\hat{P}}
,\label{CovariantPoissonBracketsDFT1}\\
\{Q'^{\hat{M}}, \Xi_{\hat{N}}^{\nabla}\}=&-\Gamma_{\hat{N}\hat{Q}}^{\ \ \ \hat{M}}Q'^{\hat{Q}}
,\\
\{{\wha{P}'_{\hat{I}}},\Xi_{\hat{N}}^{\nabla}\}=&{W_{\hat{N}\hat{I}}{}^{\hat{J}}\wha{P}'_{\hat{J}}}
,\\
\{{\wha{Q}'^{\hat{I}}}, \Xi_{\hat{N}}^{\nabla}\}=&-{W_{\hat{N}\hat{J}}{}^{\hat{I}}\wha{Q}'^{\hat{J}}}
,\\
\{\Xi_{\hat{M}}^{\nabla},\Xi_{\hat{N}}^{\nabla}\}
=&-\frac{1}{2}R_{\hat{M}\hat{N}\hat{S}}{}^{\hat{R}}(Q'^{\hat{S}}+\eta^{\hat{S}\hat{T}}P'_{\hat{T}})(P'_{\hat{R}}-\eta_{\hat{R}\hat{U}}Q'^{\hat{U}})\cr
&~~~~~~~~-\frac{1}{2}{\bar{R}_{\hat{M}\hat{N}\hat{I}}{}^{\hat{J}}(\wha{Q}'^{\hat{I}}+\eta^{\hat{I}\hat{K}}\wha{P}'_{\hat{K}})(\wha{P}'_{\hat{J}}-\eta_{\hat{J}\hat{L}}\wha{Q}'^{\hat{L}})}.\label{CovariantPoissonBracketsDFT5}
\end{align}

The canonical transformation $\phi$ acts on the DFT basis $Q'^{\hat{M}},\wha{Q}'^{\hat{I}},P'_{\hat{M}}$ and $\wha{P}'_{\hat{I}}$ as,
\begin{gather}
\phi(Q'^{\hat{M}})=E_{\hat{I}}{}^{\hat{M}}\wha{Q}'^{\hat{I}}, \quad \phi(\wha{Q}'^{\hat{I}})=E^{\hat{I}}{}_{\hat{M}}Q'^{\hat{M}},\\
\phi(P'_{\hat{M}})=E^{\hat{I}}{}_{\hat{M}}\wha{P}'_{\hat{I}},\quad \phi(\wha{P}'_{\hat{I}})=E_{\hat{I}}{}^{\hat{M}}P'_{\hat{M}}.\label{DFTCanonicalTransfPhi}
\end{gather}

We see that the Poisson brackets \eqref{CovariantPoissonBracketsDFT1} -- \eqref{CovariantPoissonBracketsDFT5} are of the same form as \eqref{CovariantPoissonBrakets2} -- \eqref{CovariantPoissonBrakets6} and thus, the Poisson bracket of the generalized vectors in DFT basis with $\Xi_{\hat{M}}^{\nabla}$ realizes the covariant derivative: 
\begin{align}
\{V^{\hat{M}}P'_{\hat{M}}, \Xi_{\hat{N}}^{\nabla}\}=\nabla_{\hat{N}}V^{\hat{M}}P'_{\hat{M}},\quad \{\wha{V}^{\hat{I}}\wha{P}'_{\hat{I}}, \Xi_{\hat{N}}^{\nabla}\}=\nabla_{\hat{N}}\wha{V}^{\hat{I}}\wha{P}'_{\hat{I}}. 
\end{align}

\subsubsection{Pre-QP-structure and gauge algebra\label{covQstructures}}

Since the covariant coordinate $\Xi_{\hat{M}}^{\nabla}$ realizes the background covariant derivative, we can formulate the pre-Q-sturcture on the covariantized pre-QP-manifold. The pre-Q-structure written with $\Xi_{\hat{M}}^{\nabla}$ realizes the generalized Lie derivatives in background covariant form. The simplest 
Hamiltonian function is 
\begin{align}
\Theta_{0}^{\nabla}=\eta^{\hat{M}\hat{N}}\Xi_{\hat{M}}^{\nabla}P'_{\hat{N}}. 
\end{align}
The derived bracket from this Hamiltonian function defines the covariant generalized Lie derivative ${\cal L}^\nabla_{\Lambda}$ with generalized vector $\Lambda=\Lambda^{\hat{M}}(X)P'_{\hat{M}}$ on a generalized vector $V=V^{\hat{M}}(X)P'_{\hat{M}}$ on the background
\begin{align}
{\cal L}_{\Lambda}^{\nabla}V\equiv -\{\{{\Lambda},\Theta_{0}^{\nabla}\},{V}\}
 ~,
\end{align}
where the covariant generalized Lie derivative is given by replacing the derivative in the generalized Lie derivative by $\nabla_{\hat M}$:
\begin{align}
{\cal L}_{\Lambda}^{\nabla}V^{\hat{M}}=&\Lambda^{\hat{N}}\nabla_{\hat{N}}V^{\hat{M}}+(\nabla^{\hat{M}}\Lambda_{\hat{N}}-\nabla_{\hat{N}}\Lambda^{\hat{M}})V^{\hat{N}}~.
\label{BCovGeneralizedLieDerivative}
\end{align}

We define the canonically transformed Hamiltonian function using $\phi$ introduced in the previous subsection: 
\begin{align}
\wha{\Theta}_{0}^{\nabla}=\phi(\Theta_{0}^{\nabla})={\eta^{\hat{I}\hat{J}}E_{\hat{I}}{}^{\hat{M}}}\Xi_{\hat{M}}^{\nabla}{\wha{P}'_{\hat{J}}}.\label{BackgroundHamiltonian}
\end{align}
Since the canonical transformation $\phi$ maps the general tangent frame to intermediate frame as in \eqref{DFTCanonicalTransfPhi}, the transformed Hamiltonian $\wha{\Theta}_{0}^{\nabla}$ realizes the covariant generalized Lie derivative for generalized vectors ${\wha{\Lambda}=\wha{\Lambda}^{\hat{I}}(X)\wha{P}'_{\hat{I}}}$ and ${\wha{V}=\wha{V}^{\hat{I}}(X)\wha{P}'_{\hat{I}}}$ in the intermediate $O(D,D)$ basis as the derived bracket: 

\begin{align}
{\cal L}_{{\wha{\Lambda}}}^{\nabla}{\wha{V}}\equiv -\{\{{\wha{\Lambda}},{\wha{\Theta}_{0}^{\nabla}}\},{\wha{V}}\}~. 
\end{align}
where
\begin{align}
{\cal L}_{{\wha{\Lambda}}}^{\nabla}{\wha{V}^{\hat{I}}}=&{\wha{\Lambda}^{\hat{J}}}\nabla_{\hat{J}}V^{\hat{I}}+({\nabla^{\hat{I}}\wha{\Lambda}_{\hat{J}}}-{\nabla_{\hat{J}}\wha{\Lambda}^{\hat{I}}}){\wha{V}^{\hat{J}}}.\label{interCovGeneralizedLieDerivative}
\end{align}
Here, we want to make some remarks.
First, note that the vielbein postulate guarantees the equivalence between the generalized Lie derivatives (\ref{BCovGeneralizedLieDerivative}) and (\ref{interCovGeneralizedLieDerivative}) defined on the two different frames.
Second, once the covariant generalized Lie derivative is defined, we can formulate
the generalized torsion in DFT on the pre-QP-manifold.
The generalized torsion of the background is then
\begin{align}
T(V_1,V_2)=\{\{V_1,\Theta_0^\nabla-\Theta_0\}, V_2\}={\cal T}_{\hat P\hat N}{}^{\hat M}V_{1}^{\hat{P}}V_{2}^{\hat{N}}P'_{\hat M}
\end{align}
where 
\begin{align}
{\cal T}_{\hat P\hat N}{}^{\hat M}=\Gamma_{\hat P\hat N}{}^{\hat M}-\Gamma_{\hat N\hat P}{}^{\hat M}+\Gamma^{\hat M}{}_{\hat P\hat N}
\end{align}
Finally, note that we can consider other possibilities of Hamiltonian functions written with $Q'$ and ${\wha{Q}'}$, but in our discussion here for the DFT case it is sufficient to consider the generalized vectors in the $P'$ and ${\wha{P}'}$ sector of the DFT basis.

The closure condition of the generalized Lie derivative \eqref{interCovGeneralizedLieDerivative} is the weak master equation \eqref{WeakMasterEQ} for generalized vectors:
\begin{align}
\{\{\{\{{\wha{\Theta}_{0}^{\nabla}},{\wha{\Theta}_{0}^{\nabla}}\},{\wha{V}_{1}}\}, {\wha{V}_{2}}\},{\wha{V}_{3}}\}=0. 
\end{align}
The above weak master equation leads to the following condition for the spin connection ${W_{\hat{M}\hat{I}}{}^{\hat{J}}}$ and arbitrary generalized vectors ${\wha{V}_{1}},\wha{V}_{2}$ and ${\wha{V}_{3}}$, 
\begin{align}
&-2(\partial^{\hat{M}}\wha{V}_{1}^{\hat{J}}\wha{V}_{2\hat{J}}\partial_{\hat{M}}\wha{V}_{3}^{\hat{I}}-2\partial^{\hat{M}}\wha{V}_{1}^{[\hat{J}}\partial_{\hat{M}}\wha{V}_{2}^{\hat{I}]}\wha{V}_{3\hat{J}})\cr
&-2\eta^{\hat{I}\hat{H}}\Big(2\Omega_{[\hat{H}\hat{J}]\hat{K}}-3W_{[\hat{H}\hat{J}\hat{K}]}\Big)E^{\hat{K}\hat{M}}\Big[\partial_{\hat{M}}\wha{V}_{1}^{\hat{L}}\wha{V}_{2\hat{L}}\wha{V}_{3}^{\hat{J}}-\partial_{\hat{M}}\wha{V}_{1}^{\hat{L}}\wha{V}_{2}^{\hat{J}}\wha{V}_{3\hat{L}}+\wha{V}_{1}^{\hat{J}}\partial_{\hat{M}}\wha{V}_{2}^{\hat{L}}\wha{V}_{3\hat{L}}\Big]\cr
&+2\Big(2\Omega_{[\hat{L}\hat{J}]\hat{K}}-3W_{[\hat{L}\hat{J}\hat{K}]}\Big)E^{\hat{K}\hat{M}}\Big[\partial_{\hat{M}}\wha{V}_{1}^{\hat{I}}\wha{V}_{2}^{\hat{L}}\wha{V}_{3}^{\hat{J}}-\wha{V}_{1}^{\hat{L}}\partial_{\hat{M}}\wha{V}_{2}^{\hat{I}}\wha{V}_{3}^{\hat{J}}+\wha{V}_{1}^{\hat{L}}\wha{V}_{2}^{\hat{J}}\partial_{\hat{M}}\wha{V}_{3}^{\hat{I}}\Big]\cr
&-3!\wha{V}_{1}^{\hat{I}}\wha{V}_{2}^{\hat{J}}\wha{V}_{3}^{\hat{K}}\Big[2R_{[\hat{I}\hat{J}\hat{K}\hat{L}]}-W_{\hat{H}[\hat{I}\hat{J}}W^{\hat{H}}{}_{\hat{K}\hat{L}]}-2(2W_{[\hat{I}\hat{J}}{}^{\hat{H}}-2\Omega_{[\hat{I}\hat{J}}{}^{\hat{H}})W_{\hat{H}\hat{K}\hat{L}]}\Big]\cr
&=0.\label{CovariantQPClosure}
\end{align}
We discuss this condition order by order in the differential on generalized vectors. Then, the following conditions are sufficient to satisfy \eqref{CovariantQPClosure}.
\begin{align}
{\partial^{\hat{M}}\wha{V}_{1}^{\hat{J}}\wha{V}_{2\hat{J}}\partial_{\hat{M}}\wha{V}_{3}^{\hat{I}}-2\partial^{\hat{M}}\wha{V}_{1}^{[\hat{J}}\partial_{\hat{M}}\wha{V}_{2}^{\hat{I}]}\wha{V}_{3\hat{J}}}=0,\label{CCforCovLieDer1}\\
(-2{\Omega_{[\hat{I}\hat{J}]\hat{K}}}+3{W_{[\hat{I}\hat{J}\hat{K}]}}){E^{\hat{K}\hat{M}}}\partial_{\hat{M}}{\wha{V}_{1}^{\hat{L}}}=0,\label{CCforCovLieDer2}\\
{2R_{[\hat{I}\hat{J}\hat{K}\hat{L}]}-W_{\hat{H}[\hat{I}\hat{J}}W^{\hat{H}}{}_{\hat{K}\hat{L}]}-2(2W_{[\hat{I}\hat{J}}{}^{\hat{H}}-2\Omega_{[\hat{I}\hat{J}}{}^{\hat{H}})W_{\hat{H}\hat{K}\hat{L}]}}=0,\label{CCforCovLieDer3}
\end{align}
The first condition is the closure condition of the generalized Lie derivative and it is satisfied with the section condition.
The second and third condition can be solved for various cases. Here we just show that the solutions for ordinary DFT and DFT$_{\rm WZW}$ are included. 

\begin{enumerate}
\item[1)] The second condition \eqref{CCforCovLieDer2} is satisfied by taking, 
\begin{align}
3{W_{[\hat{I}\hat{J}\hat{K}]}}=2{\Omega_{[\hat{I}\hat{J}]\hat{K}}}.\label{3W=2Omega}
\end{align}
Then, the third condition
becomes
\begin{align}
{R_{[\hat{I}\hat{J}\hat{K}\hat{L}]}}+{W^{\hat{H}}{}_{[\hat{I}\hat{J}}W_{|\hat{H}|\hat{K}\hat{L}]}}=0.\label{RelationOfRandW}
\end{align}
This choice \eqref{3W=2Omega} corresponds to DFT$_{\rm WZW}$, by imposing $3{W_{[\hat{I}\hat{J}\hat{K}]}}={F_{\hat{I}\hat{J}\hat{K}}}$. 
Then, the condition \eqref{RelationOfRandW} is satisfied by the Jacobi identity for the structure constant. 
\item[2)] The second condition \eqref{CCforCovLieDer2} can be satisfied by
\begin{align}
{W_{\hat{I}[\hat{J}\hat{K}]}}={\Omega_{\hat{I}[\hat{J}\hat{K}]}},\label{2ndSolution}
\end{align}
with the section condition for ${E_{\hat{I}}{}^{\hat{M}}}$ and ${\wha{V}_{1}}$, 
\begin{align}
{\Omega_{\hat{K}\hat{I}\hat{J}}E^{\hat{K}\hat{M}}}\partial_{\hat{M}}{\wha{V}_{1}^{\hat{N}}}=&
{E_{\hat{J}\hat{Q}}}\eta^{\hat{P}\hat{M}}\partial_{\hat{P}}{E_{\hat{I}}{}^{\hat{Q}}}\partial_{\hat{M}}{\wha{V}_{1}^{\hat{N}}}=0.
\end{align}
With this choice of spin connection \eqref{2ndSolution}, the Riemann tensor vanishes: ${R_{\hat{I} \hat{J} \hat{K} \hat{L}}}=0$, so the third condition becomes
\begin{align}
\eta^{\hat{M}\hat{N}}\partial_{\hat{M}}{E_{[\hat{I}}{}^{\hat{P}}}\partial_{\hat{N}}{E_{\hat{J}}{}^{\hat{Q}} E_{\hat{K}\hat{P}}E_{\hat{L}]\hat{Q}}}=0.\label{2ndcaseThiredCondition}
\end{align}
With the section condition on the background vielbein, then \eqref{2ndcaseThiredCondition} is satisfied. 
\end{enumerate}

\subsubsection{Canonical transformations and generalized flux\label{BGCovCanonicalTransf}}
In section \ref{CanonicalTransfSec}, we discussed the canonical transformations of the Hamiltonian function and introduced the generalized fluxes. 
With a similar discussion, we can consider the canonical transformation on the pre-QP-manifold on general backgrounds. A list of the canonical transformations intertwining between the various frames is given in appendix \ref{Notations}.

The canonical transformation is generated by degree 2 functions. 
The possible canonical transformation functions made from $P', {\wha{P}'}$ and ${\bar{P}'}$ are
\begin{gather}
{A}:={A^{\hat{I}\hat{M}}}P'_{\hat{M}}{\wha{P}'_{\hat{I}}},\quad {\wha{A}}:={\wha{A}^{\hat{\bar{A}}\hat{J}}}{\wha{P}'_{\hat{J}}\bar{P}'_{\hat{\bar{A}}}},\quad {{\cal A}}:={{\cal A}^{\hat{\bar{A}}\hat{M}}}P'_{\hat{M}}{\bar{P}'_{\hat{\bar{A}}}},\label{CanonicalTransfFunc2-1}\\
u:=u_{\hat{P}}{}^{\hat{M}}\eta^{\hat{N}\hat{P}}P'_{\hat{M}}P'_{\hat{N}},\quad {\wha{u}}:={\wha{u}_{\hat{I}}{}^{\hat{J}}\eta^{\hat{K}\hat{I}}\wha{P}'_{\hat{J}}\wha{P}'_{\hat{K}}},\quad {\bar{u}}:={\bar{u}_{\hat{\bar{A}}}{}^{\hat{\bar{B}}}\eta^{\hat{\bar{C}}\hat{\bar{A}}}\bar{P}'_{\hat{\bar{B}}}\bar{P}'_{\hat{\bar{C}}}}. \label{CanonicalTransfFunc2-2}
\end{gather}
Note that ${A^{\hat{I}\hat{M}}}$, ${\wha{A}^{\hat{\bar{A}}\hat{J}}}$ and ${{\cal A}^{\hat{\bar{A}}\hat{M}}}$ are $GL(2D)$ matrices and have no correspondence to vielbeine ${E_{\hat{I}}{}^{\hat{M}},\wha{E}_{\hat{\bar{A}}}{}^{\hat{I}}}$ and ${\cal E}_{\hat{\bar{A}}}{}^{\hat{M}}$ in general. 
The transformation rules for $P',{\wha{P}'}$ and $\Xi_{\hat{M}}^{\nabla}$ are,
\begin{align*}
e^{\delta_{{A}}}P'_{\hat{M}}=&\sum_{n=0}^{\infty}\frac{(-1)^{{n}}}{(2n)!}(\kappa^{n})_{\hat{M}}{}^{\hat{N}}P'_{\hat{N}}+\sum_{n=0}^{\infty}\frac{(-1)^{{n}}}{(2n+1)!}(\kappa^{n})_{\hat{M}}{}^{\hat{N}}\eta_{\hat{N}\hat{P}}{A^{\hat{I}\hat{P}}\wha{P}'_{\hat{I}}},\\
e^{\delta_{{A}}} {\wha{P}'_{\hat{I}}}=&\sum_{n=0}^{\infty}\frac{(-1)^{{n}}}{(2n)!}{(\lambda^{n})_{\hat{I}}{}^{\hat{J}}}{\wha{P}'_{\hat{J}}}-\sum_{n=0}^{\infty}\frac{(-1)^{{n}}}{(2n+1)!}{(\lambda^{n})_{\hat{I}}{}^{\hat{J}}}{A_{\hat{J}}{}^{\hat{M}}}P'_{\hat{M}},\\
e^{\delta_{{A}}}\Xi_{\hat{M}}^{\nabla}=&\Xi_{\hat{M}}^{\nabla}-\frac{1}{2}\nabla_{\hat{M}}{A^{\hat{K}\hat{N}}}({A}^{-1})_{\hat{N}}{}^{\hat{J}}{\wha{P}'_{\hat{J}}}{\wha{P}'_{\hat{K}}}+\frac{1}{2}\nabla_{\hat{M}}{A^{\hat{K}\hat{N}}}({A^{-1}})_{\hat{N}}{}^{\hat{J}}e^{\delta_{{A}}}({\wha{P}'_{\hat{J}}})e^{\delta_{{A}}}({\wha{P}'_{\hat{K}}}),
\end{align*}
where the matrices $\kappa$ and $\lambda$ are defined as $\kappa_{\hat{M}}{}^{\hat{N}}={A_{\hat{I}\hat{M}}A^{\hat{J}\hat{N}}}$ and ${\lambda_{\hat{I}}{}^{\hat{J}}}={A_{\hat{I}\hat{N}}A^{\hat{J}\hat{N}}}$, respectively. By identifying ${A}$ with the vielbein, ${A_{\hat{I}}{}^{\hat{M}}}={E_{\hat{I}}{}^{\hat{M}}}$, $\kappa$ and $\lambda$ become $\delta_{\hat{M}}{}^{\hat{N}}$ and ${\delta_{\hat{I}}{}^{\hat{J}}}$, respectively. In this case, the canonical transformation for ${A}=\theta E$ are written as 
\begin{align}
e^{\theta\delta_{E}}P'_{\hat{M}}=&P'_{\hat{M}}\cos\theta+{E^{\hat{I}}{}_{\hat{M}}}{\wha{P}'_{\hat{I}}}\sin\theta,\\
e^{\theta\delta_{E}} {\wha{P}'_{\hat{I}}}=&-{E_{\hat{I}}{}^{\hat{M}}}P'_{\hat{M}}\sin\theta+{\wha{P}'_{\hat{I}}}\cos\theta. 
\end{align}
The coordinate $\Xi_{\hat{M}}^{\nabla}$ is invariant under the canonical transformation.
\begin{align}
e^{\theta\delta_{E}}\Xi_{\hat{M}}^{\nabla}=\Xi_{\hat{M}}^{\nabla}.
\end{align}
Then, the canonical transformation $e^{\theta\delta_{E}}$ of the Hamiltonian function ${\wha{\Theta}_{0}^{\nabla}}$ is
\begin{align}
e^{\theta\delta_{E}}{\wha{\Theta}}_{0}^{\nabla}={\eta^{\hat{I}\hat{J}}E_{\hat{I}}{}^{\hat{M}}}\Xi_{\hat{M}}^{\nabla}(-{E_{\hat{J}}{}^{\hat{N}}}P'_{\hat{N}}\sin\theta+{\wha{P}'_{\hat{J}}}\cos\theta).
\end{align}
{Especially, for $\theta=-\frac{\pi}{2}$ we obtain}
\begin{align}
e^{-\frac{\pi}{2}\delta_{E}}{\wha{\Theta}_{0}^{\nabla}}=\Theta_{0}^{\nabla}.
\end{align}

Applying the similar discussion to ${\wha{A}}$, we can introduce the fluctuation vielbein ${\wha{E}_{\hat{\bar{A}}}{}^{\hat{I}}}$. When we take ${\wha{A}_{\hat{\bar{A}}}{}^{\hat{I}}}=\frac{\pi}{2}{\wha{E}_{\hat{\bar{A}}}{}^{\hat{I}}}$, we obtain the canonical transformation rules as follows.
\begin{align}
e^{{\pi\over2}\delta_{{\wha{E}}}}{\wha{P}'_{\hat{I}}}=&{\wha{E}^{\hat{\bar{B}}}{}_{\hat{I}}\bar{P}'_{\hat{\bar{B}}}},\\
e^{{\pi\over2}\delta_{{\wha{E}}}} {\bar{P}'_{\hat{\bar{A}}}}=&-{\wha{E}_{\hat{\bar{A}}}{}^{\hat{I}}\wha{P}'_{\hat{I}}},\\
e^{{\pi\over2}\delta_{{\wha{E}}}}\Xi_{\hat{M}}^{\nabla}
=&\Xi_{\hat{M}}-\frac{1}{2}{\cal E}^{\hat{\bar{C}}}{}_{\hat{M}}\wha{\Omega}^{\nabla}_{\hat{\bar{C}}\hat{\bar{A}}\hat{\bar{B}}}{\wha{E}^{\hat{\bar{A}}}{}_{\hat{I}}\wha{E}^{\hat{\bar{B}}}{}_{\hat{J}}\wha{P}'^{\hat{I}} \wha{P}'^{\hat{J}}}
+\frac{1}{2}{\cal E}^{\hat{\bar{C}}}{}_{\hat{M}}\wha{\Omega}^{\nabla}_{\hat{\bar{C}}\hat{\bar{A}}\hat{\bar{B}}}{\bar{P}'^{\hat{\bar{A}}} \bar{P}'^{\hat{\bar{B}}}}.
\end{align}
Here we have defined a tensor ${\wha{\Omega}^{\nabla}_{\hat{\bar{A}}\hat{\bar{B}}\hat{\bar{C}}}}:={\cal E}_{\hat{\bar{A}}}{}^{\hat{M}}\nabla_{\hat{M}}{\wha{E}_{\hat{\bar{B}}}{}^{\hat{I}}\wha{E}_{\hat{\bar{C}}\hat{I}}}$. This tensor is just the covariantized Weitzenb\"ock connection ${\wha{\Omega}_{\hat{\bar{A}}\hat{\bar{B}}\hat{\bar{C}}}}$, which can be decomposed as
\begin{align}
{\wha{\Omega}^{\nabla}_{\hat{\bar{A}}\hat{\bar{B}}\hat{\bar{C}}}}=\wtil{\Omega}_{\hat{\bar{A}}\hat{\bar{B}}\hat{\bar{C}}}+{{\cal E}_{\hat{\bar{A}}}{}^{\hat{M}}W_{\hat{M}\hat{I}\hat{J}}\wha{E}_{\hat{\bar{B}}}{}^{\hat{I}}\wha{E}_{\hat{\bar{C}}}{}^{\hat{J}}}
\end{align}

The canonical transformation of the Hamiltonian function $\bar{\Theta}^{\nabla}_{0}$ is
\begin{align}
e^{\frac{\pi}{2}\delta_{{\wha{E}}}}{\wha{\Theta}_{0}^{\nabla}}=&{\cal E}_{\hat{\bar{A}}}{}^{\hat{M}}\Xi_{\hat{M}}^{\nabla}{\bar{P}'^{\hat{\bar{A}}}}+\frac{1}{3!}{\wha{{\cal F}}_{\hat{\bar{A}}\hat{\bar{B}}\hat{\bar{C}}}}{\bar{P}'^{\hat{\bar{A}}}\bar{P}'^{\hat{\bar{B}}}\bar{P}'^{\hat{\bar{C}}}}-\frac{1}{2}{\wha{\Omega}^{\nabla}_{\hat{\bar{A}}\hat{\bar{B}}\hat{\bar{C}}}}{\wha{E}^{\hat{\bar{B}}}{}_{\hat{I}}\wha{E}^{\hat{\bar{C}}}{}_{\hat{J}}\wha{P}'^{\hat{I}}\wha{P}'^{\hat{J}}\bar{P}'^{\hat{\bar{A}}}}
\end{align}
where ${\wha{{\cal F}}_{\hat{\bar{A}}\hat{\bar{B}}\hat{\bar{C}}}}:=3{\wha{\Omega}^{\nabla}_{[\hat{\bar{A}}\hat{\bar{B}}\hat{\bar{C}}]}}$. 
The generalized flux can be calculated by the derived bracket similarly as \eqref{GeneralizedFluxQP1}, as
\begin{align}
{\wha{{\cal F}}_{\hat{\bar{A}}\hat{\bar{B}}\hat{\bar{C}}}}
=-\{\{{\{{e^{\frac{\pi}{2}\delta_{\tilde{E}}}{\wha{\Theta}_{0}^{\nabla}},\bar{P}'_{\hat{\bar{A}}}}\},{\bar{P}'_{\hat{\bar{B}}}}\},\bar{P}'_{\hat{\bar{C}}}}\}
.
\end{align}
Since 
\begin{align}
\braket{{\cal E}_{\hat{\bar{C}}},{\cal L}^{\nabla}_{{\cal E}_{\hat{\bar{A}}}}{\cal E}_{\hat{\bar{B}}}}=&-\{{\cal E}_{\hat{\bar{C}}},\{\{{\cal E}_{\hat{\bar{A}}},\Theta_{0}^{\nabla}\},{\cal E}_{\hat{\bar{B}}}\}\}\cr
=&-\{{\wha{E}_{\hat{\bar{C}}}{}^{\hat{K}}E_{\hat{K}}{}^{\hat{P}}}P'_{\hat{P}},\{\{{\wha{E}_{\hat{\bar{A}}}{}^{\hat{I}}E_{\hat{I}}{}^{\hat{M}}}P'_{\hat{M}},\Theta_{0}^{\nabla}\},{\wha{E}_{\hat{\bar{B}}}{}^{\hat{J}}E_{\hat{J}}{}^{\hat{N}}}P'_{\hat{N}}\}\}\cr
=&-\{e^{-\frac{\pi}{2}\delta_{E}}{e^{-\frac{\pi}{2}\delta_{\wha{E}}}\bar{P}'_{\hat{\bar{C}}}},\{\{e^{-\frac{\pi}{2}\delta_{E}}{e^{-\frac{\pi}{2}\delta_{\wha{E}}}\bar{P}'_{\hat{\bar{A}}}},\Theta_{0}^{\nabla}\},e^{-\frac{\pi}{2}\delta_{E}}{e^{-\frac{\pi}{2}\delta_{\wha{E}}}\bar{P}'_{\hat{\bar{B}}}}\}\}\cr
=&-e^{-\frac{\pi}{2}\delta_{{\wha{E}}}}e^{-\frac{\pi}{2}\delta_{E}}\{{\bar{P}'_{\hat{\bar{C}}}},\{\{{\bar{P}'_{\hat{\bar{A}}}},e^{\frac{\pi}{2}\delta_{{\wha{E}}}}e^{\frac{\pi}{2}\delta_{E}}\Theta_{0}^{\nabla}\},{\bar{P}'_{\hat{\bar{B}}}}\}\}\cr
=&-e^{-\frac{\pi}{2}\delta_{{\wha{E}}}}e^{-\frac{\pi}{2}\delta_{E}}\{{\bar{P}'_{\hat{\bar{C}}}},\{\{{\bar{P}'_{\hat{\bar{A}}}},e^{\frac{\pi}{2}\delta_{\tilde{E}}}{\wha{\Theta}_{0}^{\nabla}}\},{\bar{P}'_{\hat{\bar{B}}}}\}\}\cr
=&{\wha{{\cal F}}_{\hat{\bar{A}}\hat{\bar{B}}\hat{\bar{C}}}}.
\end{align}

\subsubsection{Pre-Bianchi identities}
Now we can consider the pre-Bianchi identity for DFT on covariantized pre-QP-manifold.  
To define the $\cal B$ function \eqref{BianchiFunction}, we take the Hamiltonian function with general fluxes as,
\begin{align}
\Theta_{F}={\cal E}_{\hat{\bar{A}}}{}^{\hat{M}}\Xi_{\hat{M}}^{\nabla}{\bar{P}'^{\hat{\bar{A}}}}+\frac{1}{3!}\mathscr{F}_{\hat{\bar{A}}\hat{\bar{B}}\hat{\bar{C}}}{\bar{P}'^{\hat{\bar{A}}}\bar{P}'^{\hat{\bar{B}}}\bar{P}'^{\hat{\bar{C}}}}+\frac{1}{2}{\mathscr{G}_{\hat{\bar{A}}\hat{I}\hat{J}}}{\bar{P}'^{\hat{\bar{A}}}\wha{P}'^{\hat{I}}\wha{P}'^{\hat{J}}},
\end{align}
and for $\Theta_{0}$, we take ${\wha{\Theta}_{0}^{\nabla}}$ given in {\eqref{BackgroundHamiltonian}}. 
Since, as we have seen before, the canonical transformations $e^{\frac{\pi}{2}\delta_{{\wha{E}}}}e^{\frac{\pi}{2}\delta_{E}}$ generate the fluctuation on the background Hamiltonian $\Theta^\nabla_0$, we choose  
the $\alpha$ in eq.{\eqref{BianchiFunction}} as $\frac{\pi}{2}{\wha{E}}$. Then, the ${\cal B}$ function is given by
\begin{align}
{\cal B}\Big(\Theta_{F},&\wha{\Theta}_{0}^{\nabla},e^{\frac{\pi}{2}\delta_{{\wha{E}}}}\Big)\cr
=&(2\partial_{\hat{N}}{\cal E}_{\hat{\bar{A}}}{}^{\hat{M}}{\cal E}_{\hat{\bar{B}}}{}^{\hat{N}}+{\cal E}^{\hat{\bar{C}}\hat{M}}\mathscr{F}_{\hat{\bar{C}}\hat{\bar{A}}\hat{\bar{B}}}+2{\Gamma_{\hat{I}\hat{K}\hat{J}}}{E^{\hat{K}\hat{M}}\wha{E}_{\hat{\bar{A}}}{}^{\hat{I}}\wha{E}_{\hat{\bar{B}}}{}^{\hat{J}}}-{\wha{\Omega}^{\nabla\hat{M}}{}_{\hat{\bar{A}}\hat{\bar{B}}}})\Xi_{\hat{M}}^{\nabla}{\bar{P}'^{\hat{\bar{A}}}\bar{P}'^{\hat{\bar{B}}}}\nonumber\\
&+({\cal E}^{\hat{\bar{C}}\hat{M}}{\mathscr{G}_{\hat{\bar{C}}\hat{I}\hat{J}}}+{\wha{\Omega}^{\nabla \hat{M}}{}_{\hat{\bar{D}}\hat{\bar{C}}}\wha{E}^{\hat{\bar{D}}}{}_{\hat{I}}\wha{E}^{\hat{\bar{C}}}{}_{\hat{J}}})\Xi_{\hat{M}}^{\nabla}{\wha{P}'^{\hat{I}}\wha{P}'^{\hat{J}}}\nonumber\\
&+\Big(-\frac{1}{3}{\cal E}_{\hat{\bar{A}}}{}^{\hat{M}}\partial_{\hat{M}}\mathscr{F}_{\hat{\bar{B}}\hat{\bar{C}}\hat{\bar{D}}}+\frac{1}{4}\mathscr{F}^{\hat{\bar{E}}}{}_{\hat{\bar{A}}\hat{\bar{B}}}\mathscr{F}_{\hat{\bar{E}}\hat{\bar{C}}\hat{\bar{D}}}\nonumber\\
&+{\Gamma_{\hat{J}\hat{I}\hat{K}}\wha{\Omega}^{\nabla}_{\hat{N}\hat{\bar{A}}\hat{\bar{B}}}E^{\hat{I}\hat{N}}\wha{E}^{\hat{J}}{}_{\hat{\bar{C}}}\wha{E}^{\hat{K}}{}_{\hat{\bar{D}}}}-\frac{1}{4}{\wha{\Omega}^{\nabla}_{\hat{M}\hat{\bar{A}}\hat{\bar{B}}}\wha{\Omega}^{\nabla\hat{M}}{}_{\hat{\bar{C}}\hat{\bar{D}}}}\Big){\bar{P}'^{\hat{\bar{A}}}\bar{P}'^{\hat{\bar{B}}}\bar{P}'^{\hat{\bar{C}}}\bar{P}'^{\hat{\bar{D}}}}\nonumber\\
&+\Big({\cal E}_{\hat{\bar{A}}}{}^{\hat{M}}\nabla_{\hat{M}}{\mathscr{G}_{\hat{\bar{B}}\hat{I}\hat{J}}}+\frac{1}{2}\mathscr{F}^{\hat{\bar{E}}}{}_{\hat{\bar{A}}\hat{\bar{B}}}{\mathscr{G}_{\hat{\bar{E}}\hat{I}\hat{J}}}-{\mathscr{G}_{\hat{\bar{A}}\hat{I}}{}^{\hat{K}}\mathscr{G}_{\hat{\bar{B}}\hat{J}\hat{K}}}\nonumber\\
&-{\Gamma_{\hat{\bar{A}}\hat{\bar{C}}\hat{\bar{B}}}{\cal E}^{\hat{\bar{C}}\hat{N}}\wha{\Omega}^{\nabla}_{\hat{N}\hat{\bar{D}}\hat{\bar{E}}}\wha{E}^{\hat{\bar{D}}}{}_{\hat{I}}\wha{E}^{\hat{\bar{E}}}{}_{\hat{J}}}+\frac{1}{2}{\wha{\Omega}^{\nabla}_{\hat{M}\hat{\bar{A}}\hat{\bar{B}}}\wha{\Omega}^{\nabla\hat{M}}{}_{\hat{\bar{C}}\hat{\bar{D}}}\wha{E}^{\hat{\bar{C}}}{}_{\hat{I}}\wha{E}^{\hat{\bar{D}}}{}_{\hat{J}}}\Big){\bar{P}'^{\hat{\bar{A}}}\bar{P}'^{\hat{\bar{B}}}\wha{P}'^{\hat{I}}\wha{P}'^{\hat{J}}}\nonumber\\
&+\frac{1}{4}({\mathscr{G}^{\hat{\bar{A}}}{}_{\hat{I}\hat{J}}\mathscr{G}_{\hat{\bar{A}}\hat{K}\hat{L}}-\tilde{\Omega}^{\nabla}_{\hat{M}\hat{\bar{A}}\hat{\bar{B}}}\tilde{\Omega}^{\nabla\hat{M}}{}_{\hat{\bar{C}}\hat{\bar{D}}}\tilde{E}^{\hat{\bar{A}}}{}_{\hat{I}}\tilde{E}^{\hat{\bar{B}}}{}_{\hat{J}}\tilde{E}^{\hat{\bar{C}}}{}_{\hat{K}}\tilde{E}^{\hat{\bar{D}}}{}_{\hat{L}}}){\wha{P}'^{\hat{I}}\wha{P}'^{\hat{J}}\wha{P}'^{\hat{K}}\wha{P}'^{\hat{L}}},
\end{align}
{where $\Gamma_{\hat{\bar{A}}\hat{\bar{B}}\hat{\bar{C}}}=\Gamma_{\hat{I}\hat{J}\hat{K}}\wha{E}_{\hat{\bar{A}}}{}^{\hat{I}}\wha{E}_{\hat{\bar{B}}}{}^{\hat{J}}\wha{E}_{\hat{\bar{C}}}{}^{\hat{K}}$.} 
The pre-Bianchi identity $\mathcal{B}=0$ gives the following relations:
\begin{align}
&2\partial_{\hat{N}}{\cal E}_{\hat{\bar{A}}}{}^{\hat{M}}{\cal E}_{\hat{\bar{B}}}{}^{\hat{N}}+{\cal E}^{\hat{\bar{C}}\hat{M}}\mathscr{F}_{\hat{\bar{C}}\hat{\bar{A}}\hat{\bar{B}}}+2{\Gamma_{\hat{I}\hat{K}\hat{J}}}{E^{\hat{K}\hat{M}}\wha{E}_{\hat{\bar{A}}}{}^{\hat{I}}\wha{E}_{\hat{\bar{B}}}{}^{\hat{J}}}-{\wha{\Omega}^{\nabla\hat{M}}{}_{\hat{\bar{A}}\hat{\bar{B}}}}=0,\label{BGpreBianchi1}\\
&{\cal E}^{\hat{\bar{C}}\hat{M}}{\mathscr{G}_{\hat{\bar{C}}\hat{I}\hat{J}}}+{\wha{\Omega}^{\nabla \hat{M}}{}_{\hat{\bar{D}}\hat{\bar{C}}}\wha{E}^{\hat{\bar{D}}}{}_{\hat{I}}\wha{E}^{\hat{\bar{C}}}{}_{\hat{J}}}\label{BGpreBianchi2}=0,\\
&\frac{1}{3}{\cal E}_{\hat{\bar{A}}}{}^{\hat{M}}\partial_{\hat{M}}\mathscr{F}_{\hat{\bar{B}}\hat{\bar{C}}\hat{\bar{D}}}-\frac{1}{4}\mathscr{F}^{\hat{\bar{E}}}{}_{\hat{\bar{A}}\hat{\bar{B}}}\mathscr{F}_{\hat{\bar{E}}\hat{\bar{C}}\hat{\bar{D}}}-{\Gamma_{\hat{J}\hat{I}\hat{K}}\wha{\Omega}^{\nabla}_{\hat{N}\hat{\bar{A}}\hat{\bar{B}}}E^{\hat{I}\hat{N}}\wha{E}^{\hat{J}}{}_{\hat{\bar{C}}}\wha{E}^{\hat{K}}{}_{\hat{\bar{D}}}}+\frac{1}{4}{\wha{\Omega}^{\nabla}_{\hat{M}\hat{\bar{A}}\hat{\bar{B}}}\wha{\Omega}^{\nabla\hat{M}}{}_{\hat{\bar{C}}\hat{\bar{D}}}}=0,\label{BGpreBianchi3}\\
&{\cal E}_{\hat{\bar{A}}}{}^{\hat{M}}\nabla_{\hat{M}}{\mathscr{G}_{\hat{\bar{B}}\hat{I}\hat{J}}}+\frac{1}{2}\mathscr{F}^{\hat{\bar{E}}}{}_{\hat{\bar{A}}\hat{\bar{B}}}{\mathscr{G}_{\hat{\bar{E}}\hat{I}\hat{J}}}-{\mathscr{G}_{\hat{\bar{A}}\hat{I}}{}^{\hat{K}}\mathscr{G}_{\hat{\bar{B}}\hat{J}\hat{K}}}\nonumber\\
&~~~~~~~~~~-{\Gamma_{\hat{\bar{A}}\hat{\bar{C}}\hat{\bar{B}}}{\cal E}^{\hat{\bar{C}}\hat{N}}\wha{\Omega}^{\nabla}_{\hat{N}\hat{\bar{D}}\hat{\bar{E}}}\wha{E}^{\hat{\bar{D}}}{}_{\hat{I}}\wha{E}^{\hat{\bar{E}}}{}_{\hat{J}}}+\frac{1}{2}{\wha{\Omega}^{\nabla}_{\hat{M}\hat{\bar{A}}\hat{\bar{B}}}\wha{\Omega}^{\nabla\hat{M}}{}_{\hat{\bar{C}}\hat{\bar{D}}}\wha{E}^{\hat{\bar{C}}}{}_{\hat{I}}\wha{E}^{\hat{\bar{D}}}{}_{\hat{J}}}=0,\label{BGpreBianchi4}\\
&{\mathscr{G}^{\hat{\bar{A}}}{}_{\hat{I}\hat{J}}\mathscr{G}_{\hat{\bar{A}}\hat{K}\hat{L}}-\tilde{\Omega}^{\nabla}_{\hat{M}\hat{\bar{A}}\hat{\bar{B}}}\tilde{\Omega}^{\nabla\hat{M}}{}_{\hat{\bar{C}}\hat{\bar{D}}}\tilde{E}^{\hat{\bar{A}}}{}_{\hat{I}}\tilde{E}^{\hat{\bar{B}}}{}_{\hat{J}}\tilde{E}^{\hat{\bar{C}}}{}_{\hat{K}}\tilde{E}^{\hat{\bar{D}}}{}_{\hat{L}}}=0.\label{BGpreBianchi5}
\end{align}

By solving the first and second equations \eqref{BGpreBianchi1} and \eqref{BGpreBianchi2}, we obtain the local expressions of $\mathscr{F}_{\hat{\bar{A}}\hat{\bar{B}}\hat{\bar{C}}}$ and $\mathscr{G}_{\hat{\bar{A}}\hat{I}\hat{J}}$. The third and fourth equation \eqref{BGpreBianchi3} and \eqref{BGpreBianchi4} are the generalized Bianchi identities for $\mathscr{F}_{\hat{\bar{A}}\hat{\bar{B}}\hat{\bar{C}}}$ and $\mathscr{G}_{\hat{\bar{A}}\hat{I}\hat{J}}$, respectively. The last equation \eqref{BGpreBianchi5} is trivially satisfied due to \eqref{BGpreBianchi2}. These equations correspond to the pre-Bianchi identities \eqref{preBianchi1} - \eqref{preBianchi5}, i.e., if we take flat background, these equations reduce to the pre-Bianchi identities of the original DFT.

\subsection{Application to DFT$_{\rm WZW}$}
In this section, we apply our discussions to DFT$_{\rm WZW}$ and specify the pre-QP-structure.
We assume the background space as a group manifold $G$, so we can regard the coordinate $\wha{P}'_{\hat{I}}$ of its tangent space $TG$ as the generator of the Lie algebra of $G$ by the injection map $j'^{*}(\wha{P}'_{\hat{I}})=T_{\hat{I}}$. Then, the derived bracket of $\wha{P}'_{\hat{I}}$ should reproduce the Lie bracket: 
\begin{align}
-\{\{{\wha{P}'_{\hat{I}}}, {\wha{\Theta}_{0}^{\nabla}}\},{\wha{P}'_{\hat{J}}}\}={j'_{*}[T_{I},T_{J}]_{\rm Lie}}.\label{Liebracket} 
\end{align}
The left hand side is calculated as,
\begin{align}
-\{\{{\wha{P}'_{\hat{I}}}, {\wha{\Theta}_{0}^{\nabla}}\},{\wha{P}'_{\hat{J}}}\}=&({W^{\hat{K}}{}_{\hat{I}\hat{J}}}+2{W_{[\hat{I}\hat{J}]}{}^{\hat{K}}}){\wha{P}'_{\hat{K}}},
\end{align}
{and the right hand side is written by definition of a Lie algebra as
\begin{align}
{j'_{*}[T_{I},T_{J}]_{\rm Lie}}={F_{\hat{I}\hat{J}}{}^{\hat{K}}\wha{P}'_{\hat{K}}}.
\end{align}}
Thus, the condition \eqref{Liebracket} {leads the condition of the spin connection: }
\begin{align}
{W^{\hat{K}}{}_{\hat{I}\hat{J}}}+2{W_{[\hat{I}\hat{J}]}{}^{\hat{K}}}={F_{\hat{K}\hat{I}}{}^{\hat{J}}}.\label{SpinConnection}
\end{align}
This condition is solved by
\begin{align}
{W_{\hat{I}\hat{J}}{}^{\hat{K}}}=\frac{1}{3}{F_{\hat{I}\hat{J}}{}^{\hat{K}}}\label{DFTwzwSpinConnection}
\end{align}
and this solution is just the one proposed in the DFT$_{\rm WZW}$ model. 
We give the brief summary of DFT$_{\rm WZW}$ in the appendix.

With this spin connection \eqref{DFTwzwSpinConnection}, the derived bracket with $\wha{\Theta}_{0}^{\nabla}$ reproduces the generalized Lie derivative (\ref{CovGeneralizedLieDerivative}) of DFT$_{\rm WZW}$  as
\begin{align}
-\{\{\wha{\Lambda},\wha{\Theta}_{0}^{\nabla}\},\wha{V}\}=\Lambda^{\hat{J}}D_{\hat{J}}V^{\hat{I}}+(D^{\hat{I}}\Lambda_{\hat{J}}-D_{\hat{J}}\Lambda^{\hat{I}})V^{\hat{J}}+F^{\hat{I}}{}_{\hat{J}\hat{K}}\Lambda^{\hat{J}}V^{\hat{K}}.
\end{align}

Since the generalized Lie derivative is realized in superfield formalism, we can rewrite the generalized flux \eqref{GeneralizedFluxOnGroupMFD} of DFT$_{\rm WZW}$ by $\wha{\Theta}_{0}^{\nabla}$. 
\begin{align}
\braket{{\cal E}_{\hat{\bar{C}}}, {\cal L}^{\nabla}_{{\cal E}_{\hat{\bar{A}}}}{\cal E}_{\hat{\bar{B}}}}=-\{{\cal E}_{\hat{\bar{C}}},\{\{{\cal E}_{\hat{\bar{A}}},\wha{\Theta}_{0}^{\nabla}\},{\cal E}_{\hat{\bar{B}}}\}\}=\wtil{F}_{\hat{\bar{A}}\hat{\bar{B}}\hat{\bar{C}}}+F_{\hat{\bar{A}}\hat{\bar{B}}\hat{\bar{C}}}. \label{GeneralizedFluxDFTwzw}
\end{align}

We have already discussed the closure condition of the generalized Lie derivative in section \ref{covQstructures}. The weak master equation for arbitrary generalized vectors is written in \eqref{CCforCovLieDer1}--\eqref{CCforCovLieDer3}. The first equation \eqref{CCforCovLieDer1} is just a section condition for fluctuation fields. In DFT$_{\rm WZW}$ case, the spin connection is required to satisfiy $3W_{[\hat{I}\hat{J}\hat{K}]}=2\Omega_{[\hat{I}\hat{J}]\hat{K}}=F_{\hat{I}\hat{J}\hat{K}}$ in \eqref{DFTwzwSpinConnection}. The equation \eqref{CCforCovLieDer2} is satisfied with this condition, and \eqref{CCforCovLieDer3} leads the Jacobi identity of the structure constant $F_{\hat{I}\hat{J}\hat{K}}$. Thus, the weak master equation yields the section condition and the Jacobi identity as the closure condition of generalized Lie derivative of DFT$_{\rm WZW}$ \eqref{CCforDFTwzw}.

The pre-Bianchi identity is of the same form and we obtain \eqref{BGpreBianchi1}--\eqref{BGpreBianchi5} where the spin connection satisfies \eqref{DFTwzwSpinConnection}.

\section{Summary and Conclusion}
In this paper we formulated the algebraic structure of DFT on a pre-QP-manifold in an $O(D,D)$ covariant form and gave the closure condition of the corresponding generalized Lie derivative.
On the pre-QP-manifold we have more freedom to accommodate the DFT algebra: the  classical master equation is relaxed to a weaker derived bracket, i.e., the weak master equation, which 
defines the closure condition for the generalized Lie derivative. The weak master equation is in fact the condition of the Leibniz identity on the pre-QP-manifold. 

Introducing a local flat frame, we defined the generalized vielbein and investigated the canonical transformations and the generalized flux on the pre-QP-manifold.   
The definition
 of the generalized vielbein as an $O(D,D)$ covariant canonical transformation provides a picture which unifies all fluxes to a single generalized flux
 and gives a systematic way to formulate the Bianchi identities: The generalized Bianchi identities for all possible fluxes can be read off from a single pre-Bianchi identity. 

We have also shown that the GSS compactification fits to this formalism. To formulate the GSS compactification we introduce the intermediate coordinate. With this coordinate we can introduce the GSS twist by a canonical transformation intertwining between the general coordinate frame and the intermediate frame.  The fluctuation vielbein is then obtained as the canonical transformation from the intermediate frame to the local flat frame. This splitting gives the right description of the generalized flux in gauged DFT (GDFT). 

One advantage of the superfield formulation is its background independence as can be seen, e.g., from the special structure of the derived bracket. It shows that 
all information on the background is completely contained in the Hamiltonian 
function $\wha{\Theta}_{\rm GSS}$ of the intermediate frame. 
From the geometrical point of view,
 it is natural to formulate the geometry by using connection and covariant derivative. Therefore, in the last section, we developed the covariantized pre-QP-manifold to formulate the background geometry and gave a consistent theory with the Hamiltonian $\wha{\Theta}^\nabla$ instead of 
 $\wha{\Theta}_{\rm GSS}$. 
 
One important observation is the algebraic property of the 
$\Xi^{\nabla}$ coordinate. It shows that 
the Poisson structure is preserved in original $PQ$ coordinates as well as in the primed DFT basis. Note that the coordinate $\Xi^{\nabla}$ is fixed by the requirement of conservation of the P-structure and the vielbein postulate.

 We have shown that the familiar geometric objects are obtained from the pre-QP-manifold through certain identifications.
Thus, we have also shown the application of the superfield formulation to the group manifold case. A construction of DFT on the group manifold has been intensively discussed in \cite{Hassler:2017yza} in the wider context of Poisson-Lie T-duality, which contains both abelian and non-abelian T-duality as special cases. The solution of the weak master equation in section 4 reduces consistently to the DFT$_{{\rm WZW}}$ theory discussed in \cite{Hassler:2017yza}. 

Finally, we discuss the relation of our approach to GSS compactification and DFT on group manifolds in supermanifold formulation.
In section \ref{GSSasCanonicalTransf} we derived the GSS twist in terms of a canonical transformation. 
There, a GSS twist matrix $U_{\hat{M}}{}^{\hat{N}}(\mathbb{Y})$ is introduced via the canonical transformation of the degree 2 function $A=\frac{\pi}{2}U$ in \eqref{canonicalIntermediate}. 
The GSS twisted vielbein is \eqref{QPGSSanzats1} and the twisted Hamiltonian function is \eqref{GSSTwistedHamiltonian}.

On the other hand, the GSS compactified DFT can be regarded as a covariantized DFT whose background manifold is a generalized twisted torus. 
From this point of view, the background vielbein $E_{\hat{I}}{}^{\hat{M}}$ is identified with the GSS twist matrix $U_{\hat{I}}{}^{\hat{M}}(\mathbb{Y})$. 
With this identification, the total vielbein ${\cal E}_{\hat{\bar{A}}}{}^{\hat{M}}$ is identified with the GSS twisted vielbein. 
In our approach, this corresponds to the GSS twist matrix introduced by the canonical transformation $e^{\delta_{A}}$ in section \ref{BGCovCanonicalTransf}. 
We identify the transformation function $A_{\hat{I}}{}^{\hat{M}}$ with the GSS twist matrix $E_{\hat{I}}{}^{\hat{M}}(\mathbb{Y})$.
Then, the GSS twisted vielbein is obtained as
\begin{align}
{\cal E}_{\hat{\bar{A}}}{}^{\hat{M}}(X)P'^{\hat{M}}=e^{-\frac{\pi}{2}\delta_{E}}({\widehat{E}_{\hat{\bar{A}}}{}^{\hat{I}}(\mathbb{X})\wha{P}'_{\hat{I}}})={\widehat{E}_{\hat{\bar{A}}}{}^{\hat{I}}(\mathbb{X})E_{\hat{I}}{}^{\hat{M}}(\mathbb{Y})P'_{\hat{M}}}. 
\end{align}
In this case, the GSS twisted Hamiltonian function \eqref{BackgroundHamiltonian} becomes, 
\begin{align}
{\wha{\Theta}_{\rm GSS}^{\nabla}}
=&{U_{\hat{I}}{}^{\hat{M}}}\Xi_{\hat{M}}{\wha{P}'^{\hat{I}}}+\frac{1}{3!}{F_{\hat{I}\hat{J}\hat{K}}\wha{P}'^{\hat{I}}\wha{P}'^{\hat{J}}\wha{P}'^{\hat{K}}}\cr
&+\frac{1}{4}{U_{\hat{I}}{}^{\hat{M}}}\Gamma_{\hat{M}\hat{N}}{}^{\hat{P}}(Q'^{\hat{N}}+\eta^{\hat{N}\hat{Q}}P'_{\hat{Q}})(P'_{\hat{P}}-\eta_{\hat{P}\hat{R}}Q'^{\hat{R}}){\wha{P}'^{\hat{I}}}-\frac{1}{3!}{F_{\hat{I}\hat{J}\hat{K}}\wha{Q}'^{\hat{I}}\wha{Q}'^{\hat{J}}\wha{P}'^{\hat{K}}}.\label{GSSHamiltonianOnDFTG}
\end{align}

The difference between the Hamiltonean function \eqref{GSSTwistedHamiltonian} and the one given above \eqref{GSSHamiltonianOnDFTG} is due to the fact that the former one was not written in a covariant form, while the latter one is covariant. Explicitly, we have
 
\begin{align}
\wha{\Theta}_{\rm GSS}^{\nabla}-\wha{\Theta}_{\rm GSS}=&\frac{1}{2}\Big(\frac{1}{2}U_{\hat{I}}{}^{\hat{P}}\Gamma_{\hat{P}\hat{M}\hat{N}}+\wtil{\Omega}_{\hat{I}\hat{J}\hat{K}}U^{\hat{J}}{}_{\hat{M}}U^{\hat{K}}{}_{\hat{N}}\Big)P'^{\hat{M}}P'^{\hat{N}}\wha{P}'^{\hat{I}}\cr
&-\frac{1}{4}U_{\hat{I}}{}^{\hat{M}}\Gamma_{\hat{M}\hat{N}\hat{P}}Q'^{\hat{N}}Q'^{\hat{P}}\wha{P}'^{\hat{I}}-\frac{1}{3!}F_{\hat{I}\hat{J}\hat{K}}\wha{Q}'^{\hat{I}}\wha{Q}'^{\hat{J}}\wha{P}'^{\hat{K}}. 
\end{align}
However, these terms do not affect the generalized Lie derivative and we obtain the same formula for the generalized flux from the connection in covariantized formulation.

\section*{Acknowledgments}
The authors would like to thank G. Aldazabal, P. Bouwknegt, and K. Yoshida
for stimulating discussions and lectures.
We also acknowledge F. Hassler for useful comments.  
The authors also would like to acknowledge H. Muraki, M. A. Heller, Y. Kaneko, S. Takezawa, K. Miura and S. Sekiya for valuable discussions. 
N.I.\ would like to thank Branislav Jur\v{c}o and the Charles University in Prague for the permission to stay as a visiting scientist and their warm hospitality.

\appendix

\section{Notation\label{Notations}}
\renewcommand{\theequation}{A.\arabic{equation} }
\setcounter{equation}{0}
In this appendix, we summarize the notations. 
We list up indices and generalized vectors in each frame in table \ref{table1}. 
We also show the corresponding canonical transformations. 
\begin{center}
\begin{table}[ht]
\includegraphics[width=17cm]{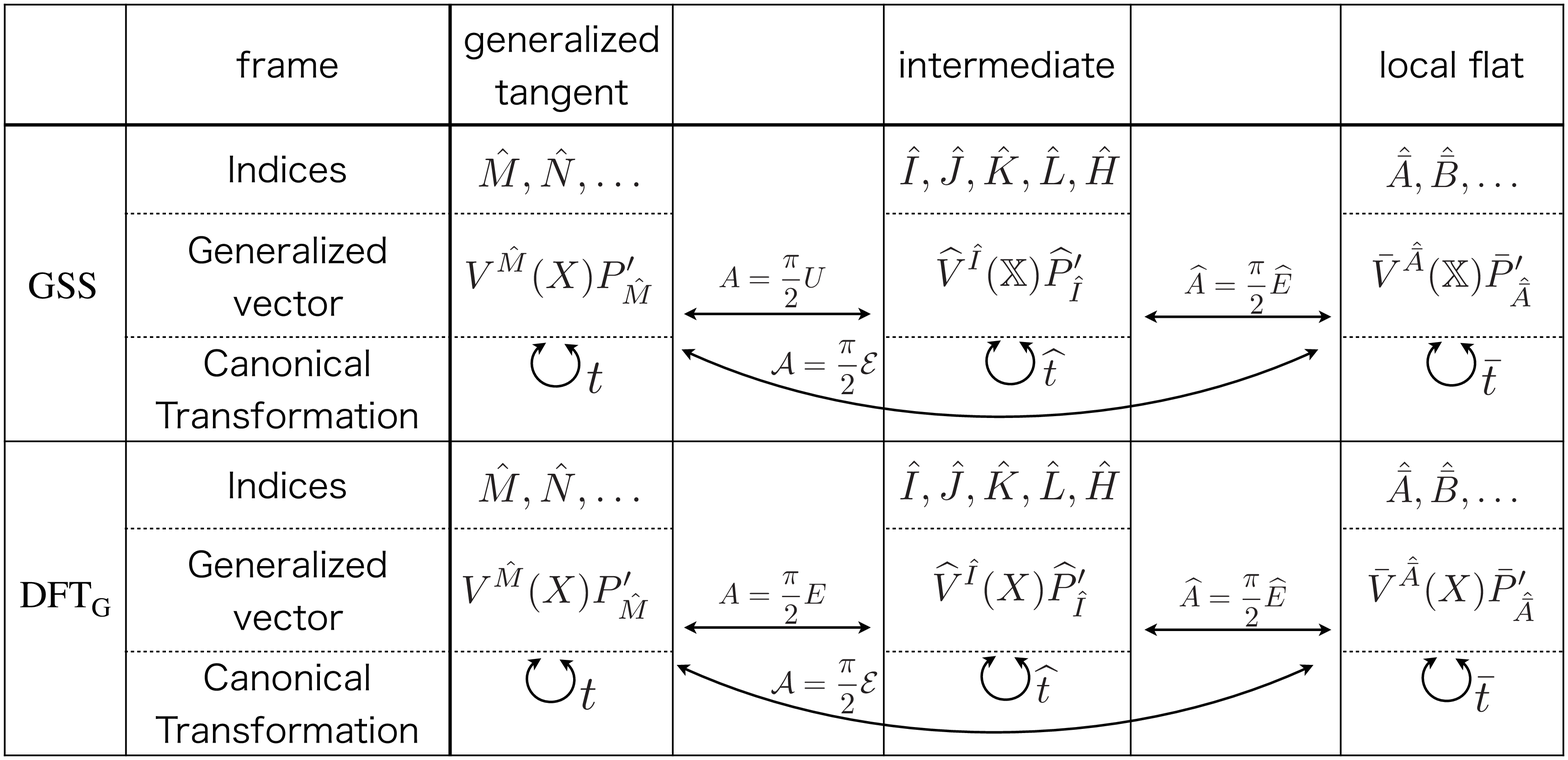}
\caption{Notations}
\label{table1}
\end{table}
\end{center}

We use indices $\hat{M},\hat{N}{}, \dots =1,2,\dots, 2D$ for generalized tangent frame, $\hat{I},\hat{J},\hat{K},\hat{L},\hat{H}=1,2,\dots, 2D$ for intermediate frame, and $\hat{\bar{A}},\hat{\bar{B}},\dots=1,2,\dots, 2D$ for local flat frame. 
We use common indices in GSS and DFT on the group manifold (DFT$_{\rm G}$). 

The generalized vectors on the pre-QP-manifold are denoted by $V^{\hat{M}}P'_{\hat{M}}, \wha{V}^{\hat{I}}\wha{P}'_{\hat{I}}$ and $\bar{V}^{\hat{\bar{A}}}\bar{P}'_{\hat{\bar{A}}}$ in general tangent frame, intermediate frame and local flat frame, respectively. 
In GSS, generalized vectors depend only on the external space coordinate $\mathbb{X}$ in the intermediate frame and the local flat frame. 
On the other hand, DFT in the covariantized case does not restrict the coordinate dependence of generalized vectors in each frame, in general. 

The arrows with the canonical transformation functions note the relations of each frame by the corresponding canonical transformations.

\section{Canonical transformation of $\Theta_{0}$ by $\theta {\cal E}$ and pre-Bianchi identity}
\renewcommand{\theequation}{B.\arabic{equation}}
\setcounter{equation}{0}
In section \ref{CanonicaTransfByE}, we discussed the canonical transformation by the degree 2 function ${\cal E}$ in \eqref{CanonicalTransfFunc}. In order to see the correspondence to original DFT structures, we restricted the transformation function ${\cal E}_{\hat{\bar{A}}}{}^{\hat{M}}$ to an element of $O(D,D)$ and discussed the canonical transformation $e^{\frac{\pi}{2}\delta_{\cal E}}$. For general $\theta$ case, we obtain the twisted Hamiltonian function in following form. 
\begin{align}
e^{\theta\delta_{\cal E}}\Theta_{0}=& \eta^{\hat{M}\hat{N}}\Xi_{\hat{M}} (P'_{\hat{N}}\cos\theta+{\cal E}^{\hat{\bar{D}}}{}_{\hat{N}}\bar{P}'_{\hat{\bar{D}}}\sin\theta)\nonumber\\
&+\Big[\Big(-\frac{1}{2} \eta^{\hat{M}\hat{N}} \eta^{\hat{\bar{A}}\hat{\bar{B}}}\partial_{\hat{M}}{\cal E}_{\hat{\bar{A}}}{}^{\hat{S}}{\cal E}_{\hat{\bar{B}}}{}^{\hat{R}}\cos\theta \sin^{2}\theta\Big) P'_{\hat{R}} P'_{\hat{S}} P'_{\hat{N}}\nonumber\\
&+ \Big(-\eta^{\hat{M}\hat{S}}\partial_{\hat{M}}E_{\hat{\bar{A}}}^{\ \hat{T}}\eta^{\hat{\bar{A}}[\hat{\bar{B}}}{\cal E}^{\hat{\bar{D}}]}{}_{\hat{T}}{\cal E}_{\hat{\bar{B}}}{}^{\hat{R}}\sin\theta\cos^{2}\theta -\frac{1}{2} \eta^{\hat{M}\hat{N}}\partial_{\hat{M}}{\cal E}_{\hat{\bar{A}}}{}^{\hat{S}}\eta^{\hat{\bar{A}}\hat{\bar{B}}}{\cal E}_{\hat{\bar{B}}}{}^{\hat{R}} {\cal E}^{\hat{\bar{D}}}{}_{\hat{N}} \sin^{3}\theta\Big) P'_{\hat{R}} P'_{\hat{S}}\bar{P}'_{\hat{\bar{D}}}\nonumber\\
& +\Big(-\frac{1}{2} \eta^{\hat{M}\hat{R}}\partial_{\hat{M}}{\cal E}_{\hat{\bar{A}}}{}^{\hat{T}}\eta^{\hat{\bar{A}}\hat{\bar{D}}}{\cal E}^{\hat{\bar{C}}}{}_{\hat{T}}  + \eta^{\hat{M}\hat{N}}\partial_{\hat{M}}{\cal E}_{\hat{\bar{A}}}{}^{\hat{T}}\eta^{\hat{\bar{A}}[\hat{\bar{B}}}{\cal E}^{\hat{\bar{C}}]}{}_{\hat{T}} {\cal E}_{\hat{\bar{B}}}{}^{\hat{R}} {\cal E}^{\hat{\bar{D}}}{}_{\hat{N}} \Big)P'_{\hat{R}}\bar{P}'_{\hat{\bar{C}}}\bar{P}'_{\hat{\bar{D}}}\sin^{2}\theta\cos\theta \nonumber\\
& +\Big(\frac{1}{2}\eta^{\hat{M}\hat{N}} \eta^{\hat{\bar{A}}\hat{\bar{B}}}\partial_{\hat{M}}{\cal E}_{\hat{\bar{A}}}{}^{\hat{T}}{\cal E}^{\hat{\bar{C}}}{}_{\hat{T}} {\cal E}^{\hat{\bar{D}}}{}_{\hat{N}}\sin^{3}\theta\Big) \bar{P}'_{\hat{\bar{B}}} \bar{P}'_{\hat{\bar{C}}}\bar{P}'_{\hat{\bar{D}}}\Big].
\end{align}

From a Hamiltonian function 
\begin{align}
\Theta_{F}=&\bar{\rho}^{\ \hat{M}}_{\hat{\bar{A}}}\Xi_{\hat{M}}\bar{P}'^{\hat{\bar{A}}}+\rho_{\hat{N}}^{\ \hat{M}}\Xi_{\hat{M}}P'^{\hat{N}}+\frac{1}{3!}{\cal F}_{\hat{\bar{A}}\hat{\bar{B}}\hat{\bar{C}}}\bar{P}'^{\hat{\bar{A}}} \bar{P}'^{\hat{\bar{B}}}\bar{P}'^{\hat{\bar{C}}}+\frac{1}{2}\Phi_{\hat{\bar{C}}\hat{M}\hat{N}}P'^{\hat{M}} P'^{\hat{N}}\bar{P}'^{\hat{\bar{C}}}\nonumber\\
&+\frac{1}{2}\Delta_{\hat{\bar{A}}\hat{\bar{B}}\hat{M}}P'^{\hat{M}}\bar{P}'^{\hat{\bar{A}}}\bar{P}'^{\hat{\bar{B}}}+\frac{1}{3!}\Psi_{\hat{M}\hat{N}\hat{P}}P'^{\hat{M}}P'^{\hat{N}}P'^{\hat{P}}
\end{align}
we obtain the pre-Bianchi identity from the following function ${\cal B}(\Theta_{F},\Theta_{0},e^{\theta\delta_{{\cal E}}})$. 
\begin{align}
{\cal B}&(\Theta_{F},\Theta_{0},e^{\theta\delta_{{\cal E}}})\cr
=&(\bar{\rho}_{\hat{\bar{A}}}{}^{\hat{M}}\bar{\rho}^{\hat{\bar{A}}\hat{N}}+\rho_{\hat{P}}{}^{\hat{M}}\rho^{\hat{P}\hat{N}}-\eta^{\hat{M}\hat{N}})\Xi_{\hat{M}}\Xi_{\hat{N}}\cr
&+(2\partial_{\hat{N}}\bar{\rho}_{\hat{\bar{A}}}{}^{\hat{M}}\bar{\rho}_{\hat{\bar{B}}}{}^{\hat{N}}+\bar{\rho}^{\hat{\bar{C}}\hat{M}}{\cal F}_{\hat{\bar{C}}\hat{\bar{A}}\hat{\bar{B}}}+\rho^{\hat{N}\hat{M}}\Delta_{\hat{\bar{A}}\hat{\bar{B}}\hat{N}}-\Omega^{\hat{M}}{}_{\hat{R}\hat{S}}{\cal E}_{\hat{\bar{A}}}{}^{\hat{R}}{\cal E}_{\hat{\bar{B}}}{}^{\hat{S}}\sin^{2}\theta)\Xi_{\hat{M}}\bar{P}'^{\hat{\bar{A}}}\bar{P}'^{\hat{\bar{B}}}\cr
&+(2\partial_{Q}\rho_{\hat{N}}{}^{\hat{M}}\rho_{\hat{P}}{}^{\hat{Q}}+\bar{\rho}_{\hat{\bar{A}}}{}^{\hat{M}}\Phi^{\hat{\bar{A}}}{}_{\hat{N}\hat{P}}+\rho^{\hat{R}\hat{M}}\Psi_{\hat{R}\hat{N}\hat{P}}+\Omega^{\hat{M}}{}_{\hat{N}\hat{P}}\sin^{2}\theta)\Xi_{\hat{M}}P'^{\hat{N}} P'^{\hat{P}}\cr
&-2(\bar{\rho}^{\hat{\bar{B}}\hat{M}}\Delta_{\hat{\bar{B}}\hat{\bar{A}}\hat{N}}+\partial_{\hat{N}}\bar{\rho}{}^{\hat{P}}_{\hat{\bar{A}}}\rho_{\hat{P}}{}^{\hat{M}}-\partial_{\hat{P}}\rho_{\hat{N}}{}^{\hat{M}}\bar{\rho}_{\hat{\bar{A}}}{}^{\hat{P}}-\rho^{\hat{P}\hat{M}}\Phi_{\hat{\bar{A}}\hat{P}\hat{N}}+{\cal E}_{\hat{\bar{A}}}{}^{\hat{S}}\Omega^{\hat{M}}{}_{\hat{N}\hat{S}}\sin\theta\cos\theta)\Xi_{\hat{M}}P'^{\hat{N}}\bar{P}'^{\hat{\bar{A}}}\cr
&+\Big(-\frac{2}{3!}\bar{\rho}{}^{\hat{M}}_{\hat{\bar{A}}}\partial_{\hat{M}}{\cal F}_{\hat{\bar{B}}\hat{\bar{C}}\hat{\bar{D}}}+\frac{3}{4}{\cal F}^{\hat{\bar{E}}}{}_{\hat{\bar{A}}\hat{\bar{B}}}{\cal F}_{\hat{\bar{E}}\hat{\bar{C}}\hat{\bar{D}}}+\frac{1}{4}\Delta_{\hat{\bar{A}}\hat{\bar{B}}\hat{M}}\Delta_{\hat{\bar{C}}\hat{\bar{D}}}{}^{\hat{M}}-\frac{1}{4}\Omega_{\hat{\bar{E}}\hat{\bar{A}}\hat{\bar{B}}}\Omega^{\hat{\bar{E}}}{}_{\hat{\bar{C}}\hat{\bar{D}}}\sin^{4}\theta\Big)\bar{P}'^{\hat{\bar{A}}}\bar{P}'^{\hat{\bar{B}}}\bar{P}'^{\hat{\bar{C}}}\bar{P}'^{\hat{\bar{D}}}\cr
&+\Big(-\bar{\rho}_{\hat{\bar{A}}}{}^{\hat{P}}\partial_{\hat{P}}\Phi_{\hat{\bar{B}}\hat{M}\hat{N}}+\frac{1}{2}{\cal F}_{\hat{\bar{A}}\hat{\bar{B}}\hat{\bar{C}}}\Phi^{\hat{\bar{C}}}{}_{\hat{M}\hat{N}}-\rho_{\hat{M}}{}^{\hat{P}}\partial_{\hat{P}}\Delta_{\hat{\bar{A}}\hat{\bar{B}}\hat{N}}-\Phi_{\hat{\bar{A}}}{}^{\hat{P}}{}_{\hat{M}} \Phi_{\hat{\bar{B}}\hat{P}\hat{N}}-\Delta_{\hat{\bar{C}}\hat{\bar{A}}\hat{M}}\Delta^{\hat{\bar{C}}}{}_{\hat{\bar{B}}\hat{N}}\cr
&~~~~~+\frac{1}{2}\Delta_{\hat{\bar{A}}\hat{\bar{B}}}{}^{\hat{P}}\Psi_{\hat{P}\hat{M}\hat{N}}-\frac{1}{2}\Omega_{\hat{P}\hat{M}\hat{N}}\Omega^{\hat{P}}{}_{\hat{\bar{A}}\hat{\bar{B}}}\sin^{4}\theta+\Omega^{\hat{P}}{}_{\hat{M}\hat{\bar{A}}}\Omega_{\hat{P}\hat{N}\hat{\bar{B}}} \sin^{2}\theta\cos^{2}\theta\Big)P'^{\hat{M}}P'^{\hat{N}}\bar{P}'^{\hat{\bar{A}}}\bar{P}'^{\hat{\bar{B}}}\cr
&+\Big(\bar{\rho}_{\hat{\bar{A}}}{}^{\hat{N}}\partial_{\hat{N}}\Delta_{\hat{\bar{B}}\hat{\bar{C}}\hat{M}}-\frac{3}{2}{\cal F}^{\hat{\bar{D}}}{}_{\hat{\bar{A}}\hat{\bar{B}}} \Delta_{\hat{\bar{D}}\hat{\bar{C}}\hat{M}}\cr
&~~~~~+\Phi_{\hat{\bar{A}}\hat{N}\hat{M}}\Delta_{\hat{\bar{B}}\hat{\bar{C}}}{}^{\hat{N}}-\frac{2}{3!}\rho_{\hat{M}}^{\ \hat{N}}\partial_{\hat{N}}{\cal F}_{\hat{\bar{A}}\hat{\bar{B}}\hat{\bar{C}}}-\Omega^{\hat{N}}{}_{\hat{M}\hat{\bar{C}}} \Omega_{\hat{N}\hat{\bar{A}}\hat{\bar{B}}}\sin^{3}\theta\cos\theta\Big)P'^{\hat{M}}\bar{P}'^{\hat{\bar{A}}}\bar{P}'^{\hat{\bar{B}}}\bar{P}'^{\hat{\bar{C}}}\cr
&+\Big(\frac{2}{3!}\bar{\rho}_{\hat{\bar{A}}}{}^{\hat{Q}}\partial_{\hat{Q}}\Psi_{\hat{M}\hat{N}\hat{P}}-\Phi^{\hat{\bar{B}}}{}_{\hat{M}\hat{N}}\Delta_{\hat{\bar{B}}\hat{\bar{A}}\hat{P}}\cr
&~~~~~+\frac{3}{2}\Phi_{\hat{\bar{A}}\hat{Q}\hat{M}}\Psi^{\hat{Q}}{}_{\hat{N}\hat{P}}-\rho_{\hat{M}}{}^{\hat{Q}}\partial_{\hat{Q}}\Phi_{\hat{\bar{A}}\hat{N}\hat{P}}+\Omega^{\hat{R}}{}_{\hat{M}\hat{N}}\Omega_{\hat{R}\hat{P}\hat{\bar{A}}}\sin^{3}\theta\cos\theta\Big)P'^{\hat{M}}P'^{\hat{N}}P'^{\hat{P}}\bar{P}'^{\hat{\bar{A}}}\cr
&+\Big(\frac{1}{4}\Phi_{\hat{\bar{A}}\hat{M}\hat{N}}\Phi^{\hat{\bar{A}}}{}_{\hat{P}\hat{Q}}+\frac{3}{4}\Psi_{\hat{R}\hat{M}\hat{N}}\Psi^{\hat{R}}{}_{\hat{P}\hat{Q}}-\frac{2}{3!}\rho_{\hat{M}}{}^{\hat{R}}\partial_{\hat{R}}\Psi_{\hat{N}\hat{P}\hat{Q}}-\frac{1}{4}\Omega_{\hat{R}\hat{M}\hat{N}}\Omega^{\hat{R}}{}_{\hat{P}\hat{Q}}\sin^{4}\theta\Big)P'^{\hat{M}}P'^{\hat{N}}P'^{\hat{P}}P'^{\hat{Q}}.\cr
\end{align}

\section{Classical master equation for $\bar{\Theta}_{0}^{\nabla}$}
\renewcommand{\theequation}{C.\arabic{equation}}
\setcounter{equation}{0}
We discussed the weak master equation for $\wha{\Theta}_{0}^{\nabla}$ in section \ref{covQstructures}. It is based on the calculation of $\{\wha{\Theta}_{0}^{\nabla},\wha{\Theta}_{0}^{\nabla}\}$. Here, we show the full form of $\{\wha{\Theta}_{0}^{\nabla},\wha{\Theta}_{0}^{\nabla}\}$. 
\begin{align}
\{\wha{\Theta}_{0}^{\nabla},\wha{\Theta}_{0}^{\nabla}\}=&\eta^{\hat{M}\hat{N}}\Xi_{\hat{M}}\Xi_{\hat{N}}\nonumber\\
&+\Big(-2\Omega_{\hat{I}\hat{J}\hat{K}}+2W_{\hat{I}\hat{J}\hat{K}}+W_{\hat{K}\hat{I}\hat{J}}\Big)E^{\hat{K}\hat{M}}\Xi_{\hat{M}}\wha{P}'^{\hat{I}}\wha{P}'^{\hat{J}}\nonumber\\
&+\Gamma^{\hat{M}}{}_{\hat{Q}\hat{P}}\Xi_{\hat{M}}(Q'^{\hat{Q}}P'^{\hat{P}}+P'^{\hat{Q}}P'^{\hat{P}}-Q'^{\hat{Q}}Q'^{\hat{P}}-P'^{\hat{Q}}Q'^{\hat{P}})-W^{\hat{M}}{}_{\hat{I}\hat{J}}\Xi_{\hat{M}}\wha{Q}'^{\hat{I}}\wha{Q}'^{\hat{J}}\nonumber\\
&-\Big[W^{\hat{H}}{}_{\hat{I}\hat{J}}\Big(\Omega_{\hat{K}\hat{L}\hat{H}}-W_{\hat{K}\hat{L}\hat{H}}\Big)+\frac{1}{2}R_{\hat{I}\hat{J}\hat{K}\hat{L}}-\frac{1}{4}W_{\hat{M}\hat{I}\hat{J}}W^{\hat{M}}{}_{\hat{K}\hat{L}}\Big]\bar{P}'^{\hat{I}}\bar{P}'^{\hat{J}}\bar{P}'^{\hat{K}}\bar{P}'^{\hat{L}}\nonumber\\
&+\Big[-W^{\hat{H}}{}_{\hat{K}\hat{L}}\Big(\Omega_{\hat{I}\hat{J}\hat{H}}+W_{\hat{I}\hat{H}\hat{J}}\Big)+\frac{1}{2}R_{\hat{I}\hat{J}\hat{K}\hat{L}}+\frac{1}{2}W_{\hat{M}\hat{I}\hat{J}}W^{\hat{M}}{}_{\hat{K}\hat{L}}\Big]\wha{P}'^{\hat{I}}\wha{P}'^{\hat{J}}\wha{Q}'^{\hat{K}}\wha{Q}'^{\hat{L}}\nonumber\\
&+\frac{1}{2}\Big[-2\Big(\Omega_{\hat{I}\hat{J}\hat{H}}+W_{\hat{I}\hat{H}\hat{J}}\Big)H^{\hat{H}\hat{N}}\Gamma_{\hat{N}\hat{M}\hat{N}}+R_{\hat{I}\hat{J}\hat{M}\hat{N}}+\Gamma_{\hat{P}\hat{M}\hat{N}}W^{\hat{P}}{}_{\hat{I}\hat{J}}\Big]\nonumber\\
&\hspace{5mm}\times\wha{P}'^{\hat{I}}\wha{P}'^{\hat{J}}(P'^{\hat{M}}-Q'^{\hat{M}})(Q'^{\hat{N}}+P'^{\hat{N}})\nonumber\\
&+\frac{1}{4}W_{\hat{M}\hat{I}\hat{J}}W^{\hat{M}}{}_{\hat{K}\hat{L}}\wha{Q}'^{\hat{I}}\wha{Q}'^{\hat{J}}\wha{Q}'^{\hat{K}}\wha{Q}'^{\hat{L}}\nonumber\\
&+\frac{1}{2}\Gamma_{\hat{P}\hat{N}\hat{M}}W^{\hat{P}}{}_{\hat{I}\hat{J}}\wha{Q}'^{\hat{I}}\wha{Q}'^{\hat{J}}(P'^{\hat{M}}-Q'^{\hat{M}})(Q'^{\hat{N}}+P'^{\hat{N}})\nonumber\\
&+\frac{1}{4}\Gamma_{\hat{R}\hat{P}\hat{N}}\Gamma^{\hat{R}}{}_{\hat{Q}\hat{M}}(P'^{\hat{M}}-Q'^{\hat{M}})(P'^{\hat{N}}-Q'^{\hat{N}})(Q'^{\hat{P}}+P'^{\hat{P}})(Q'^{\hat{Q}}+P'^{\hat{Q}}).\label{ClassicalMasterEquationInGeneral}
\end{align}

\section{Double Field Theory on group manifolds\label{DFTonGroupMfd}}
\renewcommand{\theequation}{D.\arabic{equation}}
\setcounter{equation}{0}
Let us briefly recall the DFT on group manifold defined in ref.\cite{Bosque:2015jda}.

\subsection{Background vielbein, covariant derivative and fluctuation}
We consider a $2D$-dimensional group manifold $\wha{G}$ and introduce local coordinates $X^{\hat{M}}=(\tilde{x}_{M}, x^{M})$ and a doubled metric $\eta^{\hat{M}\hat{N}}$ where
 $\hat{M},\hat{N}, \cdots$ are $GL(2D)$ indices on a curved space 
of $\wha{G}$.  We also consider the local $O(D,D)$ frame by
 introducing the generalized vielbein $E_{\hat{A}}{}^{\hat{M}}$ with $O(D,D)$ invariant metric 
  $\eta^{\hat{I}\hat{J}}$, which relate to the doubled metric as
\begin{align}
\eta^{\hat{M}\hat{N}} = \eta^{\hat{I}\hat{J}}
E_{\hat{I}}{}^{\hat{M}}E_{\hat{J}}{}^{\hat{N}}~,\label{CurvedODDmetric}
\end{align}
where, $\hat{I},\hat{K}, \cdots$ are indices of the local $O(D,D)$ frame.
The generalized vielbein $E_{\hat{I}}{}^{\hat{M}}$ takes values in $GL(2D)$ 
in general. The inverse of each metric is denoted by $\eta_{\hat M\hat N}$ and $\eta_{\hat I\hat J}$. In the following we raise or lower the curved indices and the local $O(D,D)$ indices by the corresponding metric.

The local basis of the tangent space $T\wha{G}$ is denoted by
\begin{align}
\partial_{\hat{M}}=\begin{pmatrix}\tilde{\partial}^{\hat{M}}\\\partial_{\hat{M}}\end{pmatrix}.
\end{align}
Then, the vector field defined by the generalized vielbein
\begin{align}
D_{\hat{I}}=E_{\hat{I}}{}^{\hat{M}}\partial_{\hat{M}}, 
\end{align}
satisfies the commutation relation
\begin{align}
[D_{\hat{I}},D_{\hat{J}}]=F_{\hat{I}\hat{J}}{}^{\hat{K}}D_{\hat{K}}
\label{commutationrelationofgroupmanifold}
\end{align}
where 
\begin{equation}
F_{\hat{I}\hat{J}\hat{K}} = 2 \Omega_{[\hat{I}\hat{J}]\hat{K}},
\end{equation}
with 
\begin{equation}
\Omega_{\hat{I}\hat{J}\hat{K}} = D_{\hat{I}} E_{\hat{J}}{}^{\hat{M}} E_{\hat{K}\hat{M}},
\label{WeitzenGP}
\end{equation}
as in the standard DFT.
As a background vielbein, we take the left invariant vector field and thus, 
 $F_{\hat{I}\hat{J}}{}^{\hat{K}}$ becomes the structure constant of the Lie algebra of $\wha{G}$. 

We denote the 
 affine connection by $\Gamma$ and the spin connection by $W$
 and define the covariant derivative of a tensor $V_{\hat J}{}^{\hat M}$ by
\begin{align}
\nabla_{\hat{I}}V_{\hat{J}}{}^{\hat{M}}=&D_{\hat{I}}V_{\hat{J}}{}^{\hat{M}}
+E_{\hat{I}}{}^{\hat{P}}\Gamma_{\hat{P}\hat{N}}{}^{\hat{M}}V_{\hat{J}}{}^{\hat{N}}-W_{\hat{I}\hat{J}}{}^{\hat{K}}V_{\hat{K}}{}^{\hat{M}}
\end{align}
We impose that the covariant derivative of the generalized vielbein and the $O(D,D)$ metric vanish:
\begin{align}
&\nabla_{\hat{I}}E_{\hat{J}}{}^{\hat{M}}=0.
\label{VielbeinCompatibility}
\\
&\nabla_{\hat{I}}\eta_{\hat{J}\hat{K}}=0.\label{FlatEtaCompatibility}
\end{align}
The equation \eqref{VielbeinCompatibility}
gives the following relation between the affine connection and spin connection:
\begin{align}
E^{\hat{K}}_{\ \hat{M}}W_{\hat{K}\hat{I}}^{\ \ \ \hat{J}}=E_{\ \hat{M}}^{\hat{K}}\Omega_{\hat{K}\hat{I}}^{\ \ \ \hat{J}}+\Gamma_{\hat{M}\hat{P}}^{\ \ \ \ \hat{N}}E_{\hat{I}}^{\ \hat{P}}E_{\ \hat{N}}^{\hat{J}},
\end{align}
where $\Omega$ is defined in \eqref{WeitzenGP}. The equation \eqref{FlatEtaCompatibility} gives the following 
condition on the spin connection:
\begin{align}
\nabla_{\hat{I}}\eta_{\hat{J}\hat{K}}=&W_{\hat{I}\hat{J}}{}^{\hat{L}}\eta_{\hat{L}\hat{K}}+W_{\hat{I}\hat{K}}{}^{\hat{L}}\eta_{\hat{J}\hat{L}}=0.
\end{align}
Note that the covariant derivative of $\eta_{\hat{M}\hat{N}}$ automatically vanishes,
\begin{align}
\nabla_{\hat{M}}\eta_{\hat{N}\hat{P}}	=& \partial_{\hat{M}}\eta_{\hat{N}\hat{P}}-\Gamma_{\hat{M}\hat{N}}^{\ \ \ \ \hat{Q}}\eta_{\hat{Q}\hat{P}}-\Gamma_{\hat{M}\hat{P}}^{\ \ \ \ \hat{Q}}\eta_{\hat{N}\hat{Q}}=0.
\end{align}

In the following, 
we consider the case where the total vielbein ${\cal E}_{\hat{\bar{A}}}^{\ \hat{M}}$ can split into a fluctuating part $\wha{E}_{\hat{\bar{A}}}{}^{\hat{I}}$ and a background vielbein $E_{\hat{A}}{}^{\hat{M}}$ as \cite{Bosque:2015jda},
\begin{align}
{\cal E}_{\hat{\bar{A}}}^{\ \hat{M}}=\wha{E}_{\hat{\bar{A}}}{}^{\hat{I}}E_{\hat{I}}{}^{\hat{M}}, \quad {\cal E}^{\hat{\bar{A}}}{}_{\hat{M}}=\wha{E}^{\hat{\bar{A}}}{}_{\hat{I}}E^{\hat{I}}{}_{\hat{M}}. 
\end{align}
Here $\hat{\bar{A}}, \hat{\bar{B}}, \cdots$ are indicies of the double local flat  $O(1,D-1) \times O(1, D-1)$ vector. 
In DFT on the group manifold, we consider as the dynamical fields,   
 the fluctuation vielbein $\wha{E}_{\hat{\bar{A}}}{}^{\hat{I}}$ and the fluctuation part of generalized dilation $\wha{d}$ . 
The standard DFT,
 corresponds to the trivial background vielbein, i.e., $E_{\hat{I}}{}^{\hat{M}}=\delta_{\hat{I}}{}^{\hat{M}}, E^{\hat{I}}{}_{\hat{M}}=\delta^{\hat{I}}{}_{\hat{M}}$ and the total vielbein becomes the original DFT vielbein. 
 
We also split the total dilaton ${\tilde d}$ into the background part and the fluctuation part as 
\begin{align}
{\tilde d}= d +\wha{d}. 
\end{align}
The background metric $H_{\hat{I}\hat{J}}$ is written as
\begin{align}
H_{\hat{M}\hat{N}}=E_{\hat{M}}{}^{\hat{I}}S_{\hat{I}\hat{J}}E_{\hat{N}}{}^{\hat{J}}.
\end{align}
where 
\begin{align}
S_{\hat{I}\hat{J}}=\begin{pmatrix}\tilde{s}^{IJ}&0\\0&s_{IJ}\end{pmatrix}.
\end{align}
The fluctuation metric ${\cal H}_{\hat{I}\hat{J}}$ is written by fluctuation vielbeine,
\begin{align}
{\cal H}_{\hat{I}\hat{J}}=\wha{E}^{\hat{\bar{A}}}{}_{\hat{I}}S_{\hat{\bar{A}}\hat{\bar{B}}}\wha{E}^{\hat{\bar{B}}}{}_{\hat{J}}.
\end{align}

\subsection{Covariant generalized Lie derivative and generalized flux}
In DFT on group manifold, the following generalized Lie derivative and fluxes written with the covariant derivative
 are 
introduced. 
A covariant generalized Lie derivative on a generalized vector field $\wha{V}$ is
\begin{align}
{\cal L}_{\wha{\Lambda}}^{\nabla}\wha{V}^{\hat{I}}=&\wha{\Lambda}^{\hat{J}}\nabla_{\hat{J}}\wha{V}^{\hat{I}}+(\nabla^{\hat{I}}\wha{\Lambda}_{\hat{J}}-\nabla_{\hat{J}}\wha{\Lambda}^{\hat{I}})\wha{V}^{\hat{J}},\label{CovGeneralizedLieDerivative}
\end{align}
where the gauge parameter $\wha{\Lambda}$ is a generalized vector. 
The gauge transformation of the fields is generated by this generalized Lie derivative.

Then, the covariant generalized fluxes are introduced by using the above generalized Lie derivative (\ref{CovGeneralizedLieDerivative}), the total vielbein ${\cal E}_{\hat{\bar{A}}}{}^{\hat{M}}$ and the total dilaton $\tilde{d}$ as
\begin{align}
{\cal F}_{\hat{\bar{A}}\hat{\bar{B}}\hat{\bar{C}}}=&\braket{{\cal E}_{\hat{\bar{A}}}, {\cal L}^{\nabla}_{{\cal E}_{\hat{\bar{B}}}}{\cal E}_{\hat{\bar{C}}}}=\wtil{F}_{\hat{\bar{A}}\hat{\bar{B}}\hat{\bar{C}}}+F_{\hat{\bar{A}}\hat{\bar{B}}\hat{\bar{C}}},\label{GeneralizedFluxOnGroupMFD}\\
{\cal F}_{\hat{\bar{A}}}=&-e^{2\tilde{d}}{\cal L}_{{\cal E}_{\hat{\bar{A}}}}e^{-2\tilde{d}}=2D_{\hat{\bar{A}}}\tilde{d}+\wtil{\Omega}^{\hat{\bar{B}}}{}_{\hat{\bar{B}}\hat{\bar{A}}}\label{DilatonFluxOnGroupMFD}
\end{align}
where $\wtil{F}_{\hat{\bar{A}}\hat{\bar{B}}\hat{\bar{C}}}:=3\wtil{\Omega}_{[\hat{\bar{A}}\hat{\bar{B}}\hat{\bar{C}}]}$ with $\wtil{\Omega}_{\hat{\bar{A}}\hat{\bar{B}}\hat{\bar{C}}}:={\cal E}_{\hat{\bar{B}}}^{\ \hat{M}}\partial_{\hat{M}}\wha{E}_{\hat{\bar{C}}}{}^{\hat{I}}\wha{E}_{\hat{\bar{A}}\hat{I}}$ and $F_{\hat{\bar{A}}\hat{\bar{B}}\hat{\bar{C}}}:=\wha{E}_{\hat{\bar{A}}}{}^{\hat{I}}\wha{E}_{\hat{\bar{B}}}{}^{\hat{J}}\wha{E}_{\hat{\bar{C}}}{}^{\hat{K}}F_{\hat{I}\hat{J}\hat{K}}$. 

The action of DFT$_{\rm WZW}$ is rewritten using the generalized flux \cite{Bosque:2015jda}. 
\begin{align}
S=\int d^{2D}Xe^{-2\tilde{d}}\Big(&{\cal F}_{\hat{\bar{A}}}{\cal F}_{\hat{\bar{B}}}S^{\hat{\bar{A}}\hat{\bar{B}}}+\frac{1}{4}{\cal F}_{\hat{\bar{A}}\hat{\bar{C}}\hat{\bar{D}}}{\cal F}_{\hat{\bar{B}}}{}^{\hat{\bar{C}}\hat{\bar{D}}}S^{\hat{\bar{A}}\hat{\bar{B}}}-\frac{1}{12}{\cal F}_{\hat{\bar{A}}\hat{\bar{B}}\hat{\bar{C}}}{\cal F}_{\hat{\bar{A}}\hat{\bar{B}}\hat{\bar{C}}}S^{\hat{\bar{A}}\hat{\bar{D}}}S^{\hat{\bar{B}}\hat{\bar{E}}}S^{\hat{\bar{C}}\hat{\bar{F}}}\nonumber\\
&-\frac{1}{6}{\cal F}_{\hat{\bar{A}}\hat{\bar{B}}\hat{\bar{C}}}{\cal F}^{\hat{\bar{A}}\hat{\bar{B}}\hat{\bar{C}}}-{\cal F}_{\hat{\bar{A}}}{\cal F}^{\hat{\bar{A}}}\Big).\label{ActionOfDFTonGroupMFD}
\end{align}

The DFT on group manifold has been developed as DFT$_{\rm WZW}$ which is considered as a double field formalism of conformal string field theory (CSFT) for the Wess-Zumino-Witten (WZW) model \cite{Blumenhagen:2014gva}. 

From now on we take the background group manifold as the direct product of two Lie groups $\wha{G}=\wtil{G}\times G$ and the $O(D,D)$ metric $\eta_{\hat{I}\hat{J}}$ as 
\begin{align}
\eta_{\hat{I}\hat{J}}=\begin{pmatrix}
\tilde{\kappa}^{IJ}&0\\0&-\kappa_{IJ},
\end{pmatrix}
\end{align}
where $\tilde{\kappa}^{IJ}$ and $\kappa_{IJ}$ are the Killing metrics on ${\rm Lie}(\wtil{G})$ and ${\rm Lie}(G)$, respectively. Then, the structure constant of the background group manifold is written by the structure constants of ${\rm Lie}(\wtil{G})$ and ${\rm Lie}(G)$ as
\begin{align}
F_{\hat{I}\hat{J}}{}^{\hat{K}}=\begin{cases}\tilde{f}^{IJ}{}_{K}\\f_{IJ}{}^{K}\\0\quad ({\rm otherwise})\end{cases}
\end{align}
and therefore, $F_{\hat{I}\hat{J}\hat{K}}$ is totally antisymmetric. 
In DFT$_{\rm WZW}$, the generalized Lie derivative contains the Lie bracket of the background group manifold \cite{Blumenhagen:2014gva}. 
\begin{align}
{\cal L}_{\Lambda}\wha{V}^{\hat{I}}=\wha{\Lambda}^{\hat{J}}D_{\hat{J}}\wha{V}^{\hat{I}}+(D^{\hat{I}}\wha{\Lambda}_{\hat{J}}-D_{\hat{J}}\wha{\Lambda}^{\hat{I}})\wha{V}^{\hat{J}}+F^{\hat{I}}{}_{\hat{J}\hat{K}}\wha{\Lambda}^{\hat{J}}\wha{V}^{\hat{K}}. \label{GeneralizedLieDerOfDFTwzw}
\end{align}
We assume that the covariant generalized Lie derivative \eqref{CovGeneralizedLieDerivative} reproduces the generalized Lie derivative of DFT$_{\rm WZW}$ \eqref{GeneralizedLieDerOfDFTwzw}. Then, the spin connection is written by the structure constant as follows. 
\begin{align}
W_{\hat{I}\hat{J}}{}^{\hat{K}}=\frac{1}{3}F_{\hat{I}\hat{J}}{}^{\hat{K}}\label{DFTwzwSpinConnection01}. 
\end{align}

In the following, we discuss the closure condition of the generalized Lie derivative \eqref{GeneralizedLieDerOfDFTwzw} which takes the form
\begin{align}
[{\cal L}_{\wha{\Lambda}_{1}},{\cal L}_{\wha{\Lambda}_{2}}]\wha{\Lambda}_{3}^{\hat{I}}-{\cal L}_{{\cal L}_{\wha{\Lambda}_{1}}\wha{\Lambda}_{2}}\wha{\Lambda}^{\hat{I}}_{3}=&F^{\hat{L}}{}_{[\hat{J}\hat{K}}F^{\hat{I}}{}_{\hat{H}]\hat{L}}\wha{\Lambda}_{1}^{\hat{H}}\wha{\Lambda}_{2}^{\hat{J}}\wha{\Lambda}_{3}^{\hat{K}}-L^{\hat{L}}\wha{\Lambda}_{1\hat{J}}\wha{\Lambda}_{2}^{\hat{J}}L_{\hat{L}}\wha{\Lambda}_{3}^{\hat{I}}-2L_{\hat{L}}\wha{\Lambda}^{\hat{I}}_{[1}L^{\hat{L}}\wha{\Lambda}_{2]\hat{J}}\wha{\Lambda}_{3}^{\hat{J}}.
\end{align}
The right hand side vanishes and the closure condition is satisfied with the section condition and the Jacobi identity of $F_{\hat{A}\hat{B}\hat{C}}$. 
\begin{align}
D_{\hat{I}}\cdot D^{\hat{I}}\cdot =0, \quad F^{\hat{L}}{}_{[\hat{J}\hat{K}}F^{\hat{I}}{}_{\hat{H}]\hat{L}}=0. \label{CCforDFTwzw}
\end{align}

The generalized fluxes and action are defined in the same form as \eqref{GeneralizedFluxOnGroupMFD}, \eqref{DilatonFluxOnGroupMFD} and \eqref{ActionOfDFTonGroupMFD}.

\bibliographystyle{utphys}
\bibliography{Masterbib}

\end{document}